\documentclass[12pt,twoside,openright]{book}
\usepackage[pdftex]{graphicx}
\usepackage[a4paper, tmargin=3.5cm, bmargin=2.5cm, lmargin=3.75cm, rmargin=2.5cm, twoside]{geometry}
\usepackage{latexsym,amsmath,amssymb,amsfonts}             
\usepackage{rotating}                    
\usepackage{natbib}          
\usepackage{soul}                                         

\usepackage{fancyhdr}                    
\usepackage{setspace}
\usepackage[english]{babel}
\doublespace          
\usepackage{tocloft}
\usepackage{lscape}

\usepackage[pdfpagelabels=true,plainpages=false,colorlinks=true,linkcolor=black,citecolor=black,urlcolor=blue]{hyperref}
\usepackage[small,hang]{caption2}
\usepackage[compact]{titlesec}
\numberwithin{equation}{section}
\usepackage[T1]{fontenc}                    
\makeatletter
\def\cleardoublepage{\clearpage\if@twoside \ifodd\c@page\else%
    \hbox{}%
    \thispagestyle{empty}
    \newpage%
    \if@twocolumn\hbox{}\newpage\fi\fi\fi}
\makeatother
\clearpage{\pagestyle{plain}\cleardoublepage}
\newcommand{\be}{\begin{equation}}
\newcommand{\ee}{\end{equation}}
\newcommand{\bea}{\begin{eqnarray}}
\newcommand{\eea}{\end{eqnarray}}
\newcommand{\ba}{\begin{array}}
\newcommand{\ea}{\end{array}}
\newcommand{\bi}{\begin{itemize}}
\newcommand{\ei}{\end{itemize}}
\newcommand{\bc}{\begin{center}}
\newcommand{\ec}{\end{center}}
\newcommand{\bfr}{\begin{flushright}}
\newcommand{\efr}{\end{flushright}}

\begin{document}
\thispagestyle{empty}

{ \renewcommand{\baselinestretch}{1.5}
\begin{center}
\begin{spacing}{2}
\noindent{\Large \bf HIGH-HARMONIC SPECTROSCOPY OF TWO-DIMENSIONAL MATERIALS}
\end{spacing}
\end{center}

\vspace{18cm}
\begin{flushright}
{ {\LARGE\em  \textbf{Mrudul M S} ~~}}
\end{flushright}}

\cleardoublepage
\frontmatter
\newpage
\thispagestyle{empty}

\begin{titlepage}
\begin{center}
{
%


\textbf{\Large HIGH-HARMONIC SPECTROSCOPY OF TWO-DIMENSIONAL MATERIALS} \vskip1.0cm 
{\large\emph{Submitted in partial fulfillment of the requirements}} \vskip 0.03cm 
{\large\emph{of the degree of}}
\singlespacing 
\textbf {\large Doctor of Philosophy}\\
\singlespacing
 {\large\emph{by}}\\
\singlespacing \textbf{\large Mrudul M S} \vskip0.2cm
\vskip1cm
{\large Supervisor:}\\
\singlespacing \textbf{\large{Prof. Gopal Dixit}} \vskip1cm

\includegraphics[height=42mm]{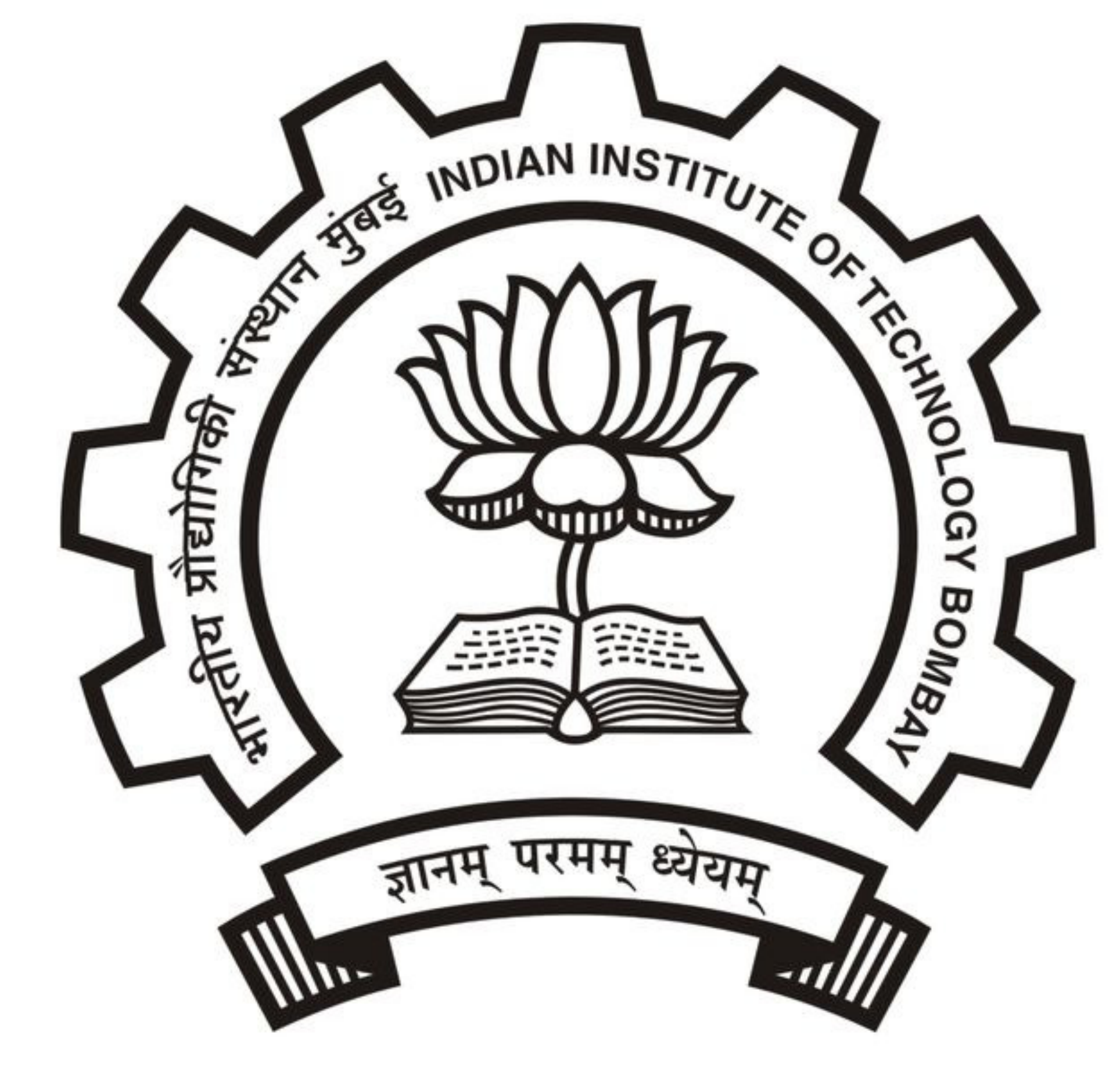}
\doublespacing
\begin{center}
\textbf{\large{DEPARTMENT OF PHYSICS\\
INDIAN INSTITUTE OF TECHNOLOGY BOMBAY\\\singlespacing
2021}}\\\singlespacing \copyright~2021 Mrudul M S. All rights reserved.
\end{center}

\fboxsep6mm
\fboxrule1.3pt
}
\end{center}
\end{titlepage}
\pagestyle{fancy}

\cleardoublepage
\newpage
\thispagestyle{empty}
\setlength{\baselineskip}{32pt}
\bigskip
\bigskip
\bigskip
\bigskip
\bigskip
\vspace*{5cm}
\begin{center}
\vspace*{0.5cm}
{\Large  \bf {Dedicated to my parents, my brother, and my wife.} } \\
\end{center}
%

\setlength{\baselineskip}{18pt}

\cleardoublepage \addcontentsline{toc}{chapter}{Thesis Approval}
\newpage
\thispagestyle{empty}
\vspace*{2cm}
\begin{center}
	\textbf{\Large\underline {Thesis Approval}}
\end{center}
\vspace{1cm}
This thesis entitled \textbf{High-harmonic Spectroscopy of Two-dimensional Materials} by \textbf{Mrudul M. S.} (Roll No. 164120007) is approved for the degree of Doctor of Philosophy in Physics.
\vskip2cm
\begin{flushright}
	\textbf{Examiners}\\
	\vspace{0.5in}
	Prof. Kamal P. Singh\\
	\vspace{0.5in}
	Prof. Sumiran Pujari\\
	\vskip1cm
	\textbf{Supervisor}\\
	\vspace{0.5in}
	Prof. Gopal Dixit
	\vskip1cm
	\textbf{Chairman}\\
	\vspace{0.5in}
	Prof. Anindya Dutta\\

\end{flushright}

\begin{flushleft}
	Date : 27/10/2021\\
	Place : Mumbai
\end{flushleft}

\cleardoublepage
\addcontentsline{toc}{chapter}{Declaration}
\newpage
\thispagestyle{empty}
 \vspace*{2cm}
\begin{center}
\textbf{\LARGE{Declaration}}
\end{center}
\vspace{1cm}
\begin{spacing}{1.3}
I declare that this written submission represents my ideas in my own words and where
others' ideas or words have been included, I have adequately cited and referenced the original
sources. I also declare that I have adhered to all principles of academic honesty and integrity
and have not misrepresented or fabricated or falsified any idea/data/fact/source in my
submission. I understand that any violation of the above will be cause for disciplinary action
by the Institute and can also evoke penal action from the sources which have thus not been
properly cited or from whom proper permission has not been taken when needed.
\vspace{2.5cm}
\begin{flushright}
Mrudul M. S. \\
(Roll No. 164120007)
\end{flushright}
\begin{flushleft}
	Date: 27/10/2021
\end{flushleft}
\end{spacing} 
\cleardoublepage
\addcontentsline{toc}{chapter}{Acknowledgements}
\newpage
\thispagestyle{empty}
\begin{center}
\vspace*{-0.4cm}
{\LARGE {\textbf{Acknowledgements}}}
\end{center}

{\setlength{\baselineskip}{8pt} \setlength{\parskip}{2pt}
\begin{spacing}{1.5}

Foremost, I would like to express my sincere gratitude to my supervisor, \textbf{Prof. Gopal Dixit}, Department of Physics, IITB, for the motivation and continuous support throughout my doctoral research. I extend my heartfelt thanks to him for introducing me into the field of ultrafast optics and supporting my research ideas. Especially, I am thankful to him for all the opportunities I have received, and the skills I have developed under his supervision during my PhD.

I want to express my sincere thanks to my RPC members, \textbf{Prof.  B. P. Singh} and \textbf{Prof.  Sumiran Pujari} of the Department of Physics, IITB, for their valuable comments and insightful questions, which helped to shape my research.

I want to express my honest gratitude to \textbf{Prof. Angel Rubio} and \textbf{Dr. Nicolas Tancogne-Dejean} of Max Planck Institute for the Structure and Dynamics of Matter, Hamburg, for inviting me as a guest researcher with financial support. My visit to Hamburg was a great learning experience. It was a wonderful opportunity to work with a well-informed researcher like Dr Nicolas Tancogne-Dejean, which helped to improve my computational skills. I sincerely thank them for their continuous support and feedbacks while working on the project and drafting the manuscript.

I want to express my sincere gratitude to \textbf{Prof. Misha Ivanov} and \textbf{Dr. \'{A}lvaro Jim\'{e}nez Gal\'{e}n} of Max Born Institute for Nonlinear Optics and Short Pulse Spectroscopy, Berlin, for their support.
I have been always fascinated by the scientific intuitions of Prof. Misha; thank you for being an inspiration to all of us. During the collaborative works carried out with them, I could improve my scientific knowledge and writing skills greatly.

I want to express my special gratitude to my teacher, \textbf{Prof. K. M. Ajith} of NITK, Surathkal, for the endless encouragement and support, without which I could have never chosen a research career.

I want to thank my labmates \textbf{Irfana N. Ansari}, \textbf{Adhip Pattanayak}, \textbf{Sucharita Giri}, \textbf{Navdeep Rana}, and \textbf{Amar Bharti} for their constant support and cooperation. Working with such cool labmates in such stressful times has been a relief. Also, I want to thank all of my friends inside and outside IITB. Furthermore, I would like to record my thanks to all the staff members of IITB for making the life inside IITB much easier.  

Last but not least, I would like to thank my family; my parents \textbf{G. Muraleedharan} and \textbf{B. Shylaja}, my brother \textbf{Midhun  M. S.}, and my wife, \textbf{Aarathy A. R.}, for their continuous encouragement and support.  
\vskip2cm
%
%

\begin{flushright}
Mrudul M S\\
Department of Physics\\
IITB
\end{flushright}
\begin{flushleft}
	Date: 27/10/2021\\
\end{flushleft}
\end{spacing}

\cleardoublepage
\addcontentsline{toc}{chapter}{Abstract}
\newpage
\thispagestyle{empty}
\begin{center}
\vspace*{-0.4cm}
{\LARGE {\textbf{Abstract}}}
\end{center}

{\setlength{\baselineskip}{8pt} \setlength{\parskip}{2pt}
\begin{spacing}{1.5}
Recent advancements in the generation of mid-infrared and terahertz laser pulses have enabled us to observe strong-field driven non-perturbative high-harmonic generation (HHG) from semiconductors, dielectrics, and semimetals. 
HHG has added another dimension to time-resolved ultrafast electron dynamics in materials with unprecedented temporal resolution. Present thesis discusses how HHG is an emerging method to probe static and dynamical properties in two-dimensional materials. In this thesis, two-dimensional materials with hexagonal symmetry are studied.

We have demonstrated that the high-harmonic spectrum encodes the fingerprints of electronic band structure and interband coupling between different bands.  
Furthermore, by analysing gapped and gapless graphene, we show how electron dynamics in a semimetal and a semiconductor are different as 
the harmonic spectrum depends differently on the polarisation of the driving laser. 
To explore the role of defects in HHG, spin-polarised vacancy defects in  hexagonal boron nitride are 
considered. It has been found that electron-electron interaction is crucial for electron dynamics in a defected solid. In all cases, we present how different symmetries of the lattice can be extracted from the harmonic spectrum. Finally, a light-driven method is proposed for observing valley-polarisation in pristine graphene, using a tailored laser pulse.  Also, a recipe is discussed to write and read valley-selective electron excitations in materials with zero bandgap and zero Berry curvature. 

\vspace{0.5in}

\end{spacing}

\cleardoublepage
\renewcommand{\contentsname}{Contents}  
\begin{spacing}{1.2}  
\tableofcontents
\cleardoublepage
\addcontentsline{toc}{chapter}{List of Figures} 
\listoffigures
\cleardoublepage
\addcontentsline{toc}{chapter}{List of Symbols and Abbreviations} 
\markboth{List of Symbols}{List of Symbols and Abbreviations}  
\chapter*{List of Symbols and Abbreviations}
\noindent {\bf Symbols}
\begin{tabbing}
aaaaaaaaaaaa \= abababababab \kill
$\mathcal{A}$\> Vector potential\\
$\mathcal{F}$\> Electric field\\
$\chi^{(n)}$\> $n^{\textrm {th}}$-order susceptibility tensor \\
$\hat{\mathcal{H}}$\> Hamiltonian \\
$\hat{\mathcal{H}}_{0}$\> Field-free Hamiltonian \\
$\hat{\mathcal{H}}^{\prime}(t)$\> Time-dependent Hamiltonian for light-matter interaction\\
$\psi$ \> Wavefunction \\
$U_{p}$\> Pondermotive energy\\
$\gamma_{k}$\> Keldysh parameter\\
$\omega$ \> Frequency \\
$\lambda$ \> Wavelength \\
$\rho_{mn}$ \> Density matrix elements \\
$\textbf{d}_{mn}$ \> Dipole matrix elements \\
$\textbf{p}_{mn}$ \> Momentum matrix elements \\
$T_2$ \> Dephasing time \\
$j(t)$ \> Total current \\
$\mathcal{I}(\omega)$ \> Intensity of HHG \\
$\eta$ \> Valley asymmetry parameter \\
\end{tabbing}
\noindent {\bf Abbreviations}
 \begin{tabbing}
aaaaaaaaaaaa \= abababababab  \kill
au \> atomic unit  \\
HHG \> High-harmonic Generation\\
2D \> two-dimensional \\
TDSE \> Time-dependent Schr\"odinger equation\\
SBE \> Semiconductor Bloch Equations\\
TDDFT \> Time-dependent density functional theory\\
h-BN \> hexagonal boron nitride \\
$V_B$ \> h-BN with boron vacancy \\
$V_N$ \> h-BN with nitrogen vacancy \\
\end{tabbing}
%

\cleardoublepage

\end{spacing}
\mainmatter
\begin{spacing}{1.5}
\chapter{Introduction}\label{Chaper1}

\section{Ultrafast Spectroscopy}
	How light interacts with matter shapes our understanding about 
	its  static and dynamical properties. 
	All  phenomena in matter, invisible to our bare eyes, can be observed using spectroscopy, 
	where light is used to interrogate the properties.  
	The invention of the first working laser in 1960 revolutionised the field of spectroscopy. 
	This is primarily due to the precise control and tunability of  laser parameters such as 
	frequency and intensity. X-ray diffraction, Raman spectroscopy, 
	infrared, ultraviolet, and visible absorption 
	spectroscopies are some of the conventional spectroscopic techniques for exploring the static 
	properties of electrons and phonons in matter.	
	
	 Interaction of light with matter can be classified as linear and nonlinear depending on the 
	 response of bound electrons to the applied electric field of the laser. When the laser field is 
	 weak compared to the characteristic field strength of the medium,  the amplitude of the 
	 electron's displacement follows the strength of the applied field linearly. 
        Following the Lorentz oscillator model, an atom can be effectively modelled as a simple 
        harmonic oscillator. In other words, when light interacts with matter, the polarisation 
        (dipole moment per unit volume) induced by the externally applied field is proportional to the 
        strength of the applied field and can be expressed as 
        $\mathcal{P}(t)$ = $\epsilon_0\chi^{(1)}\mathcal{F}(t)$.  
        Here, $\epsilon_0$ is the permittivity of the free-space and $\chi^{(1)}$ is the linear (first-order) 
 susceptibility tensor. This is the linear regime of light-matter interaction. 
The oscillator model is applicable only when the strength of an applied field is weak. 
Electrons start showing nonlinear behaviour when the externally applied field has strength comparable to the characteristic field strength. 
	
	To get a quantitative idea about  ``weak'' and ``strong'' fields, 
	we consider the characteristic electric-field strength corresponds to an electron in the 
	Bohr orbital of hydrogen, which is defined as
	\begin{equation}
		\mathcal{F}_{at} = \left(\frac{e}{4\pi \epsilon_0 a_B^2}\right).
	\end{equation}
	Here, $a_B$ is the Bohr radius and $-e$ is the electronic charge. The value 
	of $\mathcal{F}_{at}$ is 5.14$\times$10$^{9}$ V/cm, and the corresponding intensity is I$_{at}$ = 3.5 $\times$ 10$^{16}$ W/cm$^2$~\citep{boyd2020nonlinear}. These are also  
	the definitions of an atomic unit of electric field and intensity, respectively.  
	
	In contrast to ordinary light, coherent laser light is strong enough to alter the material's properties 
	significantly. When the intensity of the laser is substantially strong, then the electron's position 
	no longer follows linearly to the laser field, resulting in nonlinear optical phenomena. Depending 
	on the intensity of the laser, nonlinear optics can be classified as perturbative and 
	non-perturbative.  
	Let us write time-dependent Hamiltonian as 
	$\hat{\mathcal{H}}(t) = \hat{\mathcal{H}}_0 + \hat{\mathcal{H}'}(t)$, where 
	$\hat{\mathcal{H}}_0$ is the field-free Hamiltonian, and 
	$\hat{\mathcal{H}'}(t)$ is  time-dependent Hamiltonian for light-matter interaction. 
	As long as the strength of $\hat{\mathcal{H}'}(t)$ is very small in comparison to 
	 $\hat{\mathcal{H}}_0$,   the interaction can be treated perturbatively. 
	 Consequently,  any nonlinear optical phenomenon within perturbation theory 
	 can be understood by describing the polarisation as a power series expansion in 
	 terms of the applied electric field amplitude. 
	 In this case, the polarisation along $i$-direction is written as
	\begin{equation}\label{eq:power_series}
		\mathcal{P}_{i} = \epsilon_0 \left[ \sum_{j} \chi^{(1)}_{ij}\mathcal{F}_j + \sum_{j,k}\chi^{(2)}_{ijk}\mathcal{F}_j\mathcal{F}_k + \sum_{j,k,l}\chi^{(3)}_{ijkl}\mathcal{F}_j\mathcal{F}_k\mathcal{F}_l + \dots\right] .   
	\end{equation}
	Here, $\chi^{(n)}$ is the $n^{\rm th}$-order susceptibility tensor.

	The research area of nonlinear optics emerged soon after the invention of the first working 
	laser~\citep{maiman1960stimulated}. In 1961, Franken {\it et al.} demonstrated  
	doubling of laser frequency using a quartz crystal experiemnetally~\citep{paul2001observation}. 
	This phenomenon is termed as second-harmonic generation, a second-order nonlinear optical 
	effect. Subsequently, other perturbative nonlinear optical effects are reported, such as 
	sum-frequency generation,  difference-frequency generation, optical parametric amplification, 
	third-harmonic generation, to list a few.

        There is no assurance that the power series expansion in Eq.~(\ref{eq:power_series}) gets 
         converged. One such instance is when the laser-field intensity is comparable to the 
         characteristic intensity of the medium, and there is a possibility of photoionisation. This is a 
         non-perturbative limit of nonlinear optics where $\hat{\mathcal{H}'}(t) \gg  \hat{\mathcal{H}}_0$. 
         Such relative high intensity of the laser with ultrashort pulses are routinely generated these days in  various laboratories  across the globe. 
         
Scientists were keen on developing pulsed mode ultrafast lasers in pursuit of developing lasers with high peak powers. For a laser with particular pulse energy, the peak intensity is maximum when the pulse width is short. A laser is an ideal tool for generating such reproducible short pulses due to their temporal coherence. Experimentalists realised this method more economical to achieve desired intensities rather than increasing the pulse energy. The word ultrashort often referred to pulses shorter than picoseconds, where ultrafast optics deals with the generation, characterisation, and applications of ultrashort pulses. Over the years, scientists were able to develop laser pulses as short as few femtoseconds (1 fs = 10$^{-15}$s), and the wavelength of these laser pulses ranges from the ultraviolet, visible to the infrared part of the electromagnetic spectrum. 

In molecules, the rotational dynamics occur in the picosecond timescale, 
whereas the natural timescale of molecular dynamics, vibrational dynamics, and chemical reactions 
range from hundreds of femtoseconds to a few femtoseconds. 
Therefore, ultrashort femtosecond laser is an ideal tool for probing dynamical properties of molecules and chemical reactions in a time-resolved manner, forming the backbone of femtochemistry.

Due to the non-perturbative nature of the interaction, 
intense ultrafast pulses interact with matter substantially differently. 
There are many fascinating nonlinear optical phenomena reported 
over the years by following intense light-matter interaction. 
Some of these include above-threshold ionisation observed by Agostini and co-workers 
in 1979~\citep{agostini1979free} and double ionisation of electrons observed by Walker and co-workers 
 in 1994~\citep{walker1994precision}. Both of these phenomena are associated with photoionisation in the strong laser field. 

Among other strong-field driven processes, 
we are particularly interested in high-harmonic generation (HHG), 
observed by Ferray {\it et al.} in 1988~\citep{ferray1988multiple}. 
When an ultrashort laser interacts with matter, high-order harmonics of the incident photon energy are generated. Harmonic orders up to 33$\rm ^{rd}$ were observed from different inert gas systems using a laser of intensity 10$^{13}$ W/cm$^2$. Subsequently, scientists were able to generate harmonics up to few hundreds of orders when inert gas atoms are excited with ultrashort lasers~\citep{chang1997generation,l1993high}.

Probing electron dynamics in matter on characteristic timescale is an emerging research topic.  
Within Bohr model of atom, an electron in hydrogen takes 152 attoseconds (1 as = 10$^{-18}$ s) to complete a revolution around the orbital. 
Undoubtedly, time-resolved spectroscopy of electron dynamics requires  
laser pulses of attosecond durations. 
For this purpose, ultrafast sources in the extreme ultraviolet  and x-ray regimes should be developed.
Recent breakthroughs in ultrafast optics have shown that attosecond pulses can be generated either from a free-electron laser sources or from a table-top setup using HHG. 
Two decades ago, two independent research groups of Pierre Agostini and Ferenc Krausz 
demonstrated  the generation of attosecond  
pulses, in extreme ultraviolet energy regime, using HHG~\citep{paul2001observation,hentschel2001attosecond}. 
In recent years,  HHG becomes a powerful method to probe electron dynamics in systems 
as diverse as atoms, molecules~\citep{lein2007molecular,baker2006probing,smirnova2009high,shafir2012resolving,bruner2016multidimensional,worner2010following}  and solids ~\citep{kruchinin2018colloquium, ghimire2011observation, schubert2014sub, langer2016lightwave, lakhotia2020laser, you2017anisotropic, pattanayak2019direct, luu2015extreme, lanin2017mapping, vampa2015all, silva2018high,  bauer2018high, reimann2018subcycle, garg2016multi,lakhotia2020laser}. 
With these motivations, let us understand the theory of HHG in atoms and solids in detail in the following sections.

\section{High-Harmonic Generation in Gases}

A typical high-order harmonic spectrum consists of  three distinctive regimes as presented in Fig.~\ref{fig1.1}. 
In the perturbative regime,  intensity of harmonics  reduces consistently, and 
this behaviour is compatible with the perturbative nonlinear interaction 
where subsequent higher-order harmonics are exponentially weaker in comparison to the 
previous-order harmonics. The non-perturbative nature of nonlinear interaction manifests in the plateau regime, where nearby harmonics have comparable intensity.

The first accepted theoretical model for HHG from atoms was proposed by Corkum in 1993~\citep{corkum1993plasma}. This is also known as the semiclassical three-step model, which will be discussed in detail in the following subsection. 
Certainly, in strong-field regime, a different kind of  theory is needed where  
field-free potential acts like a perturbation. A year after the semiclassical model of Corkum, Lewenstein {\it et al.} provided  quantum mechanical model of HHG based on strong-field approximation~\citep{lewenstein1994theory}. 

\begin{figure}[t!]
	\centering
	\includegraphics[width=0.8\linewidth]{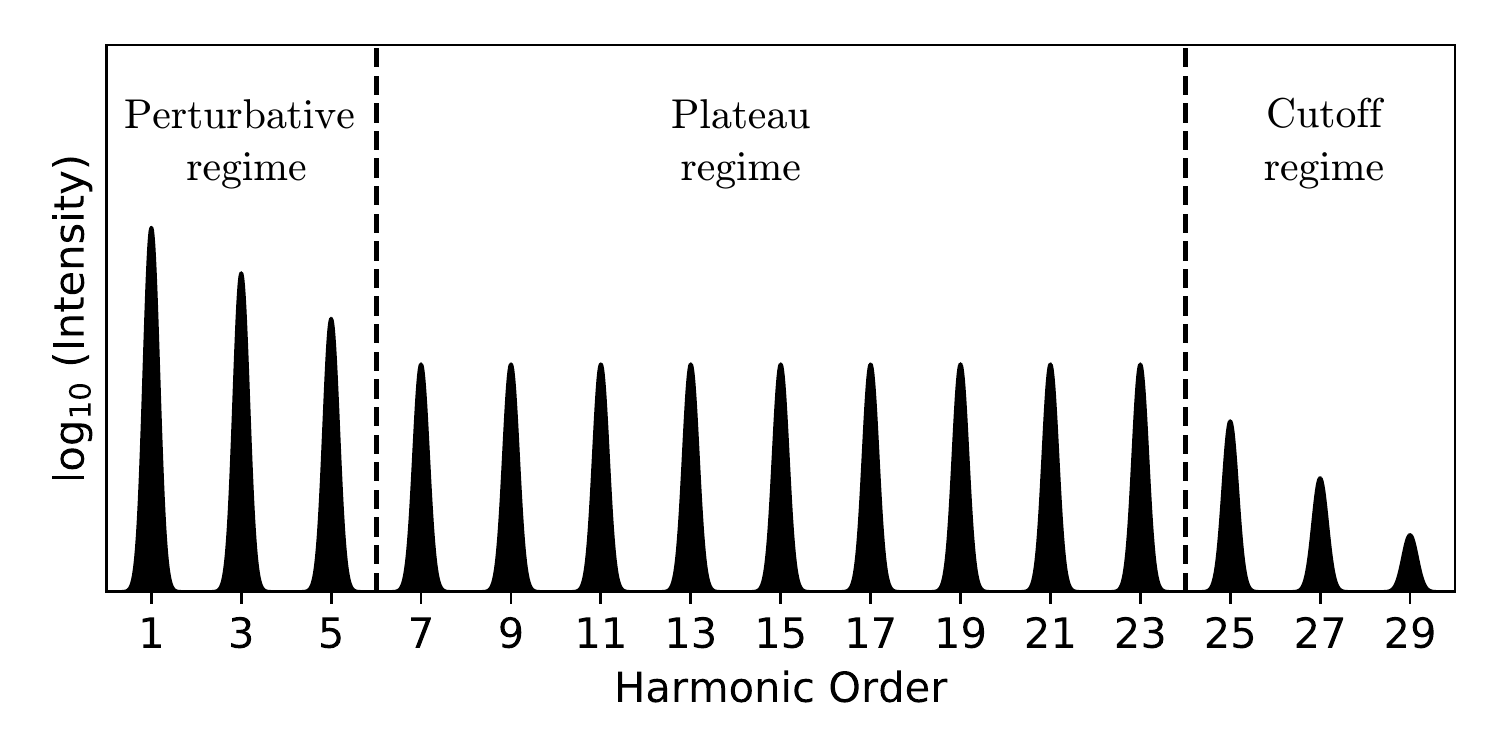}
	\caption{A typical high-order harmonic spectrum.} \label{fig1.1}
\end{figure}

\subsection{Semiclassical Three-Step Model}
An illustration of Corkum's semiclassical three-step model is presented in Fig.~\ref{fig1.2}. 
The following events can occur in the given order for a bound electron in the presence of a strong oscillating laser field. 

\begin{enumerate}
	\item An intense laser field distorts the Coulombic potential so that the bound electron can tunnel out from the lowered potential barrier.
	\item The tunneled electron, initially created with zero kinetic energy, gets accelerated by the influence of the external laser field.
	\item When the direction of the laser field reverses, the ionised electron can return and 
	recollide with the parent ion. This recollision can lead to recombination with parent ion, 
	and releases a photon of energy equal to the total energy of the electron.
\end{enumerate}
Here, the second step can be described using the classical equation of motion. Whereas, the 
first and the third steps require quantum mechanical treatment.  
In the three-step model, a single-active electron is assumed, and 
the interaction between electrons is ignored.  
Let us review  strong-field interaction in detail.

\begin{figure}[t!]
	\centering
	\includegraphics[width=\linewidth]{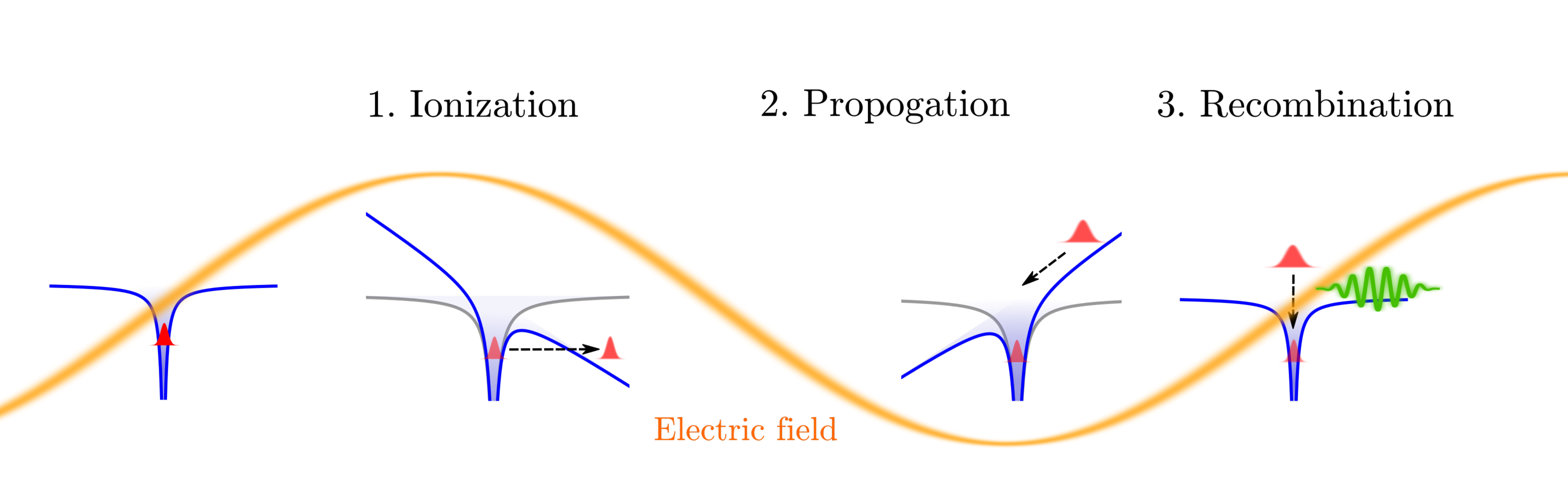}
	\caption{Schematic diagram of the three-step model for high-order harmonic generation in atoms.} \label{fig1.2}
\end{figure}

A typical feature of strong-field interaction mainly originates from the field's capability of ionisation. There are two ways in which a bound electron can be ionised, either by tunnel ionisation or a transition to continuum via absorption of one or more photons (single or multi-photon ionisation). 
The mechanism of ionisation is determined by the properties of both the laser and the matter. 
A general  approach to identify the dominant mechanism is based on 
the dimensionless parameter introduced by Keldysh:   
Keldysh parameter $\gamma_{_K} = \sqrt{I_p/2 U_p}$, where  
$I_p$ is the ionisation potential, and $U_p$ is the 
cycle-averaged kinetic energy of a free electron in the presence of a laser field, termed as pondermotive energy. 
For a laser with central frequency $\omega$, and  electric field amplitude $\mathcal{F}_0$, 
the total time-dependent electric field is defined as $\mathcal{F}(t) = \mathcal{F}_0\cos(\omega t)$. A free electron in this field experience a force which can be equated to $m\ddot{x}(t)=-e\mathcal{F}_0\cos(\omega t)$. The average kinetic energy (pondermotive energy) of a free electron can be calculated from this equation of motion and is equal to
 \begin{equation}
	U_p = \frac{1}{2}m\left\langle\dot{x}(t)^2\right\rangle = \frac{e^2\mathcal{F}_0^2}{4m\omega^2}.
\end{equation}

$\gamma_{_K}$ is also defined as the ratio between the tunnelling time of an electron and the laser period.
Here, tunnelling time is defined as the time taken by the electron to cross the barrier classically. 
When the time period of laser is longer, an electron sees a quasi-static potential barrier through which the electron can tunnel ionise. 
In terms of $\gamma_K$, it is when $\gamma_{_K} \ll  1$, which is valid for intense, long-wavelength laser pulses. On the other hand, multi-photon ionisation is the dominant mechanism  when 
$\gamma_{_K} \gg  1$.  
For getting an  idea about the value of the Keldysh parameter, 
let us consider a laser of intensity 10$^{14}$ W/cm$^2$ and wavelength 800 nm interacting with 
hydrogen atom, the corresponding value of $\gamma_{_K}$ is approximately unity~\citep{grossmann2018theoretical}.

	Once the electron is freed from the Coulombic potential, it is dominantly influenced by the strong electric-field of the laser. 
	In this interaction regime, understanding the motion of a free electron in the presence of the laser field is important. 
	At intensities higher than 10$^{18}$ W/cm$^2$, a free electron can acquire velocity comparable with the speed of light. For describing such physics, we need to resort to 
	the relativistic regime of nonlinear optics. Laser sources that can generate pulses as intense as 10$^{20}$ W/cm$^2$ have been developed till now. 
	Typically HHG is observed for laser intensity within the range of 10$^{13}$-10$^{16}$ W/cm$^2$. Therefore, it is safer to use classical equations of motion for an electron in the laser field. 
	Note that the quantum diffusion of electron wavepacket while propagating in the continuum is neglected here.

\begin{figure}[t!]
	\centering
	\includegraphics[width=\linewidth]{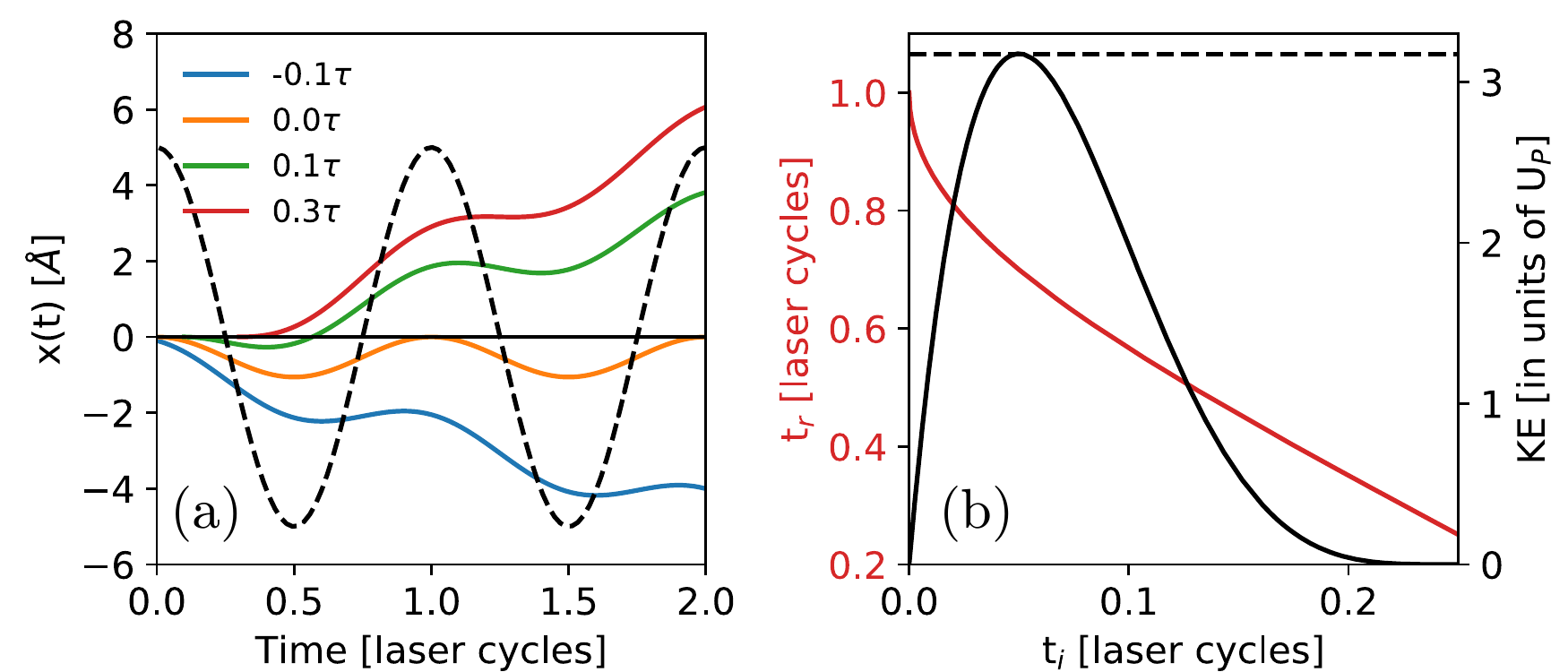}
	\caption{(a) Electron trajectories for an electron born at different times during the laser field. 
	Here, $\tau$ is the period of the laser field and $t_i$ is the time of ionisation. The dashed line shows the electric field. 
	(b) Kinetic energy of the electron at the time of recombination $t_r$ (black line) 
	as a function of time of ionisation $t_i$. The dashed vertical line is the maximum energy equal to 3.17$U_p$. The time of recombination as a function of time of ionisation (red line).} \label{fig1.3}
\end{figure}

Consider an electron bound to the nucleus at $x$ = 0. Assuming the electron is tunnel ionised at time $t_i$ with zero initial velocity [$\dot{x}(t_i)$ = 0], the position of electron at any time can be obtained by solving the equation of motion as
\begin{equation}\label{eq:traj}
	x(t) = \frac{e\mathcal{F}_0}{m \omega ^2} \left[\cos(\omega t) - \cos(\omega t_i) + \omega\sin(\omega t_i)(t-t_i)\right].
\end{equation}
Electron trajectories obtained from Eq.~(\ref{eq:traj}), for different values of ionisation times 
$t_i$ are shown in Fig.~\ref{fig1.3}(a). 
It is interesting to see that electron born at the peak of laser field returns periodically after each laser cycle. Whereas an electron born before the peak never returns to the ion core. 
For an electron born after the peak of the laser field, one born earlier, recollide at a later time. This is also evident from the relation of ionisation time and recombination time 
presented in Fig.~\ref{fig1.3}(b) (see red line). 
The total energy of the electron at the time of recombination, or the high-harmonic photon energy, is given by $\hbar \omega = I_p + \textbf{KE}$, where $\textbf{KE}$ is 
the kinetic energy of an electron at the time of recombination as a function of $t_i$, and 
plotted in Fig.~\ref{fig1.3}. 
Photon of the same energy can be generated from two different trajectories, one short and one long,
as apparent from Fig.~\ref{fig1.3}. 
The interference of these trajectories is an important consideration in high-harmonic spectroscopy. 
The maximum energy of a photon emitted during HHG is given by  
the relation $\hbar \omega_c = I_p$ + 3.17$U_p$. 
This implies that the cutoff energy is linearly proportional to the intensity of the laser field. 
The energy cutoff rule, obtained by the semiclassical model, satisfactorily explains  
the experimental observations. 
The underlying mechanism implies that the generation of high-order harmonics is a sub-cycle process, which is the main reason for the generation of attosecond pulses from HHG. 
Another implication of the semiclassical model of HHG is that the electron ionised using a circularly polarised laser never recombines to the parent ion. 
Therefore, circularly polarised harmonics can not be generated using a circularly polarised driving pulse efficiently. 
The harmonic yield decreases drastically when a small value of ellipticity in the driving pulse is introduced.

In general, a physical observable has to be defined to simulate HHG spectrum. 
In HHG, the Fourier transform of the dipole acceleration serves the purpose. 
Following Lamor's theorem of electromagnetism,  
accelerating charges or time-varying polarisation can act as a source of  electromagnetic radiation. Time-dependent dipole can be, respectively, calculated in length, velocity, and acceleration forms as
	\begin{equation}\label{eq:acceleration}
		\begin{split}
			a(t) &=   \frac{d^2}{dt^2}\left\langle \Psi(t) | \hat{x}| \Psi(t) \right\rangle, \\
		&= \frac{d}{dt} \left\langle \Psi(t) \right| \hat{p}\left| \Psi(t) \right\rangle,~~\textrm{and}  \\ 
			&= \left\langle \Psi(t) \right| -\partial_x V \left| \Psi(t) \right\rangle.
		\end{split}
	\end{equation}
Here, $\left|\Psi(t)\right\rangle$ is the time-dependent wavefunction, obtained after solving the time-dependent Schr\"odinger equation.

Now, let us discuss an important question: 
why there are only odd harmonics present when high-order harmonics are generated from centerosymmetric systems such as noble-gas atoms. 
This can be understood from a simple symmetry consideration. 
When matter interacts with a periodic laser field, the total Hamiltonian can be written as 
\begin{equation}\label{eq:Hfloq}
	\hat{\mathcal{H}}(t) = \frac{\hat{p}^2}{2m} + V(x) + ex\mathcal{F}_0\cos(\omega t). 
\end{equation}
Here, Hamiltonian is periodic in time as $\hat{\mathcal{H}}(t+\tau) = \hat{\mathcal{H}}(t)$ 
for $\tau = 2\pi/\omega$.  Following  Floquet theorem of periodically driven quantum systems, a time-periodic Hamiltonian has a wavefunction of the following form as 
\begin{equation}
	\Psi_n (x,t) = e^{-iE_nt}\psi_n(x,t).
\end{equation}
Here, $\psi_n(x,t)$ is the Floquet function which is  periodic in time [$\psi_n(x,t+\tau) = \psi_n(x,t)$].  
In the case of a symmetric potential [$V(-x) = V(x)$], Hamiltonian given in Eq.~(\ref{eq:Hfloq}) 
is invariant under generalised parity transformation $\hat{P}$, $x \rightarrow -x$ and $t \rightarrow t + T/2$. 
Therefore, Floquet states transform by $\hat{P}$ as $\hat{P}\psi_n(x,t) = \pm \psi_n(x,t)$.
 	
The contribution to the $n\rm ^{th}$-harmonic from the time-dependent dipole moment 
$d(t) = \left\langle \psi_n(t) |\hat{x}| \psi_n(t) \right\rangle$, using Fourier transform, 
is written as~\citep{grossmann2018theoretical}
\begin{equation}\label{eq:dm}
	\begin{split}
	d_m &= \frac{1}{\tau} \int_0^\tau d(t')e^{im\omega t'} dt' \\
	&= \frac{1}{\tau}\left[\int_0^{\tau/2} \left\{ d(t') \pm d(t' + \tau/2)\right\} e^{im\omega t'}dt' \right].
	\end{split}
\end{equation}
Here, + (-) is for even (odd) value of $m$. It is straightforward to show that 
\begin{equation}\label{eq:dt_relation}
	\begin{split}
	d(t + \tau/2) &= \int dx~\psi_n^*(-x,t)~x~\psi_n(-x,t) \\
	&= -\int dx~\psi_n^*(x',t)~x'~\psi_n(x',t) \\
	&= -d(t).
	\end{split}
\end{equation}
On substituting the value of Eq.~(\ref{eq:dt_relation}) to Eq.~(\ref{eq:dm}),  
we can show that there will be only odd harmonics in the spectrum for a material with inversion symmetry.

\section{High-Harmonic Generation in Solids}

Recent advancements in mid-infrared and terahertz laser sources have enabled 
the generation of strong-field driven HHG from solids, including semiconductors, dielectrics,  and nano-structures, below their damage threshold~\citep{ghimire2019, ghimire2011observation, ghimire2011redshift, zaks2012experimental, schubert2014sub,vampa2015all, vampa2015linking, hohenleutner2015real, luu2015extreme, ndabashimiye2016solid, you2017high,lanin2017mapping,sivis2017tailored, langer2018lightwave}. 
With the pioneering work of Ghimire {\it et al.}~\citep{ghimire2011observation}, 
HHG in solids offers fascinating  avenues to control, understand,
and probe light-driven carrier dynamics  in solids on attosecond timescale~\citep{silva2018high, silva2018all, bauer2018high, chacon2018observing, reimann2018subcycle, floss2018ab, mrudul2020high}.

The difference in the underlying mechanism of HHG from atoms in the gas phase 
and from solids is presented in Fig.~\ref{fig1.4}. 
Due to the considerable overlap between electron wavefunctions at nearby lattice sites, 
electron states are best described in the momentum-space using the Bloch theorem. 
A real-space picture of HHG in solids is shown in Fig.~\ref{fig1.4}(a). 
An electron ionised from a lattice site can propagate over a distance of many lattice periods in 
real-space. 
Therefore, it can recombine with any other lattice site other than its parent  ion. 
Yet, due to the lattice periodicity, HHG in solids is a coherent process. 
As the electron need not recombine with the parent lattice site, 
the ellipticity dependence of HHG in solids is entirely different. 
It is possible to generate circularly polarised harmonics using a circularly polarised driving pulse  
as reported in the pioneering work by Ghimire {\it et al.}~\citep{ghimire2011observation}. 
Another contrasting observation in solid HHG is that 
the energy cutoff depends linearly on the amplitude of the laser's electric field, whereas  
a quadratic dependence was reported in HHG from gases. 
This implies that a reciprocal-space explanation of the underlying mechanism is required. 

\begin{figure}[t!]
	\centering
	\includegraphics[width=\linewidth]{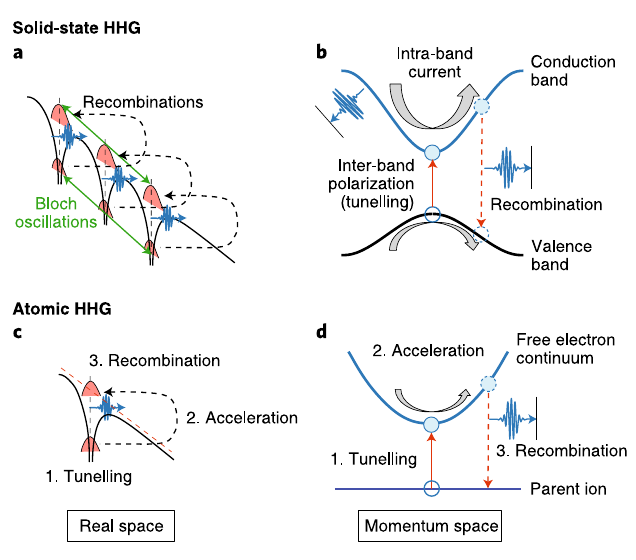}
	\caption{A comparative illustration of HHG from atoms and solids. This figure is adapted from Ref.~\citep{ghimire2019}.} \label{fig1.4}
\end{figure}

A reciprocal-space picture of the solid HHG is presented in Fig.~\ref{fig1.4}(b). 
A two-band model for a semiconductor is considered with one valence and one conduction band. 
The HHG spectrum of solid has two significant contributions: 
First one is stemming from the time-varying interband polarisation. 
An electron can be excited to the conduction band by an intense laser-field, and a hole is generated in the valence band. 
The  laser's electric field can accelerate the electrons and holes in their respective energy bands.  
This gives the second contribution: an intraband current. 
In the strong-field regime, interband and intraband contributions are coupled. The role of their contributions in  different parts of the spectrum is an interesting area to explore. 
For example, in the below band-gap region in the spectrum, only intraband current is expected to  contribute. 
Note that, there is no intraband current in atomic HHG as a hole in an atom is expected to be localised 
[see Fig.~\ref{fig1.4}(d)]. 
When an electron moves adiabatically along a particular conduction band in the absence of scattering, it diffracts from the Brillouin-zone boundaries 
with a frequency $\omega_B = e\mathcal{F}_0/\hbar a_0$ with $a_0$ as the lattice parameter. 
These diffractions from the zone  boundaries give dynamical Bloch oscillations, 
which result in linear dependence of energy cut off on the laser's electric-field amplitude in the case of HHG from solids~\citep{ghimire2011observation}. 

Unlike in gas phase, atoms within a solid possess periodicity and high-electron density. 
Moreover, high-order harmonics in solids are generated from lasers of intensities in TW/cm$^2$ orders, lower compared to the requirements of gas-phase HHG. 
With all these unique characteristics, HHG from solids offers a compact and superior source for coherent and bright attosecond pulses in extreme ultraviolet and soft x-ray energy regime~\citep{ghimire2019, luu2015extreme,  vampa2017merge, kruchinin2018colloquium}.
 
Another exciting aspect of solid HHG is that the harmonic spectrum imprints the anisotropic nature of the lattice. 
Unlike the two-band model used in Fig.~\ref{fig1.4}(b), actual materials include a large number of energy bands, and the high-harmonic spectrum arising from multiple energy bands contains rich information~\citep{hohenleutner2015real,hawkins2015effect}. 
For example, multiple plateau structure in HHG spectrum is observed due to the 
interband coupling  among different pairs 
of energy bands~\citep{ndabashimiye2016solid,wu2015high}. 
Interestingly, solid HHG is an ideal tool for studying the static and dynamical properties of valence electrons in solids.
In recent years, HHG in solids has been employed to explore several exciting processes such as 
band structure tomography~\citep{vampa2015all, lanin2017mapping, tancogne2017impact}, 
quantum phase transitions~\citep{silva2018high,  bauer2018high, silva2019topological}, the  
realisation of petahertz current in solids~\citep{luu2015extreme, garg2016multi}, and
dynamical Bloch oscillations~\citep{schubert2014sub, mcdonald2015interband}.

\section{Motivation}
In this thesis, we thoroughly analyse HHG from hexagonal two-dimensional (2D) materials. Nowadays, 2D materials are at the centre of tremendous research activities as they reveal different electronic and optical properties as compare to their  bulk counterparts. 
The realisation of atomically thin monolayer graphene has led to breakthroughs in fundamental and applied sciences~\citep{novoselov2004electric, geim2009graphene}. 
Charge carriers in graphene are described by the massless Dirac equation and exhibit exceptional transport properties~\citep{neto2009electronic}, making graphene very attractive for novel electronics applications. 
Soon after discovery of graphene, several other 2D materials have been synthesised including hexagonal boron nitride, transition metal dichalcogenides, silicene, germanene, etc.
Peculiar electron-photon interaction and electron-localisation makes 2D materials exciting candidates for studying ultrafast electron dynamics. 

Monolayer 2D materials such as graphene~\citep{yoshikawa2017high, al2014high}, transition-metal dichalcogenides~\citep{liu2017high, langer2018lightwave},  and hexagonal boron nitride (h-BN)~\citep{tancogne2018atomic,  le2018high,yu2018two} among others have been used to generate strong-field driven high-order harmonics. HHG in h-BN is well studied  when the  polarisation of the laser pulse is in-plane~\citep{le2018high, yu2018two} as well as out-of-plane~\citep{tancogne2018atomic} of the material. Using out-of-plane driving laser pulse, Tancogne-Dejean {\it et al.} have shown that atomic-like harmonics can be generated from h-BN~\citep{tancogne2018atomic}.  Also, h-BN  is used to explore the competition between atomic-like and bulk-like characteristics of  HHG~\citep{tancogne2018atomic, le2018high}. In MoS$_2$, it has been experimentally demonstrated that the generation of  high-order harmonics is more efficient in monolayer in comparison  to its bulk counterpart~\citep{liu2017high}. Moreover,  HHG from  graphene exhibits unusual dependence on the laser ellipticity~\citep{yoshikawa2017high}. Langer {\it et al.} have demonstrated the control in a light-driven change of the valley pseudospin in WSe$_2$~\citep{langer2018lightwave}. These works have shed light on the fact that 2D materials are very promising for  studying light-driven electron dynamics and for more technological applications in petahertz electronics~\citep{garg2016multi} and valleytronics~\citep{schaibley2016valleytronics}. Furthermore, using atomically thin material helps to avoid macroscopic propagation effects and reabsorption of harmonics by the medium.
In this thesis, we aim to investigate few  important questions of HHG in 
2D materials such as how the attosecond electron dynamics is different in a semimetal and a semiconductor, in a pristine and a defected material, etc.

\section{Thesis Overview}
This thesis consists of six main chapters and a bibliography. At the beginning of each chapter, a brief survey of the literature is presented.  

{\bf Chapter 1} provides an overview of ultrafast spectroscopy in perturbative and non-perturbative regimes.  Furthermore, HHG in gases and solids and their underlying mechanisms are presented.  
The classic three-step model is discussed in detail. A brief  overview of HHG in materials, especially in 2D, materials is presented. 

A detailed theoretical framework for understanding HHG in solids is given in  {\bf Chapter 2}. 
Methods to simulate HHG from model potentials in Cartesian basis and Bloch basis are discussed. 
This discussion provides reciprocal-space and real-space recollision models for HHG in solids. 
Moreover, to simulate HHG in real materials, semiconductor Bloch equations and time-dependent density functional theory are introduced.  

We discuss HHG from both monolayer and bilayer graphene  with the effect of interlayer coupling in bilayer graphene in {\bf Chapter 3}. 
Moreover, a comparison of HHG in gapped and gapless graphene and 
the role of Berry curvature in HHG from gapped graphene are investigated. 
The effects of polarisation-direction  and the ellipticity of the laser pulse on HHG are also 
presented.  
 
In {\bf Chapter 4}, we show  how tailored laser pulses can generate valley polarisation 
in a zero band-gap material such as pristine graphene with zero Berry curvature. 
Also, a recipe to read out the induced valley polarisation is presented. 

{\bf Chapter 5} discusses the possibility of high-harmonic spectroscopy of spin-polarised defects in  hexagonal boron-nitride. The role of boron and nitrogen vacancies in HHG from 
defected hexagonal boron-nitride is presented. 
Furthermore, the role of electron-electron interaction is also investigated.  

{\bf Chapter 6} provides the conclusion and the future scope of the present thesis.

\cleardoublepage
\chapter{Theoretical Framework for HHG from Solids}\label{Chapter2}
In this chapter, atomic units are used throughout unless stated otherwise.
The general recipe for a theoretical understanding of any non-relativistic quantum mechanical phenomena is to solve time-dependent Schr\"{o}dinger equation (TDSE) as  

\begin{equation}{\label{eq:TDSE}}
i \frac{ d}{dt}\Psi (\left\lbrace\textbf{r}_\mu \sigma_\mu \right\rbrace ,\left\lbrace\textbf{R}_\nu \right\rbrace,t)  = \hat{\mathcal{H}}(t)\Psi (\left\lbrace\textbf{r}_\mu\sigma_\mu\right\rbrace,\left\lbrace\textbf{R}_\nu\right\rbrace, t).
\end{equation} 
Here, $\textbf{r}_\mu \sigma_\mu$ is the notation for space and spin degrees of freedom of the $\mu^{\textrm{th}}$ electron, and $\textbf{R}_\nu$ stands for the positions of the $\nu^{\textrm{th}}$ ion-core. In Eq.~(\ref{eq:TDSE}), nuclei's kinetic-energy can be neglected since these are much heavier than electrons\footnote[1]{This approximation is known as the Born-Oppenheimer approximation.}. Consequently, the electron and nucleus degree of freedom can be separated. Throughout this work, we considered frozen nuclei and any effects arising from lattice vibrations are neglected. Therefore, our objective is to solve TDSE for electrons in the presence of light as     
\begin{equation}{\label{eq:eTDSE}}
	i \frac{ d}{dt}\Psi_e (\left\lbrace\textbf{r}_\mu\sigma_\mu\right\rbrace,t)   = [\hat{\mathcal{H}_0} + \hat{\mathcal{H}}^\prime(t)]\Psi_e (\left\lbrace\textbf{r}_\mu\sigma_\mu\right\rbrace, t) .
\end{equation} 
Here, $\hat{\mathcal{H}_0}$ is the field-free Hamiltonian and $\mathcal{\hat{H}}'(t)$ 
is the light-matter interaction Hamiltonian. The field free Hamiltonian is given by $ \hat{\mathcal{H}_0} = \hat{T}_e + \hat{V}_{e-i} + \hat{W}_{e-e}$. Here, $\hat{T}_e = \sum_{\mu=1}^N -\nabla_\mu^2/2$ is the  kinetic energy operator, $\hat{V}_{e-i} = \sum_{\mu=1}^N v(\textbf{r}_\mu)$ is the Coulomb attraction between electron and ion-core, and $\hat{W}_{e-e} = \frac{1}{2}\sum_{\mu\neq \nu}^N w(|\textbf{r}_\mu-\textbf{r}_\nu|)$ is the electron-electron Coulomb interaction. It is essential to point out that the ion-core coordinates enter in the electronic Hamiltonian as parameters in $\hat{V}_{e-i}$. Therefore, $\hat{V}_{e-i}$ is the term that distinguishes different electronic systems.  The time-dependent 
interaction Hamiltonian is given by  
$\mathcal{\hat{H}}'(t) = -\sum_\mu \textbf{r}_\mu \cdot \mathcal{F}(t) $, where $\mathcal{F}(t)$ is the 
electric field. 
Owing to the large wavelength of the field compared to the lattice parameters, dipole approximation is employed in this thesis. Therefore,  the spatial dependence of the field is neglected.

The $W_{e-e}$ term essentially makes Eq.~(\ref{eq:eTDSE}) a $6N\times6N$ coupled differential equation, which is not exactly solvable except for few systems such as hydrogen. 
Moreover, in the strong-field limit, the magnitude of the interaction 
Hamiltonian $\hat{\mathcal{H}}'(t)$ is comparable to $\hat{\mathcal{H}_0}$, which prevents us from using perturbative treatments. The way to solve such a problem is by numerically propagating TDSE with reliable approximations.

This chapter is designed as follows: In Section~\ref{section:2.1}, we discuss about solving TDSE 
for a model periodic potential.  In Section~\ref{section:2.2}, we will analyse HHG 
spectrum for a model periodic potential and discuss the basic mechanism.  In Section~\ref{section:2.3}, we discuss how the theoretical methods can be extended to realistic solids.

\section{HHG from a Model Potential}\label{section:2.1}
In this section, we neglect electron-electron interaction potential [W$_{e-e}$ in Eq.~(\ref{eq:eTDSE})].  To model the periodic electron-nuclear (V$_{e-i}$) interaction in solids, we use an one-dimensional periodic potential and corresponding  TDSE is written as 

\begin{equation}\label{eqn:IP-TDSE}
	i\frac{ d}{ dt} \left| \psi(t) \right\rangle = \left[\hat{\mathcal{H}}_0 + \hat{\mathcal{H}}'(t) \right]\left| \psi(t) \right\rangle.
\end{equation}
Here, $\hat{\mathcal{H}}_0 = -\nabla^2/2 + V(\hat{x})$ with $V(x)$ as a periodic potential. The laser polarization is considered along the crystal axis.

The Mathieu-type model potential is expressed as
\begin{equation}
	V(x) = -V_0 [1 + \cos(2\pi x /a)].
\end{equation}
Here, $a$ is the lattice parameter, and $-2V_0$ is the depth of the potential as shown in Fig.~\ref{fig2.1}(a). This is a commonly used potential for modelling HHG from solids~\citep{wu2015high,liu2017wavelength,ikemachi2017trajectory}.

In the following subsections, we show two different approaches to solve TDSE for model periodic potential. We use a lattice parameter, $a$ = 8 a.u. and $V_0$ = 0.37 a.u. for the study, similar to Refs.~\citep{wu2015high, liu2017wavelength, liu2017time, ikemachi2017trajectory}.

\subsection{Cartesian Basis}\label{subsection:Cartesian}

In this approach, the numerical methods developed for atoms are extended for a periodic potential. The method is adapted from Ref.~\citep{liu2017time}. $\hat{\mathcal{H}}_0$ is diagonalised in cartesian space to obtain the eigen-spectrum of the periodic lattice as 

\begin{equation}
	\hat{\mathcal{H}}_0  \psi_n(x)  =  \mathcal{E}_n  \psi_n(x) .
\end{equation}
The coordinate space within the region [-800 a.u., 800 a.u.] is considered, which corresponds to 200 lattice periods. The grid-spacing of 0.25 a.u. is used. For a wide-band-gap semiconductor, all valence-band electronic-states are filled, and all conduction-band electronic-states are empty. Owing to the low excitation probability considered in the present work, we choose the highest occupied (ho) eigenstate as the ground-state wave function [$\phi(x,t=0) = \psi_{ho}(x)$]. The light-matter interaction is modelled in length gauge as $\hat{\mathcal{H}}'(t) = \hat{x}\mathcal{F}(t)$. Time-dependent wave function $\phi(x,t)$ is obtained by solving TDSE numerically using split-operator method~\citep{hermann1988split} with a time-step 0.03 a.u. At each time-step, the wave function is multiplied with a mask function of width 100 a.u. to avoid unphysical reflection from the boundaries.

The expectation value of the derivative of current, $j(t)$ at any particular time, using acceleration form is proportional to

\begin{equation}
	\frac{d}{dt}j(t) \propto \int \frac{\partial V}{\partial x} ~|\psi(x,t)|^2~dx .
\end{equation}
The high-harmonic spectrum can be calculated as 
\begin{equation}
	\mathcal{I}(\omega) = \left|\mathcal{FT}\left[\frac{ d}{dt} j(t)\right] \right|^2 .
\end{equation}
Here, $\mathcal{FT}$ stands for Fourier transform. To obtain high-harmonic spectrum, a linearly polarised laser pulse of wavelength 3.2 $\mu$m and intensity 0.8 TW/cm$^2$ is used in the simulation. The pulse contains eight optical cycles with sine-square envelope. The values of wavelength and intensity mimic the values used in the experiment performed by Ghimire et al.~\citep{ghimire2011observation}. To avoid material damage, long wavelength and lower intensity in comparison to HHG from
gases are used. The corresponding HHG spectrum obtained is presented in Fig.~\ref{fig2.1}(b) (violet).

\begin{figure}[t!]
	\centering
	\includegraphics[width=\linewidth]{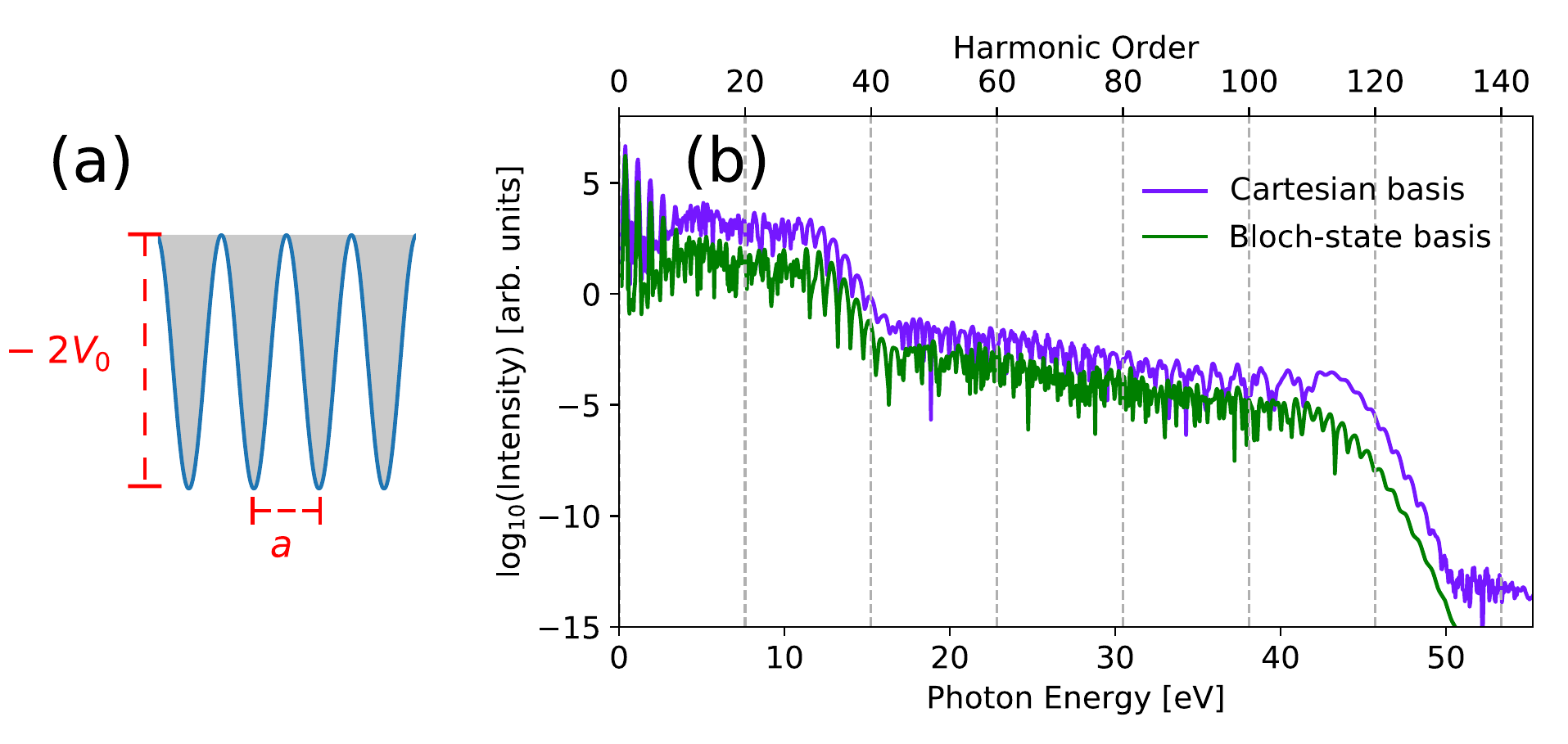}
	\caption{(a) Mathieu-type potential. (b) High-harmonic spectrum simulated for the Mathieu-type potential simulated in Cartesian (violet) and Bloch-state basis (green). Here, the laser has an intensity of 0.8 TW/cm$^2$ and a wavelength of 3.2 $\mu$m.} \label{fig2.1}
\end{figure}

\subsection{Bloch-State Basis}
In this subsection, we make use of the periodicity of the lattice by using Bloch functions as basis. To preserve the lattice-periodicity in the total Hamiltonian, we consider $\hat{\mathcal{H}}'(t)$ in velocity  gauge as $\hat{\mathcal{H}}'(t) = \hat{p} \cdot \mathcal{A}(t)$. 
Here, $\mathcal{A}(t)$ is the vector potential of the driving laser-field and it is related to electric 
field  $\mathcal{F}(t)$ as $\mathcal{F} = -\partial \mathcal{A}/\partial t $. 
The eigenstates of $\hat{\mathcal{H}}_0$ is obtained in Bloch-state basis as 
\begin{equation}\label{eq:Bloch}
	\hat{\mathcal{H}}_0 \left|\phi_{nk} \right\rangle = \mathcal{E}_{nk} \left| \phi_{nk} \right\rangle .
\end{equation}
Here, $n$ is the band index and $k$ is the wavevector. Solution to Eq.~(\ref{eq:Bloch}) is Bloch functions defined as $\phi_{nk}(x) = e^{ikx}u_{nk}(x)$ and $u_{nk}(x)$ has the periodicity of the lattice.

The Bloch functions at different {\bf k}-points can be propagated independently within dipole approximation. For a particular {\bf k}-point, the time-dependent wave function can be expanded in Bloch-state basis as~\citep{wu2015high},

\begin{equation}\label{eqn:Bloch-expn}
	\left| \psi_{k}(t) \right\rangle = \sum_n \alpha_{nk}(t)\left| \phi_{nk} \right\rangle.
\end{equation}  

Substituting Eq.~(\ref{eqn:Bloch-expn}) in TDSE [Eq.~(\ref{eqn:IP-TDSE})] 
within velocity-gauge and multiplying with $\left\langle \phi_{mk} \right|$ yields
\begin{equation}\label{eqn:BlochTDSE}
	i \frac{\partial}{\partial t}\alpha_{nk} = \alpha_{nk}\mathcal{E}_{nk} + \mathcal{A}(t)\sum_m p^{nm}_k \alpha_{mk}. 
\end{equation}
Here, $p_{nm}^k$ is the momentum-matrix elements defined as $p_{nm}^k = \left\langle \phi_{nk} \right| \hat{p} \left| \phi_{mk} \right\rangle$. We solve Eq.~(\ref{eqn:BlochTDSE}) 
using a fourth-order Runge-Kutta method with a time-step of 0.01 a.u~\citep{korbman2013quantum}. The high-harmonic  spectrum is calculated as 
\begin{equation}
	\mathcal{I}(\omega) = \left|\mathcal{FT}\left[ \frac{d}{dt} \int j(k, t) ~dk \right] \right|^2 .
\end{equation}
Initially all the valence bands are expected to be filled. However, since we are only considering low excitation probability, we have assumed only 3$\%$ occupation near the highest occupied valence band. Other deeply bound valence bands are neglected in the calculation. The spectra were converged for the number of conduction bands. Similar approximations are used in Ref.~\citep{wu2015high}.

Finally, we calculated the HHG spectrum for the potential and laser parameters used in subsection~\ref{subsection:Cartesian} and presented in Fig.~\ref{fig2.1}(b)(green). The HHG spectrum calculated by both the methods agree quite well and have important features. In the next section, we discuss how the mechanism of HHG can be understood from this simple model.

\section{Mechanism of HHG in Solids}\label{section:2.2}

\begin{figure}[t!]
	\centering
	\includegraphics[width=\linewidth]{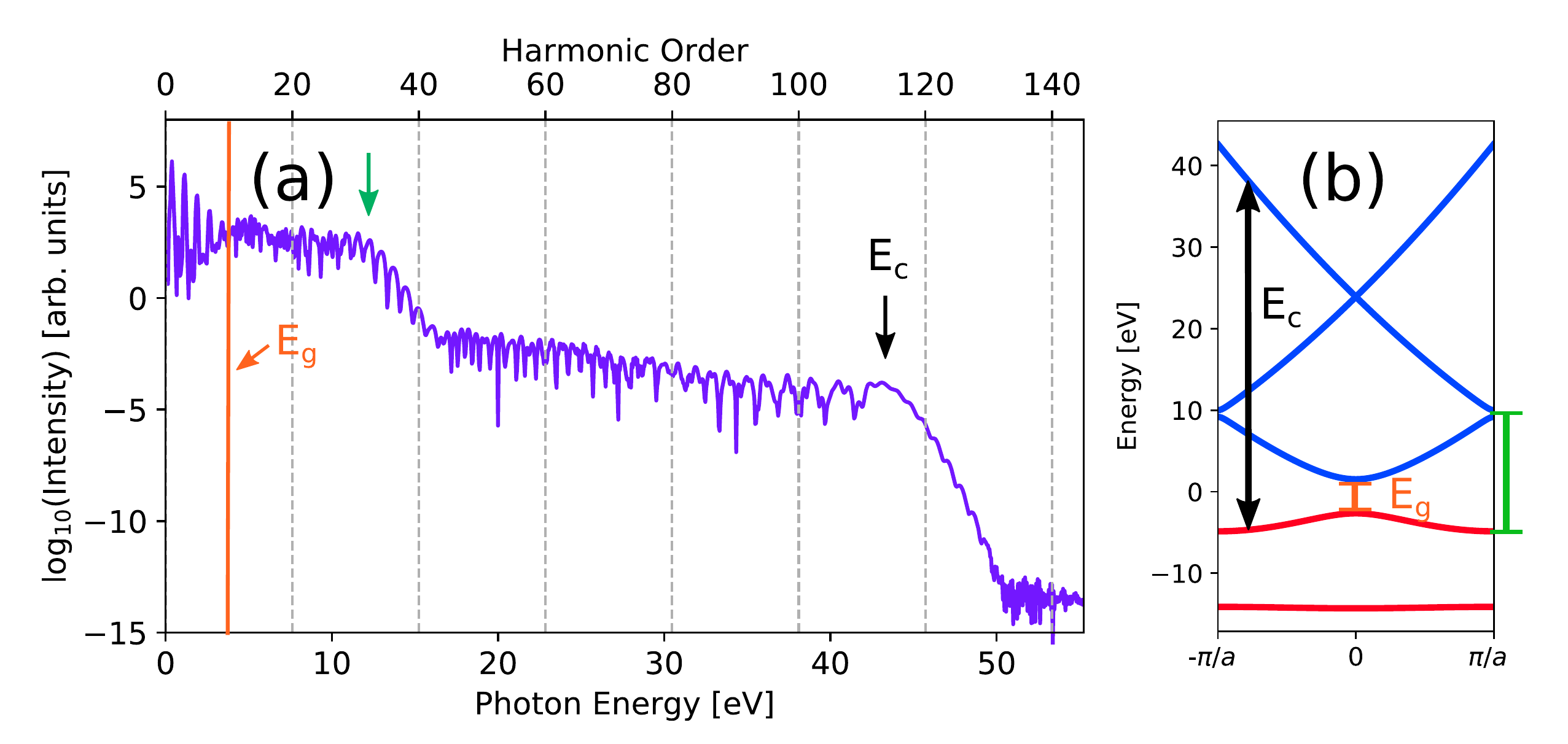}
	\caption{(a) The simulated high-harmonic  spectrum for the Mathieu-type potential, for the same laser parameters used in ~\ref{fig2.1}(b). The  vertical orange line indicates the minimum band-gap (E$_g$). (b) The electronic band-structure corresponds to the Mathieu-type potential. 
		Valence  and conduction bands are shown in red and blue colours, respectively.} \label{fig2.2}
\end{figure}

In this section, we do a detailed investigation of the underlying mechanism of HHG in solids. Unlike a simple three-step recollision model in gases, the mechanism of HHG is quite different in solids. In subsection~\ref{ssection:reciprocal}, we show how the HHG spectrum can be understood from the electronic band-structure.
In subsection~\ref{ssection:real}, we show how a real-space recollision model can be developed to understand HHG in solids.

\subsection{A Reciprocal-Space Understanding}\label{ssection:reciprocal}
The simulated spectrum of solid HHG reproduces exciting features as presented in Fig.~\ref{fig2.2}(a). The major features can be listed as follows a) there are clear harmonic peaks with monotonically decreasing intensity up to particular energy, b) it exhibits both a primary and a secondary plateaus 
and a sudden transition from primary to secondary plateaus with clear cutoffs, and c) the harmonic signal in the plateau region is noisy.

Let us try to understand the mechanism of HHG from the corresponding band-structure of the Mathieu potential presented in Fig.~\ref{fig2.2}(b). The valence bands are shown in red color, whereas conduction bands are shown in blue colour. The electron can have two types of dynamics in solids: the intraband dynamics originating from an electron (hole) accelerating in a conduction (valence) band, and the contribution from the interband transitions as a result of the 
recombination of an electron from conduction band to a hole in a valence band. 

In the energy-band structure, the minimum band-gap (E$_g$) is at 4.19 eV [marked in orange color in Figs.~\ref{fig2.2}(a) and \ref{fig2.2}(b)]. In the below band-gap region,  interband transitions are not possible, and those transitions are purely intraband. In the above band-gap region, there can  be contributions from interband and intraband transitions, which interfere and gives the noisy signal. Assuming infinite dephasing time is also found out to be a reason for the noisy spectra~\citep{vampa2014theoretical}. The primary plateau lies between E$_g$ and 12.86 eV. Here, the primary energy cutoff corresponds to the maximum energy between the first conduction and valence bands. Moreover, the secondary energy cutoff (E$_c$) is at 44.23 eV, which is less than the maximum band-gap energy between third conduction and valence bands. 

These results are in good agreement with the previously published results~\citep{wu2015high,liu2017high,ikemachi2017trajectory}. The primary plateau arises due to the 
interband transition from the first conduction band  to the valence band. 
The electron can also move to the higher conduction band 
via interband tunnelling [see e.g.~\citep{hawkins2015effect,ndabashimiye2016solid,ikemachi2017trajectory}]. 
Transitions from the higher-lying conduction band  to the 
valence band  lead to the secondary plateau, e.g.~\citep{ndabashimiye2016solid}. 
Since the interband transitions between the second conduction and valence band is less probable compared to the transitions between the first conduction and valence band, the secondary plateau is about five orders of magnitude lower than the primary plateau.

\begin{figure}[t!]
	\centering
	\includegraphics[width=\linewidth]{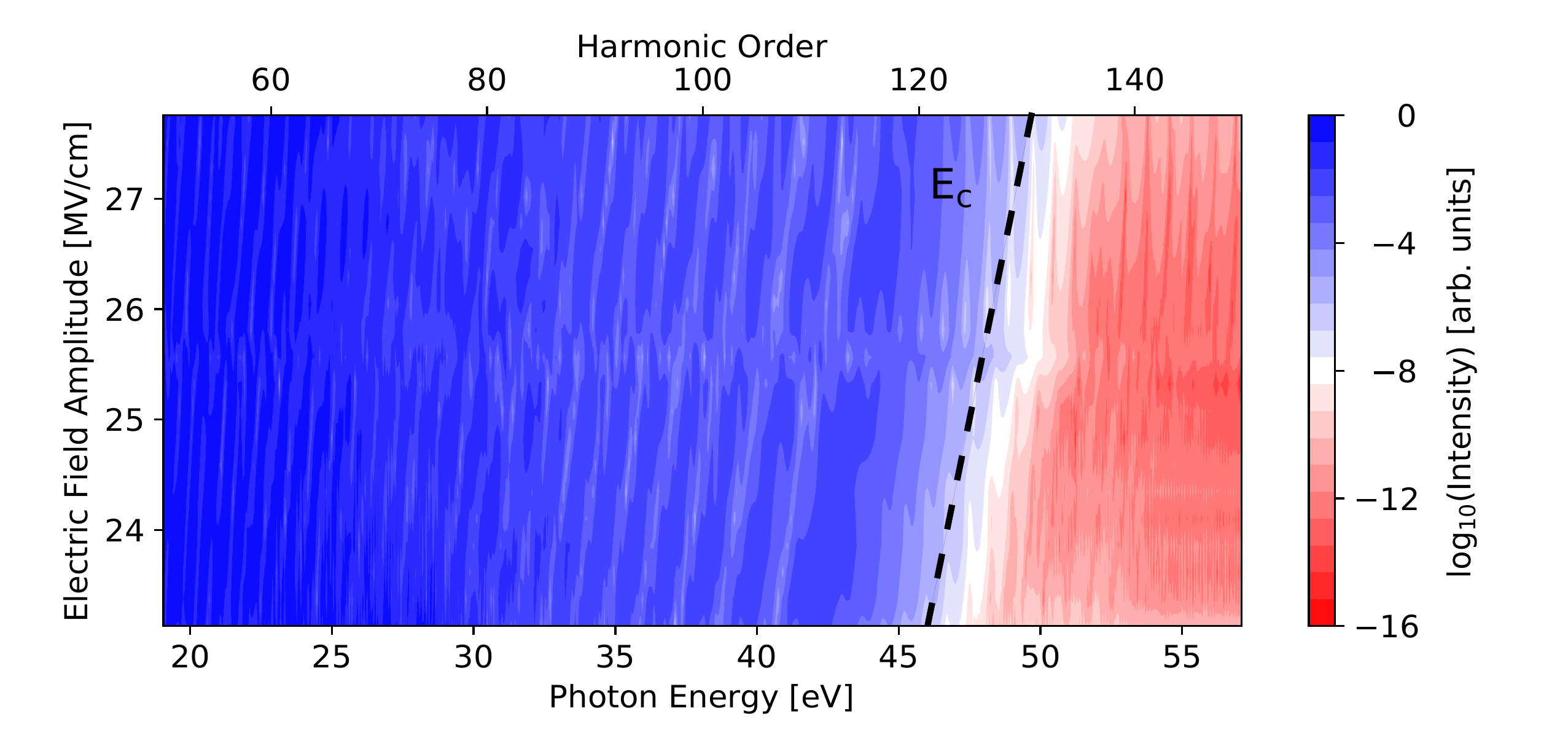}
	\caption{High-harmonic spectrum as a function of electric field strength.  
		The energy region of the secondary plateau is shown and the black dashed line indicates the energy cutoff (E$_c$).} \label{fig2.3}
\end{figure}

We have established how the band-structure information is embedded in the harmonic 
spectrum using an interband explanation. Now we analyse how the intraband mechanism plays a role in 
the electric field dependence of energy cutoff.   
High-harmonic  spectrum as a function of electric field amplitude is presented in Fig.~\ref{fig2.3}. The energy cutoff (E$_c$)  in the second plateau varies linearly with the field amplitude of driving laser as is 
typical for solids~\citep{ghimire2011observation, ghimire2014strong, wu2015high, du2017quasi}. This is in-contrast to the case of atomic-HHG, where the cutoff scales quadratically with the electric field amplitude as a consequence of the three-step-model. The reason for the linear dependence is explainable using dynamical Bloch oscillations, where the Bloch frequency is directly proportional to the electric-field amplitude~\citep{ghimire2011observation}.

\subsection{A Real-Space Recollision Model}\label{ssection:real}
Here we describe results of our numerical experiment, which allows us to  link directly HHG in solids with  real-space electron-hole recollision. We take advantage of  the 
angstrom-scale spatial resolution embedded in the harmonic signal, well established in molecules~\citep{smirnova2009high, haessler2010attosecond, lein2002role, lein2002interference, lein2007molecular, vozzi2005controlling, kanai2005quantum, odvzak2009interference, torres2010revealing, sukiasyan2010exchange}. 
The spatial information arises from half-scattering during electron-molecule recombination. 
It manifests in characteristic minima in the HHG spectra~\citep{lein2002role, lein2002interference, lein2007molecular}. 
The characteristic minima are laser-intensity independent~\citep{smirnova2009high, lein2002role, lein2002interference, lein2007molecular} and are 
associated with structure-based minima in the photorecombination cross 
sections~\citep{lein2002role, lein2002interference, lein2007molecular, vozzi2005controlling, kanai2005quantum, odvzak2009interference, torres2010revealing, sukiasyan2010exchange}, mirroring the well-known structure-related minima in photoionization. 
In diatomic molecules, the minima result from the Cohen-Fano 
interference of the two photoionization pathways originating at the two nuclei~\citep{cohen1966interference}.

Clearly, if real-space recollision between the conduction-band
electron and the valence-band hole underlies HHG in
solids, it is supposed to exhibit the same Cohen-Fano type interference
minima when the unit cell of the periodic lattice
potential has a two-centre structure. In our study, we restrict ourselves to wide band-gap materials, low-frequency drivers, and low excitation probability, where effective single-particle description is adequate.

To explore the idea, we model a semiconductor with two atom basis. In such a case, there will be two characteristic lengths for the unit-cell, the inter-atomic distance as well as the lattice constant. When the laser polarization is along the direction of the interatomic bond, we can effectively model this system using a one-dimensional bichromatic lattice potential as shown in  Fig.~\ref{fig2.4}(a) and 
can be expressed as

\begin{figure}[t!]
	\centering
	\includegraphics[width=0.7\linewidth]{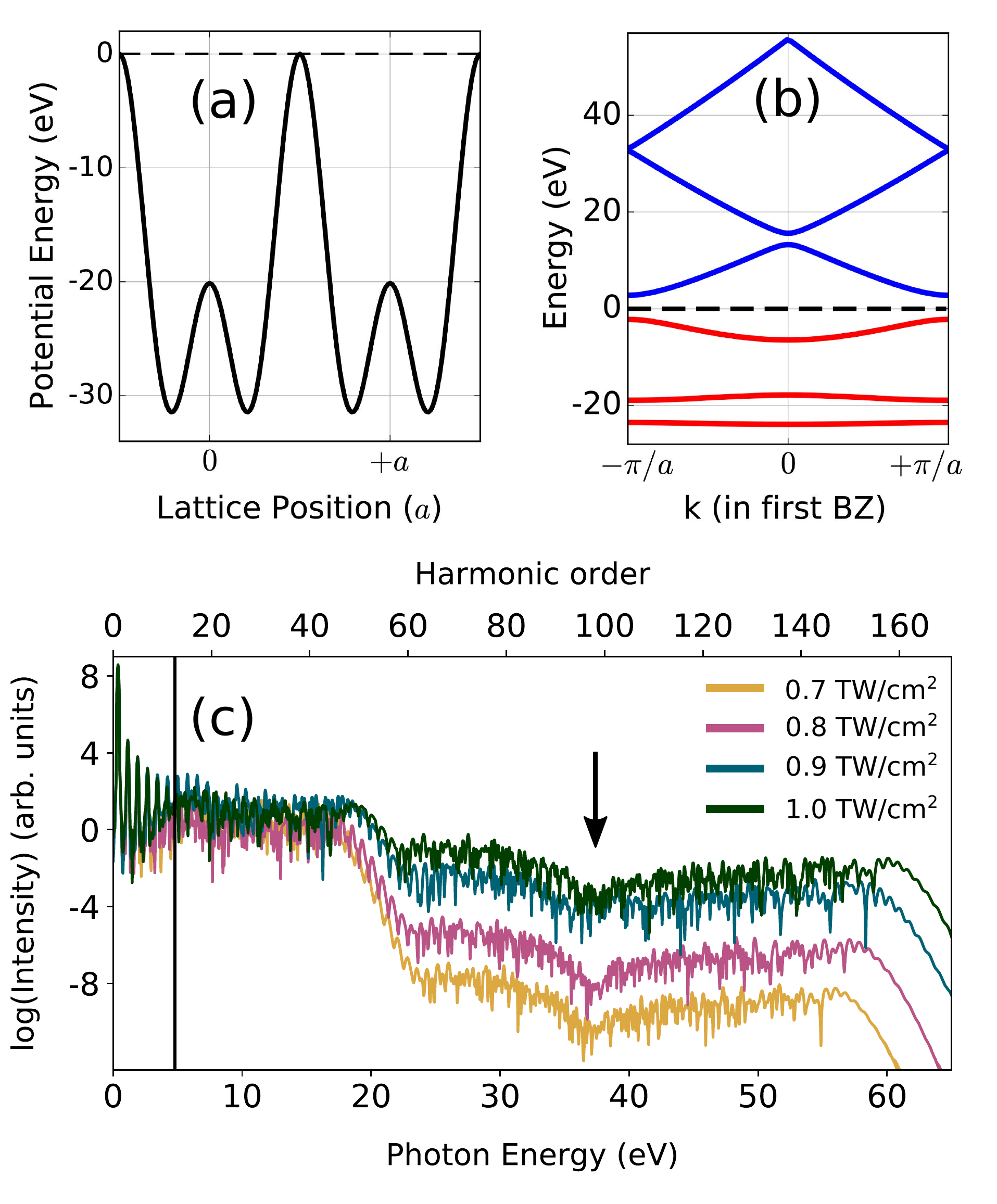}
	\caption{A bichromatic potential (a) and the corresponding band-structure with valence (red) and conduction bands (blue) marked. (b) High-harmonic  spectrum  as  a  function  of  driving  laser intensity  for  the  bichromatic  lattice. The  black  arrow represents the position of the minimum.} \label{fig2.4}
\end{figure}

\begin{equation}\label{eq02}
	V(x)  = -V_{0} \left[ (\alpha + \beta) - \alpha \cos\left(\frac{4 \pi x}{a}\right) - \beta \cos\left(\frac{2 \pi x}{a}\right)\right]. 
\end{equation} 
Here, $V_{0}$ is the potential depth, $a$ is the lattice constant. Each unit cell  has a double-well shape, with $a$ the 
distance between the unit cells [see Fig.~\ref{fig2.4}(a)], and the ratio of $\alpha$ and $\beta$ control the depth of the double-well potential.  
In the present study, 
$V_{0} = 0.37$ a.u., $a = 8$ a.u. and $\alpha = \beta = 1$ are used. Fig.~\ref{fig2.4}(b) shows the energy-band structure within the first Brillouin Zone for this bichromatic lattice. 
The minimum energy band-gap is 4.99 eV at the edge of the Brillouin  zone ($k = \pm \pi/a$).

To ensure the robustness of our findings, we have also simulated the HHG spectra  by 
solving TDSE  in real-space. 
The results obtained from two different numerical approaches, in real-space and in the Bloch state basis, show excellent agreement with 
each other.

For the bichromatic lattice, the HHG spectrum is shown in Fig.~\ref{fig2.4}(c), for an eight optical cycles linearly polarised laser pulse
with a sine-square envelope and $\lambda = 3.2~\mu$m. Spectra corresponding to the 
four different laser intensities are shown,
0.7  TW/cm$^{2}$ (yellow), 0.8 TW/cm$^{2}$ (pink),  
0.9 TW/cm$^{2}$ (blue) and 1.0 TW/cm$^{2}$ (green). 
The spectra  exhibit both a primary and a secondary plateau and 
a sharp transition from the primary  to the secondary plateau, with clear cutoffs.

The multiple-plateau structure and energy cutoff of HHG spectrum [Fig.~\ref{fig2.4}(c)] are consistent with the corresponding band-structure [Fig.~\ref{fig2.4}(b)] as explained in the subsection~\ref{ssection:reciprocal}. The intensity of the second plateau increases with the laser intensity, see Fig.~\ref{fig2.4}, reflecting higher probability of the inter-band excitation to the higher conduction band. The monotonically increasing cutoff with respect to laser intensity is also consistent with the linear cutoff dependence explained in subsection~\ref{ssection:reciprocal}.

The key feature of interest is the pronounced minimum 
in the second plateau, clearly present in Fig.~\ref{fig2.4}(c) (see black arrow). 
To identify its physical origin, we plot the 
position of the minimum as a function of the laser intensity in Fig.~\ref{fig2.5}(a). 
It shows that the position of the minimum is not sensitive to the laser intensity, just like 
the Cohen-Fano type interference minimum in HHG from molecules
~\citep{lein2002role, lein2002interference, lein2007molecular, kanai2005quantum, vozzi2005controlling, boutu2008coherent, zhou2008molecular, odvzak2009interference, torres2010revealing}.

\begin{figure}[t!]
	\centering
	\includegraphics[width=\linewidth]{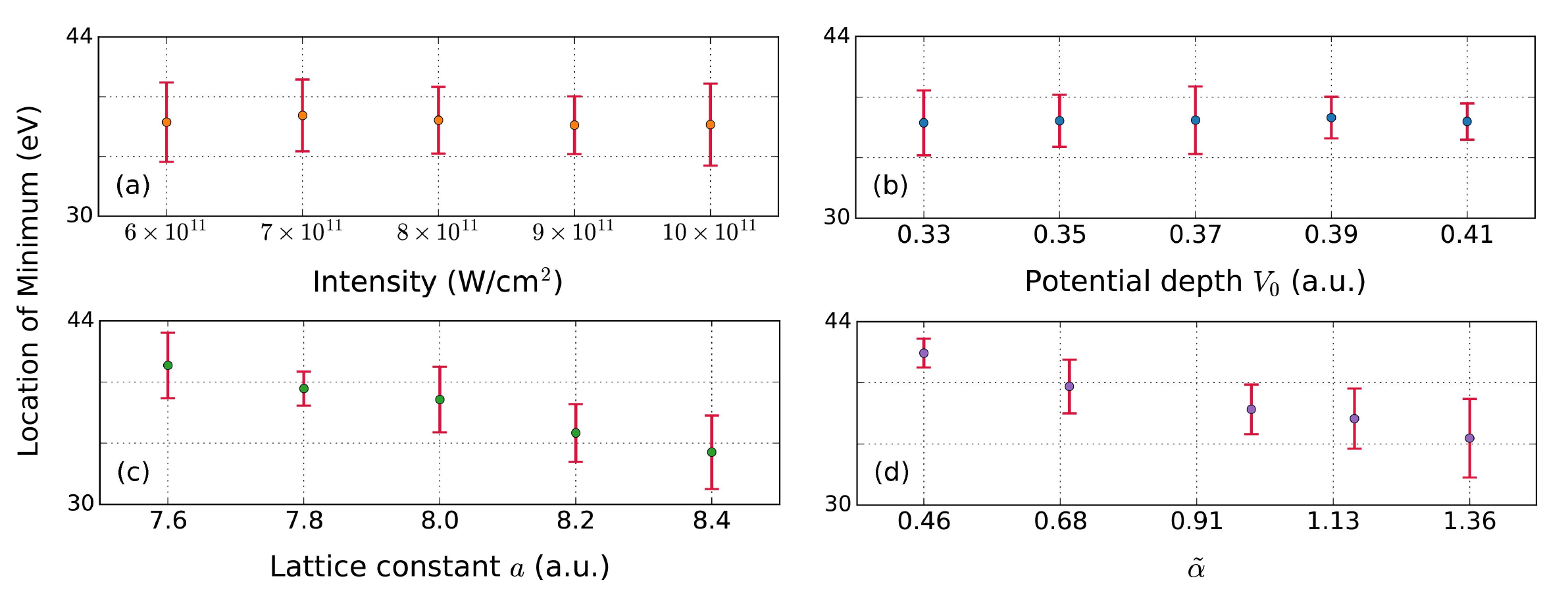}
	\caption{The variation in the position of the minimum in high-harmonic spectrum as a function of (a) intensity of the driving laser ($\lambda$ = 3.2 $\mu$m, V$_0$ = 0.37 a.u., and $a$ = 8 a.u.), (b) potential depth V$_0$ for  the  fixed  lattice  constant  ($a$ = 8  a.u.),  (c)  as a function  of  the lattice  constant  ($a$)  (for V$_0$ = 0.37 a.u.),  and  (d)  as  a  function  of the depth of the double-well  potential $\tilde{\alpha} = \alpha/\beta^2$ (for V$_0$ = 0.37 a.u., $a$ = 8 a.u.). In (b)-(d), the laser intensity I = 0.8 TW/cm$^2$ was used. The error bar represents the width of the interference minimum.}\label{fig2.5}
\end{figure}

To verify this conclusion, we look at the position of the minimum as a function of the 
parameters of the bichromatic lattice potential [see Eq.~(\ref{eq02})].  As expected 
for the Cohen-Fano type interference in radiative recombination during recollision,
the position of the minimum is independent of the depth of the bichromatic potential ($V_{0}$) 
as long as the distance between the wells does not change, see 
Fig.~\ref{fig2.5}(b). However, the position of the minimum changes  as soon as we start to vary the lattice constant $a$, see Fig.~\ref{fig2.5}(c). 
As the lattice constant is increased, the minimum shifts  towards lower photon energies  
as it should be.  
Identical observations have been reported for oriented
molecules, where the interference minimum occurs
at a lower harmonic order for larger internuclear bond-length,
(or when the aligned molecular ensemble is rotated towards the field polarization)
~\citep{lein2002role, lein2002interference}. 
As can be seen from Eq.~(\ref{eq02}), the ratio of $\alpha$ and $\beta$  
controls the depth of the double-well potential. 
Changing this ratio changes the depth of the potential barrier between 
the two wells, and thus the distance between the two peaks
of the corresponding wave function. 
For small $\tilde{\alpha} = \alpha/\beta^{2}$, 
the distance between 
the two peaks in the double-well wave function is smaller, and so the 
minimum is shifted to higher energies as evident from  Fig.~\ref{fig2.5}(d).
Note that, the harmonic spectrum from single colour  lattice (monochromatic lattice) does not 
exhibit any minimum in the spectrum.  
Therefore, analogous to structural minimum in oriented molecules, this minimum in solid HHG is related to the structure of the potential. 

In diatomic molecules, 
the structural minimum associated with photorecombination disappears when the two
nuclei are substantially different, so that the ground state is localized on a single nucleus.
The same should happen here. 

\begin{figure}[t!]
	\centering
	\includegraphics[width=0.7\linewidth]{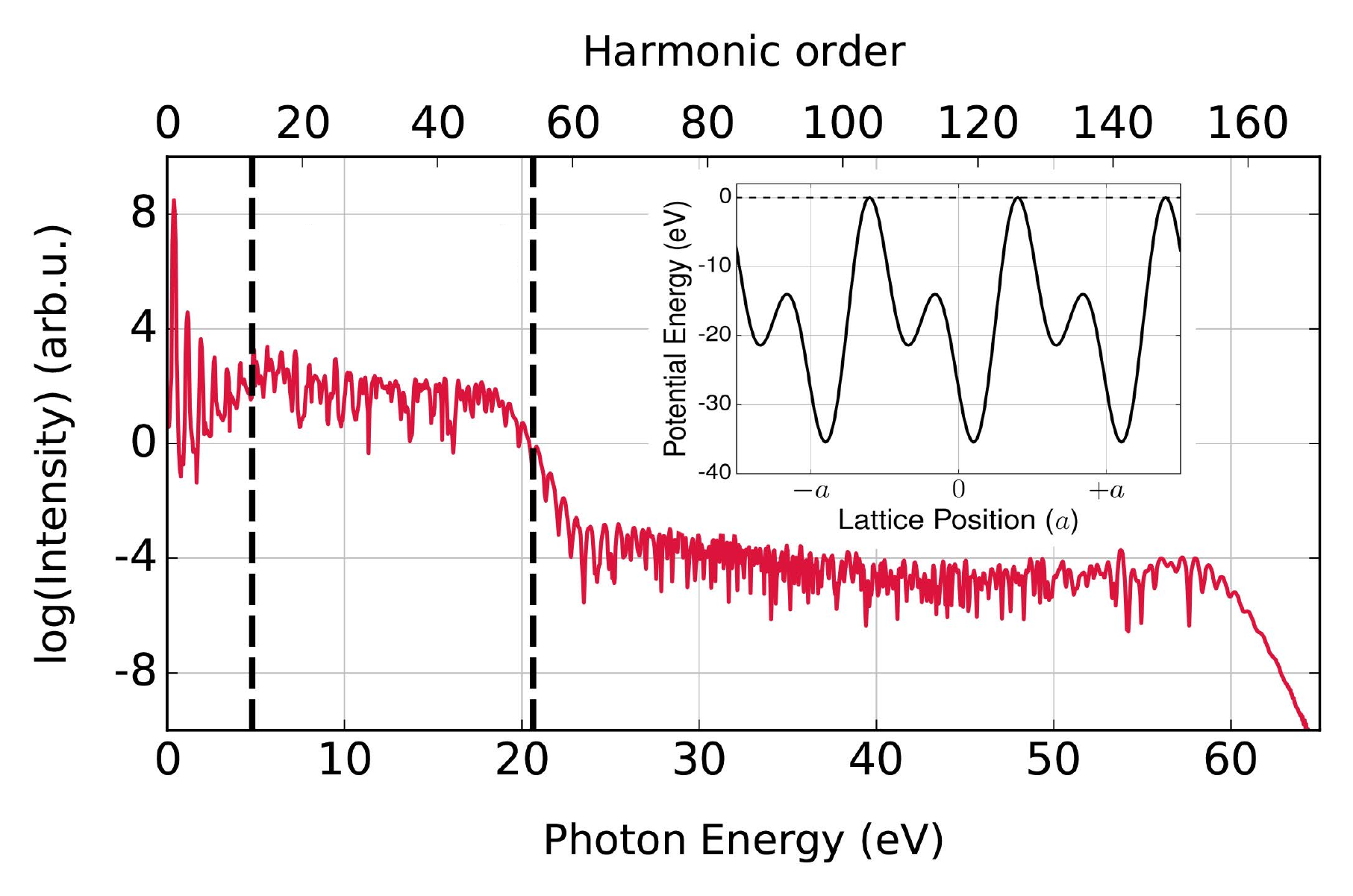}
	\caption{High-harmonic spectrum for the asymmetric bichromatic periodic lattice (shown in inset). The harmonic spectrum is obtained for intensity of 0.8 TW/cm$^2$ and wavelength of 3.2 $\mu$m. Here, the vertical dashed lines at 4.8 eV and 20.6 eV, respectively, the minimum and maximum band-gaps between the first conduction band and the valence band.} \label{fig2.6}
\end{figure}

To check this effect,  we introduce 
asymmetry into the double-well potential of the lattice as shown in  Fig.~\ref{fig2.6} (inset). 
The asymmetry is introduced by adding a 90$^{\circ}$ phase difference between the two spatial frequency 
components of the lattice. 
The corresponding harmonic spectrum is shown in Fig.~\ref{fig2.6} for I = 0.8 TW/cm$^{2}$ and 
$\lambda = 3.2~\mu$m. While the   
overall harmonic spectrum is the same as for the symmetric bichromatic potential 
[see Figs. ~\ref{fig2.4}(c) and ~\ref{fig2.6}], the 
minimum disappears. 
Therefore, the minimum in solid HHG does indeed represent the 
structural minimum in recombination, in direct analogy with HHG in molecules, providing clear 
evidence of the recollision picture of HHG in solids.

\section{HHG from Realistic Solids}\label{section:2.3}
At this stage, it is important to go beyond the  periodic model potential and understand the electron dynamics in realistic systems. This helps one to understand the symmetries, effect of dephasing, appreciate the role of interband and intraband mechanisms as well as the effect of electron-electron interactions in solid HHG. For these purposes, we introduce two approaches in this section  as
\begin{enumerate}
	\item Semiconductor Bloch equations (SBE), where the single electron TDSE can be solved for a few electronic bands of a realistic solid.  
	\item Time-dependent density functional theory (TDDFT), in which the electron-electron interaction can be also incorporated.
\end{enumerate}

\subsection{Semiconductor Bloch Equations}\label{section2.3.1}

The electron dynamics in semiconductors in the presence of strong lasers can be theoretically modelled by  SBE~\citep{haug2009quantum}. 
Recently, SBE is extensively used in high-harmonic spectroscopy of solids~\citep{vampa2014theoretical,hohenleutner2015real,jiang2018role}. 
SBE  allow us to separate interband and intraband contributions in the HHG spectrum~\citep{wu2015high}. Moreover, it helps us to understand the effect of dephasing. The adiabatic equation for Bloch electrons in Houston basis is written as 
\begin{equation}\label{EqnHouston}
	\hat{H}(t)\left|\phi^H_{n,\textbf{k}_0} \right\rangle = \mathcal{E}_n(\textbf{k}(t))\left|\phi^H_{n,\textbf{k}_0} \right\rangle.
\end{equation}
Here, $\left|\phi^H_{n,\textbf{k}_0} \right\rangle$ is the instantaneous eigenstate  for a Bloch electron 
and   $\textbf{k}(t) = \textbf{k}_0 + \mathcal{A}(t)$. 
Eq.~(\ref{EqnHouston}) is not equivalent to TDSE, but $\left\lbrace \phi^H_{n,\textbf{k}_0} \right\rbrace $ form a complete basis. The Houston basis functions are related to the Bloch basis function with crystal momentum $\textbf{k}(t)$ as 
\begin{equation}
	\left|\phi^H_{n,\textbf{k}_0}(t)\right\rangle = e^{-i\mathcal{A}\cdot\hat{\textbf{r}}}\left|\phi^B_{n,\textbf{k}(t)}\right\rangle,
\end{equation}
where $ \left\langle \hat{\textbf{r}}^{} \right.   \left|\phi^B_{n,k(t)}\right\rangle = e^{i\textbf{k}(t)\cdot\textbf{r}}u_{n,\textbf{k}(t)}(\textbf{r})$.

Let $\left| \psi(t)\right\rangle$ be the wave function, which fulfils  TDSE  as
\begin{equation}\label{eq2.3}
	i\frac{ d}{dt}\left|\psi(t)\right\rangle 
	= \left[\frac{1}{2}(\hat{\textbf{p}}+\mathcal{A}(t))^2 + V(\hat{\textbf{r}}) \right]\left|\psi(t)\right\rangle.
\end{equation}
Here, $\left| \psi(t)\right\rangle$ can be expanded in Houston basis as 
\begin{equation}\label{eq2.4}
	\left|\psi(t)\right\rangle = \sum_{n,\textbf{k}_0} a_{n,\textbf{k}_0}(t)\left|\phi^H_{n,\textbf{k}_0}(t)\right\rangle.
\end{equation}
Substituting Eq.~(\ref{eq2.4}) in Eq.~(\ref{eq2.3}) and taking an inner product with $\left\langle \phi^H_{m,\textbf{k}_0}\right| $ gives 
\begin{equation}\label{eq2.5}
	i \sum_{n,\textbf{k}_0}\left[\dot{a}_{n,\textbf{k}_0}(t)\delta_{nm}+a_{n,\textbf{k}_0}(t)\left\langle\phi^H_{m,\textbf{k}_0}\right|\partial_t  \left|\phi^H_{n,\textbf{k}_0}(t) \right\rangle \right] = \sum_{n,k_0} a_{n,\textbf{k}_0} \mathcal{E} (\textbf{k}(t))\delta_{nm}.
\end{equation}
The matrix elements, $\left\langle \phi^H_{m,\textbf{k}_0}(t) \right|\partial_t \left| \phi^H_{n,\textbf{k}_0}(t) \right\rangle$ can be simplified as,
\begin{equation}\label{eq2.6}
	\left\langle \phi^H_{m,\textbf{k}_0}(t) \right|\partial_t \left| \phi^H_{n,\textbf{k}_0}(t) \right\rangle = -\mathcal{F}(t) \cdot \left\langle u_{m,\textbf{k}(t)}\right|\nabla_\textbf{k} \left|u_{n,\textbf{k}(t)} \right\rangle.
\end{equation}
On substituting Eq.~(\ref{eq2.6}) in Eq.~(\ref{eq2.5}) and simplifying gives~\citep{wu2015high} us
\begin{equation}\label{eq2.7}
	i \partial_{t} a_{m,\textbf{\textbf{k}}_0}  = \epsilon_m^{\textbf{k}_0+\mathcal{A}(t)}a_{m,\textbf{k}_0}   -\mathcal{F}(t)\cdot\sum_{n}\textbf{d}_{mn}^{\textbf{k}_0+\mathcal{A}(t)} a_{n,\textbf{k}_0}, 
\end{equation}
where $\textbf{d}_{nm}^{\textbf{k}}$ is defined as 
$i\left\langle u_{n,\textbf{k}}\right|\nabla_{\textbf{k}} \left| u_{m,\textbf{k}} \right\rangle$. 
A set of equations equivalent to Eq.~(\ref{eq2.7}) in terms of density matrix elements can be derived by defining the density matrix elements, 
$ \rho_{mn}^{\textbf{k}_0} = a_{m,\textbf{k}_0} a{^*}_{n,\textbf{k}_0}$ and solving Eq.~(\ref{eq2.7}).
\begin{equation}
	\frac{\partial}{\partial t} \rho_{mn}^{\textbf{k}_0} = -i\mathcal{E}_{mn}^{\textbf{k}_0+\mathcal{A}(t)}\rho^{\textbf{k}_0}_{mn}+i\mathcal{F}(t)\cdot \left[\sum_l\left(\textbf{d}_{ml}^{\textbf{k}_0+\mathcal{A}(t)}\rho_{ln}^{\textbf{k}_0}-\textbf{d}_{ln}^{\textbf{k}_0+\mathcal{A}(t)}\rho_{ml}^{\textbf{k}_0}\right)\right],
\end{equation}
where  $\mathcal{E}^{\textbf{k}}_{mn}$ is the band-gap energy between $m$ and $n$ bands at $\textbf{k}$. A phenomenological term accounting for the decoherence can be added, with a constant dephasing time $T_2$, the semiconductor Bloch equations are given by
\begin{equation}\label{eqn:SBE}
	\partial_{ t} \rho _{mn}^{\textbf{k}_0}  = \left[ -i \mathcal{E}_{mn}^{\textbf{k}_0 + \mathcal{A}} -\frac{\tilde{\delta}_{mn}}{T_2} \right]\rho_{mn}^{\textbf{k}_0}
	+ 
	i \mathcal{F}\cdot\left[\sum_l \left(\textbf{d}_{ml}^{~\textbf{k}_0 + \mathcal{A}}\rho_{ln}^{\textbf{k}_0} - \textbf{d}_{ln}^{~\textbf{k}_0+\mathcal{A}}\rho_{ml}^{\textbf{k}}  \right) \right] .
\end{equation}
Here, $\tilde{\delta}_{mn}$ is defined as (1-$\delta_{mn}$). The term accounts for the band population relaxation ($T_1$) is neglected assuming $T_1$  $\gg$ $T_2$~\citep{hwang2008single}. 

The current at any $\bf{k}$-point in the Brillouin zone can be calculated as
\begin{equation}
	\begin{split}
		\textbf{J}^{\textbf{k}}(t) &= \sum_{m,n} \rho_{mn}^{\textbf{k}}(t)~\textbf{p}_{nm}^{~\textbf{k}+\mathcal{A}} \\
		&= \sum_{m\neq n} \rho_{mn}^{\textbf{k}}(t)~\textbf{p}_{nm}^{~\textbf{k}+\mathcal{A}} + \sum_{m=n} \rho_{m,n}^{\textbf{k}}(t)~\textbf{p}_{nm}^{~\textbf{k}+\mathcal{A}}\\
		&= \textbf{J}_{inter}^{\textbf{k}}(t) + \textbf{J}_{intra}^{\textbf{k}}(t). 
	\end{split}
\end{equation}
Here, $\textbf{J}_{inter}^{\textbf{k}}(t)$ and $\textbf{J}_{intra}^{\textbf{k}}(t)$ are, respectively, interband and intraband contributions to the total current, and \textbf{p}$_{nm}^\textbf{k}$ is 
momentum matrix element, which can be obtained as 
$\textbf{p}_{nm}^{\textbf{k}} = \left\langle n, \textbf{k} \left| \nabla_{\textbf{k}} \hat{\mathcal{H}}_{\textbf{k}} \right| m, \textbf{k} \right\rangle$. The off-diagonal elements of momentum and dipole-matrix elements are related as $\textbf{d}^{\textbf{k}}_{mn}$  =  $i\textbf{p}^{\textbf{k}}_{mn}/\mathcal{E}^{\textbf{k}}_{mn}$. Here, with the knowledge of band-structure, dipole and momentum matrix elements, Eq.~(\ref{eqn:SBE}) can be solved for any realistic material. 

We used length gauge due to the following reasons. In a basis involving infinite bands, any physical observable measured from SBE is gauge invariant. In contrast, for a model involving finite number of bands, the nonlinear response of semiconductors depends on the laser-gauge choice. For such a truncated basis, velocity gauge calculations have many disadvantages over length gauge, as the calculation only converges for numerous bands, and there is divergence at small frequencies.

The structure gauge choices don't affect the velocity gauge calculations, as each {\bf k}-point is treated independently. On the other hand, for the length gauge calculation, the relative phase of the wave function in the reciprocal-space is critical.  The phase of the wave function should be continuous and periodic. This problem can be fixed either by obtaining wavefunctions analytically or by using the twisted parallel transport gauge~\citep{yue2020structure}.

The high-harmonic spectrum is determined from the Fourier-transform of the time-derivative of the current as
\begin{equation}\label{eq2.23}
	\mathcal{I}(\omega) = \left|\mathcal{FT}\left(\frac{d}{dt} \left[\int_{BZ} \textbf{J}(\textbf{k},t)~\rm d\textbf{k} \right]\right) \right|^2 .
\end{equation}
Here, integration is performed over the entire Brillouin zone.

\subsection{Time-Dependent Density Functional Theory}\label{section:2.3.2}
There are situations where electron-electron interaction becomes extremely important as 
recently demonstrated by  TDDFT study of HHG~\citep{tancogne2018atomic,tancogne2018ultrafast}. In this subsection, we briefly present the theory of TDDFT for HHG in solids.  

Density functional theory (DFT) is a popular method to solve electronic structure problem. 
Our goal is to solve  TDSE with electron-electron interaction [Eq.~(\ref{eq:eTDSE})]. 
For that, we will start our discussion by reviewing the ground-state DFT, where the time-independent Schr\"odinger equation corresponding to Eq.~(\ref{eq:eTDSE}) is expressed as
\begin{equation}
	\left[ \hat{T}_e + \hat{V}_{e-i} + \hat{W}_{e-e} \right] \Psi_0(\left\{\textbf{r} \sigma \right\}) =  \mathcal{E} \Psi_0(\left\{\textbf{r} \sigma \right\}).
\end{equation}  
Here, $\Psi_0$ is the ground state wave function. 
In their seminal work, Hohenberg and Kohn showed that any quantum mechanical property of a 
many-electron system can be deduced from their ground-state density, $n_0(\textbf{r})$, without the requirement of the many-electron wave function 
$\Psi_0$~\citep{hohenberg1964inhomogeneous}. This is the foundation of DFT.  

For any non-relativistic quantum mechanical system, the form of $\hat{T}_e$ and $\hat{W}_{e-i}$ are universal. On the other hand, $\hat{V}_{e-i}$ characterise the system. Hohenberg and Kohn showed that there is a one-to-one mapping between $v_{ei}(\textbf{r})$ and $n_0(\textbf{r})$, 
whereas $\Psi_0$ is a unique functional of $n_0(\textbf{r})$. This imply that the ground-state density contains all the information of the electronic system. Consequently, any observable $\hat{O}$ is also a unique functional of $n_0(\textbf{r})$ as  $O[n_0] = \left\langle \Psi_0[n_0] \right| \hat{O} \left| \Psi_0[n_0] \right\rangle$. In particular, ground-state energy functional can be defined as
\begin{equation}
	E[n_0] = \left\langle \Psi_0[n_0] \right| \mathcal{\hat{H}}_0 \left| \Psi_0[n_0] \right\rangle.
\end{equation}
However, the exact energy functional can be found out variationally. Expressing the energy functional in terms of single-particle orbitals $\phi_\mu$, we obtain
\begin{equation}\label{eq:Efun}
	E[n] =   T_s[\left\{\phi_\mu[n]\right\}] + W_H[n] + E_{xc}[n] + \int d^3r~n(\textbf{r})v_{ei}(\textbf{r}).
\end{equation}
Here, $T_s$ is the kinetic energy of the non-interacting system with density $n$, defined as
\begin{equation}
	T_s[\left\{\phi_\mu[n]\right\}] = -\frac{1}{2}\sum_\sigma \sum_{i=\mu}^N \int d^3r\int d^3r'~\phi^*_\mu(\textbf{r}\sigma)\nabla^2 \phi_\mu(\textbf{r}\sigma),
\end{equation} 
and $W_H$ is the classical Hartree term
\begin{equation}
	W_H[n] = \frac{1}{2}\int d^3r\int d^3r'~\frac{n(\textbf{r})n(\textbf{r}')}{|\textbf{r}-\textbf{r}'|}.
\end{equation}
Finally, $E_{xc}$ is the exchange-correlation energy functional, which is by construction contains all other information about system. On minimising the energy-functional in Eq.~(\ref{eq:Efun}) using variational principle, we get Kohn-Sham (KS) equation as~\citep{kohn1965self}
\begin{equation}
	\left(-\frac{1}{2}\nabla^2 + v_s[n_0](\textbf{r})\right)\phi_\mu(\textbf{r}\sigma) = \epsilon_\mu\phi_\mu(\textbf{r}\sigma).
\end{equation}
KS equation is equivalent to single-electron Schr\"odinger equation,  where the effective single-particle potential is defined as
\begin{equation}
	v_s[n](\textbf{r}) = v_{e-i}[n](\textbf{r}) + v_H[n](\textbf{r}) + v_{xc}[n](\textbf{r}).
\end{equation}
Here, the Hartree potential is defined as 
$v_H[n](\textbf{r}) = \partial W_H[n]/\partial n(\textbf{r})$, and exchange-correlation potential is defined as  $v_{xc}[n](\textbf{r}) = \partial E_{xc}[n]/\partial n(\textbf{r})$. The solution of KS equation  is obtained self consistently.

Extending the theory of DFT to dynamical systems in not straightforward. 
The theory of TDDFT  was developed based on works done by Runge and   Gross~\citep{runge1984density}. 
The time-dependent Hamiltonian in Eq.~(\ref{eq:eTDSE}) can be seperated as $\hat{T_e} + \hat{W}_{e-e}+\hat{V}_{ext}(t)$. Here, the term corresponding to external potential  $\hat{V}_{ext}(t)$ is defined as, $\hat{V}_{ext}(t) = \hat{V}_{e-i} + \mathcal{\hat{H}}'(t)$. 
By definition, $\hat{V}_{ext}(t) = \sum_{i=\mu}^N v_{ext}(\textbf{r},t)$. The scalar external potential is assumed to be smooth and finite, which can be Taylor expanded around $t_0$. 
Runge and Gross showed that for a many-electron system evolving 
from a ground-state $\Psi(t = t_0) = \Psi_0$, 
there is a one-to-one mapping between external potential $v_{ext}(\textbf{r},t)$ and time-dependent density $n(\textbf{r}, t)$. A time-dependent KS equation analogous to single-electron TDSE can be set up as 
\begin{equation}{\label{eq:TDKS}}
	i \frac{\partial}{\partial t}\phi_\mu(\textbf{r}\sigma,t) =\left(-\frac{1}{2}\nabla^2 + v_s(\textbf{r},t)\right)\phi_\mu(\textbf{r}\sigma,t).
\end{equation}
Here, the time-dependent density is calculated from the KS orbitals as 
\begin{equation}
	n(\textbf{r},t) = \sum_\sigma \sum_{i=\mu}^N |\phi_\mu(\textbf{r}\sigma,t)|^2.
\end{equation}
To calculate the current, all the occupied KS orbitals for the periodic Hamiltonian $\phi_{n\textbf{k}}$ are propagated according to Eq.~(\ref{eq:TDKS}). The expression for current is defined as 

\begin{equation}
\textbf{j}_{\sigma}(t) =  \Re \left[\sum_{n,\textbf{k}} \int \phi^*_{n\textbf{k}}(\textbf{r}\sigma,t) \left\{ \hat{\textbf{p}} + \mathcal{A}(t) \right\} \phi_{n\textbf{k}}(\textbf{r}\sigma,t)~d^3r \right].
\end{equation}

Finally, high-harmonic  spectrum can be calculated as
\begin{equation}
	\mathcal{I}(\omega) = \left|\mathcal{FT}\left[\sum_{\sigma=\uparrow,\downarrow}\frac{d}{dt} ~\textbf{j}_\sigma(t)   \right] \right|^2.
\end{equation}
Finally, it is essential to note how $n(\textbf{r},t)$ and time-derivative of current 
(hence high-harmonic  spectrum) are related.  This relationship for a many-body system is
\begin{equation}
	\frac{d}{dt}\textbf{j}(t) = -  \int  n(\textbf{r},t) \nabla v_{ext}(\textbf{r},t) d^3 r. 
\end{equation}\label{eq:current_der}
A comprehensive derivation and explanation of the above equation were provided by Tancogne-Dejean et al.~\citep{tancogne2017impact}. It is trivial to see that the gradient term in the above equation is relevant only for the electron-nuclei potential when the dipole approximation is employed.  In such a situation, the equation reduces to acceleration form as described in Eq.~(\ref{eq:acceleration}). This also convinces the importance of atomic arrangements in attosecond electron dynamics.

\cleardoublepage
\chapter{HHG from Gapless  and Gapped Graphene}\label{Chapter3}

The realisation of an atomically-thin monolayer graphene has catalysed  a series of breakthroughs in 
fundamental and applied sciences~\citep{geim2009graphene}. 
Graphene shows unusual electronic and optical properties in comparison  to its bulk counterpart~\citep{novoselov2004electric}. 
The unique electronic structure of graphene exhibits  varieties of nonlinear optical processes~\citep{avetissian2016coherent, hendry2010coherent, kumar2013third}. 
HHG from monolayer and  few-layer graphenes has been extensively studied in the past~\citep{hafez2018extremely, avetissian2018impact, chizhova2017high, al2014high, yoshikawa2017high, zurron2019optical, taucer2017nonperturbative, liu2018driving, chen2019circularly, mikhailov2007non, gupta2003generation, mrudul2021light,sato2021high}. 
The underlying mechanism of HHG in graphene~\citep{zurron2018theory} was 
different from the one explained using a two-band model by Vampa \textit{et al.}~\citep{vampa2014theoretical}. 
The intraband current from the linear band-dispersion of graphene was expected to be the dominating mechanism~\citep{mikhailov2007non, gupta2003generation}. 
This is a consequence of the highly nonparabolic nature of the energy bands~\citep{ghimire2011observation}. 
In contrast to this prediction, the interband and intraband mechanisms in graphene are coupled ~\citep{taucer2017nonperturbative, liu2018driving, al2014high,sato2021high} except for low intensity driving fields~\citep{al2014high}. Vanishing band-gap and diverging dipole matrix elements near 
Dirac points lead to intense interband mixing of valence and conduction bands in graphene~\citep{zurron2018theory,kelardeh2015graphene}. The ellipticity dependence of HHG from graphene has been  observed experimentally~\citep{yoshikawa2017high, taucer2017nonperturbative} as well as discussed theoretically~\citep{yoshikawa2017high, taucer2017nonperturbative, liu2018driving, zurron2019optical}. 
Taucer \textit{et al.}~\citep{taucer2017nonperturbative} have demonstrated  that 
the ellipticity dependence of the harmonics in graphene is atom like. In contrast, a higher harmonic yield for a particular ellipticity was observed by Yoshikawa \textit{et al.}~\citep{yoshikawa2017high}. 
The anomalous ellipticity dependence was attributed to the strong-field interaction in the semi-metal regime~\citep{tamaya2016diabatic, yoshikawa2017high}.

Along with monolayer graphene, bilayer graphene is also attractive due to its interesting optical response~\citep{yan2012tunable}. Bilayer graphene can be made by stacking another layer of the monolayer graphene on top of the first.   
Three suitable configurations  of the bilayer graphene are possible: 
(a) AA stacking in which the second layer is placed exactly on top of the first layer; (b) 
AB stacking in which the B atom of the upper layer is placed on the top of the A atom of the lower layer, whereas the other type of atom occupies the centre of the hexagon; 
and (c) twisted bilayer in which the upper layer is rotated by an angle with respect to the lower layer. 
AB stacking, also known as the Bernel stacking, is the one that is a more energetically stable structure and mostly realised in experiments [see Fig.~\ref{fig3.1}(b)]~\citep{rozhkov2016electronic, mccann2013electronic}.
Avetissian \textit{et al.} have discussed the role of the multiphoton resonant excitations in HHG for AB stacked bilayer graphene~\citep{avetissian2013multiphoton}.  
Moreover, HHG from twisted bilayer graphene has been explored recently~\citep{ikeda2020high}. However, the comparison of HHG from monolayer and bilayer graphene; and a thorough investigation of the role of interlayer coupling are unexplored.

Another fascinating class of 2D materials is gapped graphene. There are several ways in which an arbitrary band-gap can be introduced in graphene~\citep{zhou2007substrate,castro2007biased,li2008chemically,pedersen2008graphene,pedersen2008optical}. In this chapter, we are mainly 
interested in band-gap opening due to inversion symmetry breaking. This can be experimentally achieved by growing graphene on silicon carbide substrate~\citep{zhou2007substrate}. Moreover, the symmetry of gapped graphene is identical to other 2D materials such as transition metal dichalcogenides (TMDC), and hexagonal boron nitride (h-BN). 
Unlike gapless graphene, gapped graphene has non-zero Berry curvature resulting in nontrivial topological properties. Ultrafast electron dynamics~\citep{kelardeh2021ultrafast,motlagh2020ultrafast,motlagh2019ultrafast} and HHG~\citep{dimitrovski2017high} from gapped graphene are studied in the past. Here, we compare the symmetries and polarisation dependence of HHG from gapped graphene with gapless monolayer graphene.

In this chapter, we investigate HHG from monolayer, bilayer, and gapped graphene. Symmetry and polarisation properties on the harmonic spectrum are analysed. 

\section{Numerical Methods}\label{section:1}

\begin{figure}[t!]
	\centering
	\includegraphics[width=0.8\linewidth]{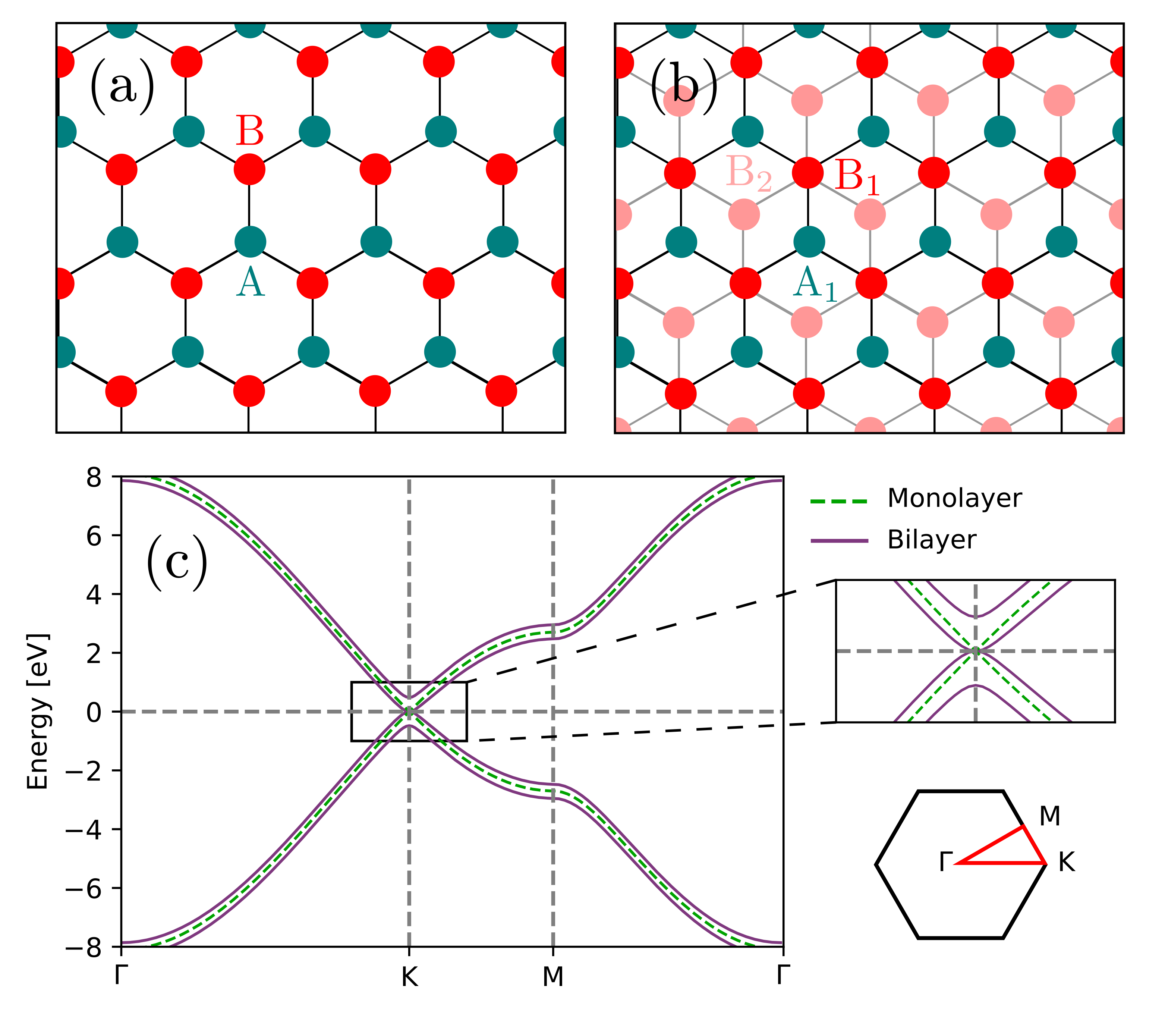}
	\caption{Monolayer and bilayer graphene (AB-stacked). 
		The top view of (a) the monolayer and (b) bilayer graphenes, respectively. 
		The carbon atoms are arranged in a honeycomb lattice with two inequivalent 
		carbon atoms (A and B). In bilayer graphene, the B atoms of the top layer are placed on 
		top of A atoms of the bottom layer. 
		(c)  The band structure of the monolayer (green) and bilayer graphene (violet).} \label{fig3.1}
\end{figure}

The real-space lattice of monolayer graphene is shown in Fig.~\ref{fig3.1}(a). 
Carbon atoms are arranged in a honeycomb lattice with a two-atom basis unit-cell. 
A and B in Fig.~\ref{fig3.1}(a) represents two inequivalent carbon atoms in a unit cell. The lattice parameter of graphene is equal to  2.46 \AA. Nearest-neighbour tight-binding approximation is implemented by only considering the $p{_z}$ orbitals of the carbon atoms. The Hamiltonian for monolayer graphene is defined as
\begin{equation}\label{eq:3.1}
	\hat{\mathcal{H}}_{0} = -t_0 f(\textbf{k})\hat{a}^{\dagger}_k \hat{b}_k + \textrm{H.c.}
\end{equation}
Here, $\hat{a}_k^{\dagger}$ ($\hat{b}_k$) is the creation (annihilation) operator associated with A (B) type of the atom in the unit cell,  
$f(\textbf{k})$ is defined as $f(\textbf{k}) = \sum_i e^{i\textbf{k}\cdot \delta_i}$, where $\delta_i$ is  the nearest neighbour vectors.  
A nearest-neighbour in-plane hopping energy $t_0$  of 2.7 eV is used~\citep{reich2002tight,trambly2010localization,moon2012energy}.  
The eigenvalues of the Hamiltonian are given by

\begin{equation}
	\mathcal{E}(\textbf{k}) = \pm t_{0}|f(\textbf{k})|.
\end{equation}

Similarly, the Hamiltonian for AB-stacked bilayer graphene can be defined as
\begin{equation}
	\hat{\mathcal{H}}_{AB} = -t_0 f(\textbf{k}) \left[\hat{a}^\dagger_{1k}\hat{b}_{1k} + \hat{a}^\dagger_{2k}\hat{b}_{2k}   \right] + t_\perp \hat{a}^\dagger_{2k}\hat{b}_{1k} +  \textrm{H.c.}
\end{equation}
Here, 1 and 2 denote the carbon atoms in the upper and lower layers, respectively. 
An inter plane hopping energy $t_\perp$ of 0.48 eV is used for an interlayer separation equal to  3.35 \AA~\citep{trambly2010localization,moon2012energy}. 
The band-structure for the bilayer graphene is given as
\begin{equation}
     \mathcal{E}(\textbf{k}) =  [\pm t_\perp \pm \sqrt{4 |f(\textbf{k})|^2 t_0^2 + t_\perp^2}]/2 .
\end{equation}

Figure~\ref{fig3.1}(c) presents the energy band-structure of both monolayer and bilayer graphene. 
The band-structure of monolayer graphene has zero band-gap and linear dispersion near 
two points, known as $\bf{K}$-points, in the Brillouin zone. 
On the other hand, bilayer graphene near $\bf{K}$-points shows a quadratic dispersion. 
Due to the zero band-gap nature, both monolayer and bilayer graphene are semi-metals. 
Here, electron-hole symmetry is considered by neglecting higher-order hopping and overlap of the orbitals.

SBE in Houston basis is solved~\citep{houston1940acceleration, krieger1986time, floss2018ab} as discussed in section~\ref{section2.3.1}. The momentum matrix elements between $\left| m,\textbf{k}\right\rangle$ and $\left| n,\textbf{k}\right\rangle$ states from the tight-binding Hamiltonian can be obtained as~\citep{pedersen2001optical} 
\begin{equation}
	\textbf{p}_{nm}(\textbf{k}) = \left\langle n,\textbf{k} \right|\nabla _\textbf{k} \hat{\mathcal{H}} _\textbf{k}\left| m,\textbf{k}\right\rangle.
\end{equation}
 
For HHG from monolayer graphene, a dephasing time within the range of 2 fs to 35 fs has been used in the past~\citep{sato2021high,liu2018driving,taucer2017nonperturbative}. 
	Moreover, a detailed investigation about dephasing time dependence on HHG from monolayer graphene has been discussed in Ref.~\citep{liu2018driving}. 
	In this chapter, a dephasing time of 10~fs is considered for monolayer, bilayer, and gapped graphene.

From the harmonic spectrum calculated using Eq.~(\ref{eq2.23}), the integrated yield for $n^{\textrm{th}}$ harmonic in $\mu$-direction can be calculated as
\begin{equation}
	\mathcal{I}^{(n)}_\mu = \int_{(n-0.5)\omega}^{(n+0.5)\omega}  \mathcal{I}_\mu (\omega^\prime) d\omega^\prime.
\end{equation}
Ellipticity of the $n^{\textrm{th}}$  harmonic can be calculated from the integrated harmonic yield as 
\begin{equation}
	|\epsilon_n| = \min\left(~\sqrt{\frac{\mathcal{I}^{(n)}_y}{\mathcal{I}^{(n)}_x}}, \sqrt{\frac{\mathcal{I}^{(n)}_x}{\mathcal{I}^{(n)}_y}} ~\right).
\end{equation}

Optical joint density of states (JDOS) estimates the number of available states for an electron to do an interband transition from the valence to conduction band absorbing an energy 
$E  ( = E_{c} - E_{v})$. JDOS is calculated as 
\begin{equation}
	{\rm JDOS}(E) = \int \frac{dS}{\left|\nabla_\textbf{k}(E_c(\textbf{k})-E_v(\textbf{k}))\right|}.
\end{equation}
The integral is performed over a constant energy surface with energy $E$ in momentum space.

\section{Results}\label{section:2}
In this chapter, a driving laser pulse with an intensity of 1$\times$10$^{11}$ W/cm$^2$ and a 
wavelength of 3.2 $\mu$m are used. The laser pulse is  eight-cycles in duration with a sin-squared  envelope.  
The intensity of the driving pulse is much below the damage threshold and lower than the one used in experimental HHG from graphene~\citep{yoshikawa2017high, taucer2017nonperturbative}. The same parameters  of the driving laser pulse are used throughout this chapter unless stated otherwise.

\subsection{HHG from Monolayer and Bilayer Graphene}\label{section:3.1}
In this section, we discuss HHG from monolayer and bilayer graphene with 
AA and AB configurations. Moreover, 
the role of interband and intraband contributions are investigated in both cases. 
The role of interlayer coupling in HHG from bilayer graphene is investigated. 
Furthermore, polarisation and ellipticity dependences of the HHG are discussed. 

\begin{figure}
	\centering
	\includegraphics[width=0.8\linewidth]{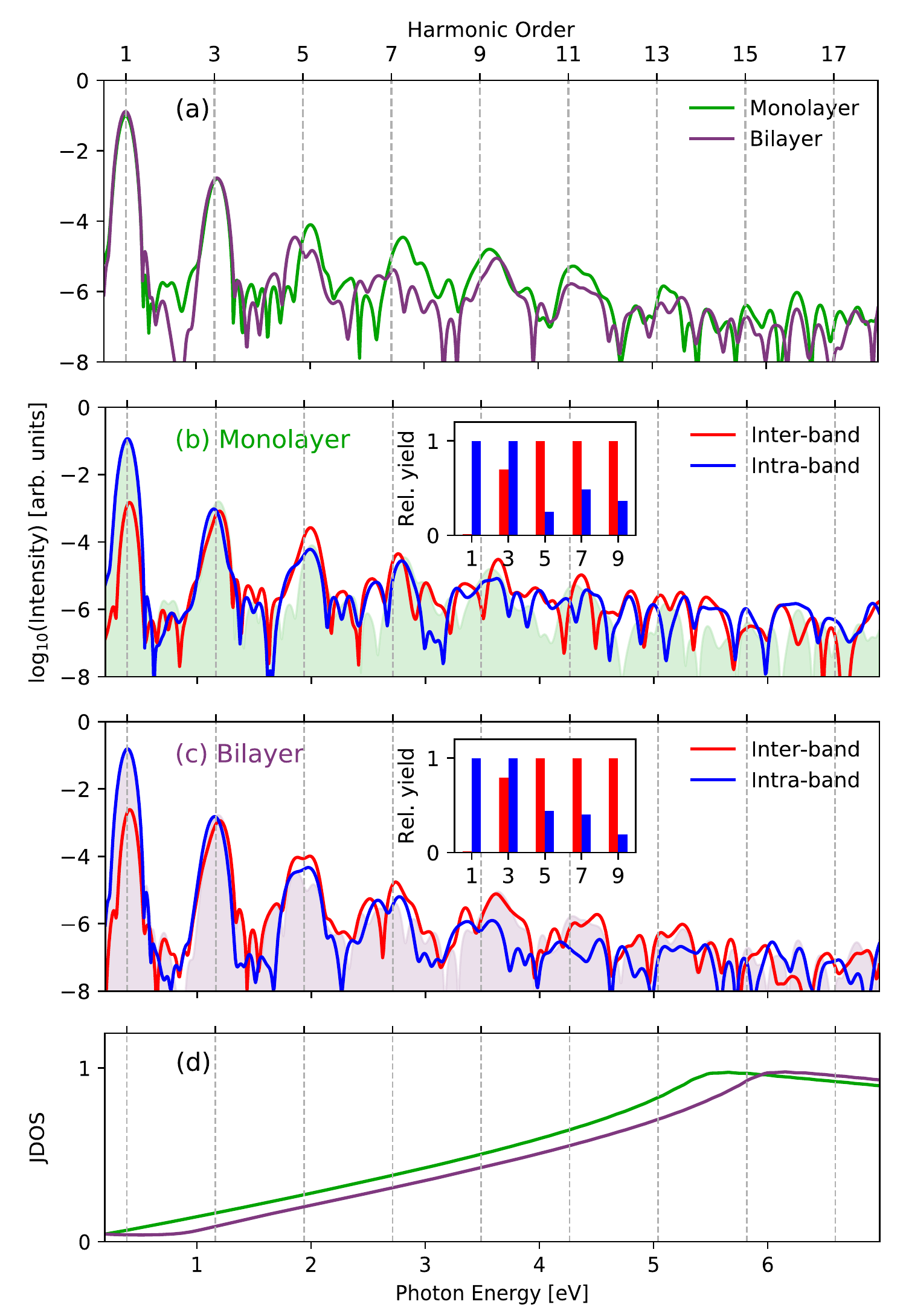}
	\caption{(a) High-harmonic spectrum of the monolayer (green) and bilayer (violet)   
		graphene with AB stacking. A linearly polarized pulse along the $\Gamma$-K direction is considered here. The high-harmonic intensity is normalised to the total number of electrons in 
		monolayer and bilayer graphenes. The interband and intraband contributions to the high-harmonic spectrum for 
		(b) monolayer and (c) bilayer graphenes. The total harmonic spectrum is also plotted for the reference. 
		The relative harmonic yield (integrated) for different orders from interband and intraband contributions is plotted in the insets of (b) and (c), respectively. (d) The normalised optical joint density of states (JDOS) of monolayer (green) and bilayer (violet) graphenes.} \label{fig3.2}
\end{figure}

Figure~\ref{fig3.2} presents the HHG spectrum of monolayer graphene and its comparison with the spectrum of the  bilayer graphene for a  linearly polarised laser pulse 
polarised along $x$-axis ($\Gamma$-K in the Brillouin zone).
Here, AB stacking of bilayer graphene is considered. The intensity of  the HHG spectrum 
is normalised with respect to  the total number of electrons in monolayer and bilayer graphenes. 
It is  apparent that the third harmonic (H3) is matching well  in both cases. 
However, harmonics higher than H3 show significantly different behaviour
as the interlayer coupling between the two layers plays a meaningful role.

The interband and intraband contributions to the total harmonic spectra for monolayer and bilayer graphene are shown in Figs.~\ref{fig3.2}(b) and (c), respectively. 
Both contributions play a  strong role to the total spectra as reflected from the figure. 
A strong interplay of interband and intraband contributions was reported for monolayer graphene~\citep{taucer2017nonperturbative, liu2018driving, al2014high}. 
Unlike the wide band-gap semiconductors~\citep{wu2015high},  the interband and intraband transitions take place at the same energy scales for both monolayer and bilayer graphenes~\citep{stroucken2011optical} due to the vanishing band-gap. 
The relative  (integrated) harmonic yield from interband and intraband contributions is plotted in the insets of Figs.~\ref{fig3.2}(b) and (c). Here, intraband contribution dominates upto H3, 
whereas interband contribution dominates for fifth (H5) and higher-order harmonics for both monolayer and bilayer graphenes. The enhanced contributions from interband transitions at higher orders can be attributed to the increased joint density of states at higher energies as shown in Fig.~\ref{fig3.2}(d).

Also, as the low-energy band structures are different for monolayer and bilayer graphenes  [Fig.~\ref{fig3.1}(c)], 
the nature of harmonic spectra is not obvious from the band-structure point of view. 
To have a better understanding of the underlying mechanism of the 
harmonic generation in both cases,  
the role of the interlayer coupling in HHG is discussed in the next subsection.

\subsubsection{Role of Interlayer Coupling in  HHG}

To understand how the interlayer coupling between two layers affects the harmonic generation in 
bilayer graphene, the harmonic spectrum as a function of interlayer coupling strength ($t_{\perp}$) is shown in Fig.~\ref{fig3.3}(a). 
Reducing the interlayer coupling strength is equivalent of moving the two layers of graphene
farther apart. 
The red dashed line in Fig.~\ref{fig3.3}(a) corresponds to the interlayer coupling used in simulations presented in Fig.~\ref{fig3.2}. 
It is evident from Fig.~\ref{fig3.3}(a) that  H5 and higher-order harmonics are sensitive 
with respect to $t_{\perp}$. 
Moreover,  different harmonic orders affected differently. 
Therefore, the yield of harmonic orders are non-linear functions of interlayer coupling.

To explore further how different hopping terms affect the HHG in bilayer graphene, 
an additional hopping term,  $t_3$, between B atoms of the top layer and A atoms of the bottom layer is introduced. The modified Hamiltonian for AB-stacked bilayer graphene can be written as
\begin{equation}
		\begin{split}
			\hat{\mathcal{H}}_{AB} =& -t_0 f(\textbf{k}) \left[\hat{a}^\dagger_{1k}\hat{b}_{1k} + \hat{a}^\dagger_{2k}\hat{b}_{2k}   \right] \\
			&+ t_\perp \hat{a}^\dagger_{2k}\hat{b}_{1k}  -t_3 f^*(\textbf{k}) \hat{a}^\dagger_{1k}\hat{b}_{2k} +\textrm{H.c.}
		\end{split}
\end{equation}
Here, a hopping energy $t_3$ of 0.3 eV is used~\citep{charlier1991first,min2007ab}.  
The corresponding harmonic spectrum is presented in Fig.~\ref{fig3.3}(b). 
It is evident from the figure that the additional interlayer coupling $t_3$ affects all the harmonics higher than H3. 
It is apparent that the interlayer coupling has a strong role in determining the non-linear response of bilayer graphene.

\begin{figure}
	\centering
	\includegraphics[width=0.8\linewidth]{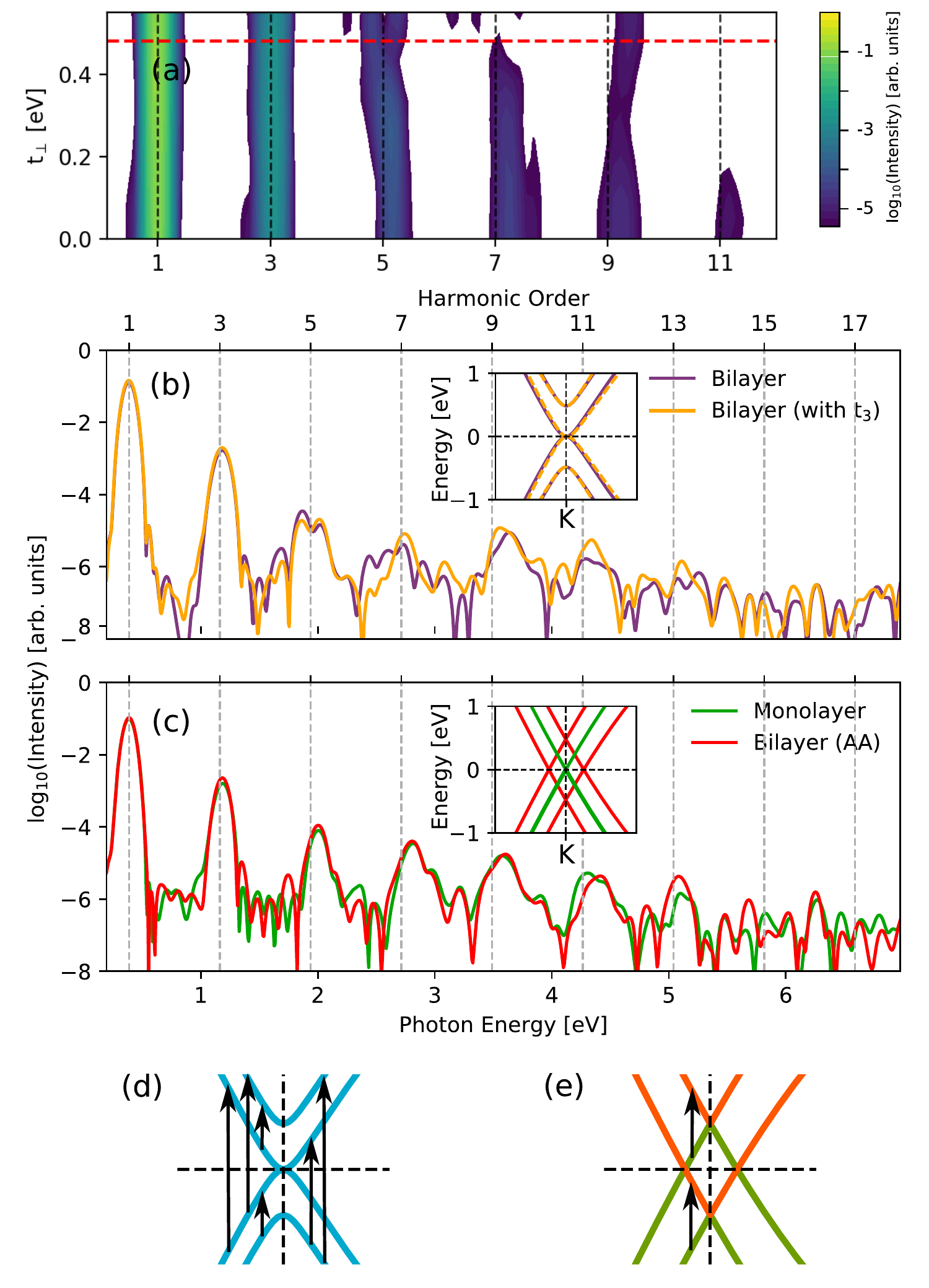}
	\caption[]{(a) High-harmonic spectrum  of bilayer graphene with AB stacking as a function of interlayer coupling ($t_{\perp}$), where $t_{\perp}$ = 0 corresponds to the HHG from monolayer graphene. The red dashed line corresponds to the actual value of $t_{\perp}$ used in the calculations. (b) HHG from bilayer graphene (in AB stacking) with $t_{\perp}$ (B$_1$-A$_2$) coupling only (violet)  and with both $t_{\perp}$ and $t_3$ (A$_1$-B$_2$) coupling (orange). (c) HHG from bilayer graphene with AA stacking (red) and monolayer graphene (green). 
		In (b) and (c), the band-structures near the $\bf{K}$-point are shown in the inset. (d) and (e) show the non-zero momentum matrix elements in AB and AA stacked bilayer graphene, respectively.}
	\label{fig3.3}
\end{figure}

Now let us discuss how HHG depends on different stacking configurations of the bilayer graphene. 
As discussed in the introduction, bilayer graphene can be realised in AA and AB stacking. 
AA stacking of  bilayer graphene is realised by stacking the monolayer precisely on  
top of the first layer. 
The top view of the AA-stacked bilayer looks exactly as a monolayer graphene [Fig.~\ref{fig3.1}(a)], 
where A$_1$ couples with A$_2$ and B$_1$ couples with B$_2$ with a coupling strength of $t_{\perp}$. 
The harmonic profile of the bilayer graphene with AA configuration matches well with the spectrum of monolayer graphene as presented in Fig.~\ref{fig3.3}(c).

The band structures near the $\bf{K}$-point for AB and AA stacked bilayer graphene are shown in the insets of Figs.~\ref{fig3.3}(b) and (c), respectively.  
For AB-stacked bilayer graphene, a slight change in band-structure results in a significant change in the spectrum [see Fig.~\ref{fig3.3}(b)]. 
On the other hand, for AA-stacked bilayer graphene, the difference in the band-structure is not reflected in the spectrum [see Fig.~\ref{fig3.3}(c)].

A better understanding about the HHG mechanism can be deduced by considering the roles of the band structure as well as the interband momentum-matrix elements.   
The energy bands of the AA-stacked bilayer graphene within nearest neighbour tight-binding approximation are given by 

\begin{equation}
	\mathcal{E}(\textbf{k}) = \pm t_{\perp} \pm t_0 |f(\textbf{k})|.
\end{equation}
This is equivalent to the shifted energy bands of monolayer graphene by $\pm t_{\perp}$. 
Also the corresponding momentum matrix elements give non-zero values only for the pairs $t_{\perp} \pm t_0 |f(\textbf{k})|$ and $-t_{\perp} \pm t_0 |f(\textbf{k})|$ as shown in Fig.~\ref{fig3.3}(e).  
On the other hand, in AB-stacked bilayer graphene, all pairs of bands have non zero momentum matrix elements near the $\bf{K}$-point as shown in Fig.~\ref{fig3.3}(d). 
The similar band dispersion and JDOS compared to monolayer graphene result  
in similar harmonic spectrum for AA-stacked bilayer graphene. 
On the other hand, in bilayer graphene, an electron in the conduction band can recombine to a hole in 
any of the valence bands near the $\bf{K}$-points as shown in Fig~\ref{fig3.3}(d). 
These different interband channels interfere and therefore generate 
the resulting harmonic spectrum for the AB-stacked bilayer graphene.

From here onward only bilayer graphene with AB stacking is considered,  
as the HHG spectra of the monolayer and bilayer graphenes with AA stacking  are identical.  
In the succeeding subsections, we explore polarisation and ellipticity dependences of the HHG 
from monolayer and bilayer graphenes. 

\subsubsection{Polarisation-Direction Dependence on the High-Harmonic Spectrum}

The vector potential corresponding to a linearly polarised laser pulse can be defined as
\begin{equation}
	\textbf{A}(t) = A_0f(t) \cos(\omega t)\left[\cos(\theta)\hat{\textbf{e}}_x + \sin(\theta) \hat{\textbf{e}}_y\right].
\end{equation}
Here, $f(t)$ is the envelope function and $\theta$ is the angle between laser polarisation and the $x$-axis ($\Gamma$-K in the Brillouin zone). 

\begin{figure}[h!]
	\centering
	\includegraphics[width=0.8\linewidth]{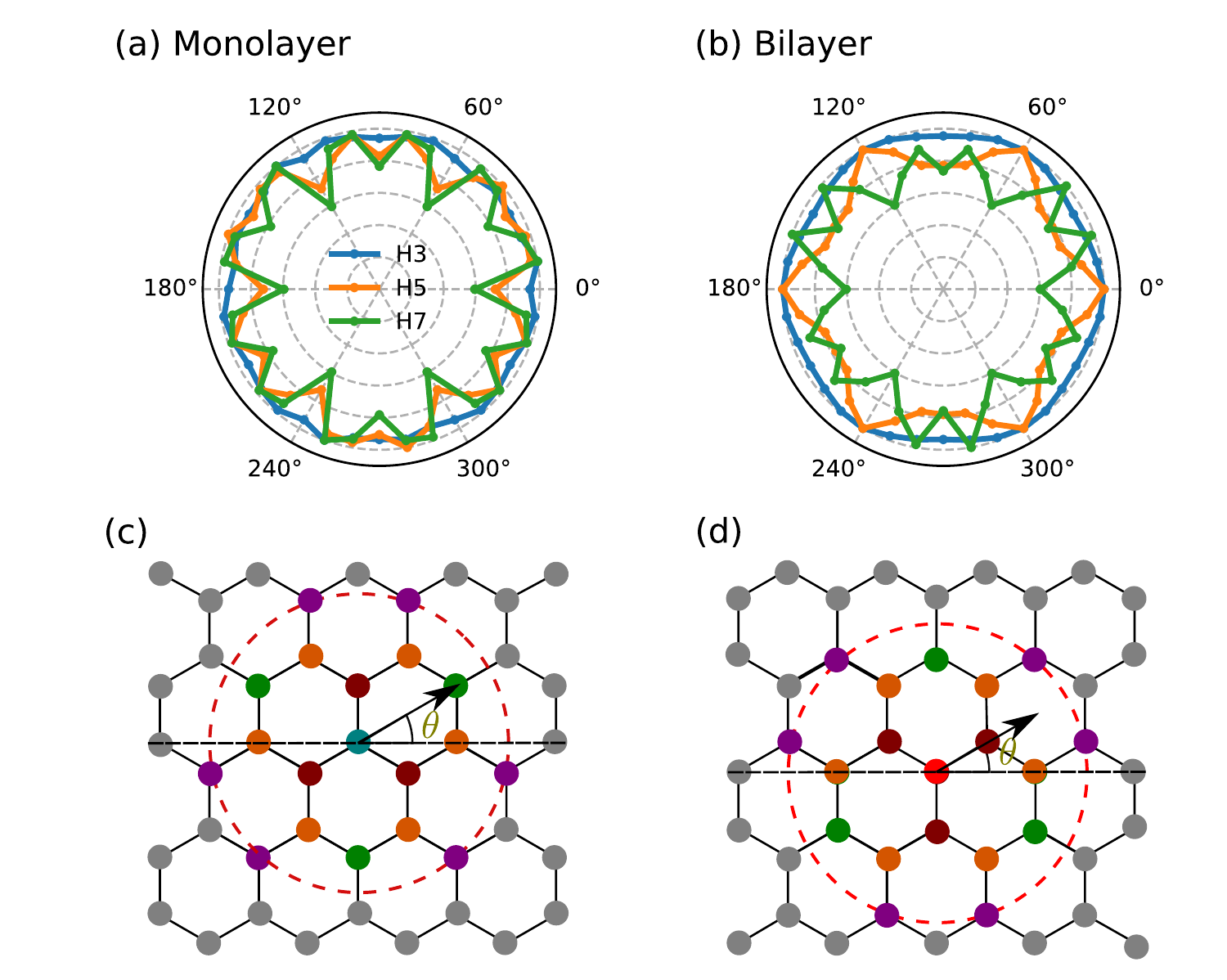}
	\caption{Polarisation dependence of the normalised harmonic yield for (a) monolayer and (b) bilayer graphenes.
		Here,  $\theta$ is  an angle between laser polarisation and the 
		$x$-axis along  $\Gamma$-K in the Brillouin zone.  
		An illustration of the semi-classical real-space model with nearest neighbours of (c) A-type and (d) B-type carbon atoms. The first, second, third, and fourth nearest  neighbours are shown using brown, orange, green and violet colours, respectively. } \label{fig3.4}.
\end{figure} 

\begin{figure}[h!]
	\centering
	\includegraphics[width=\linewidth]{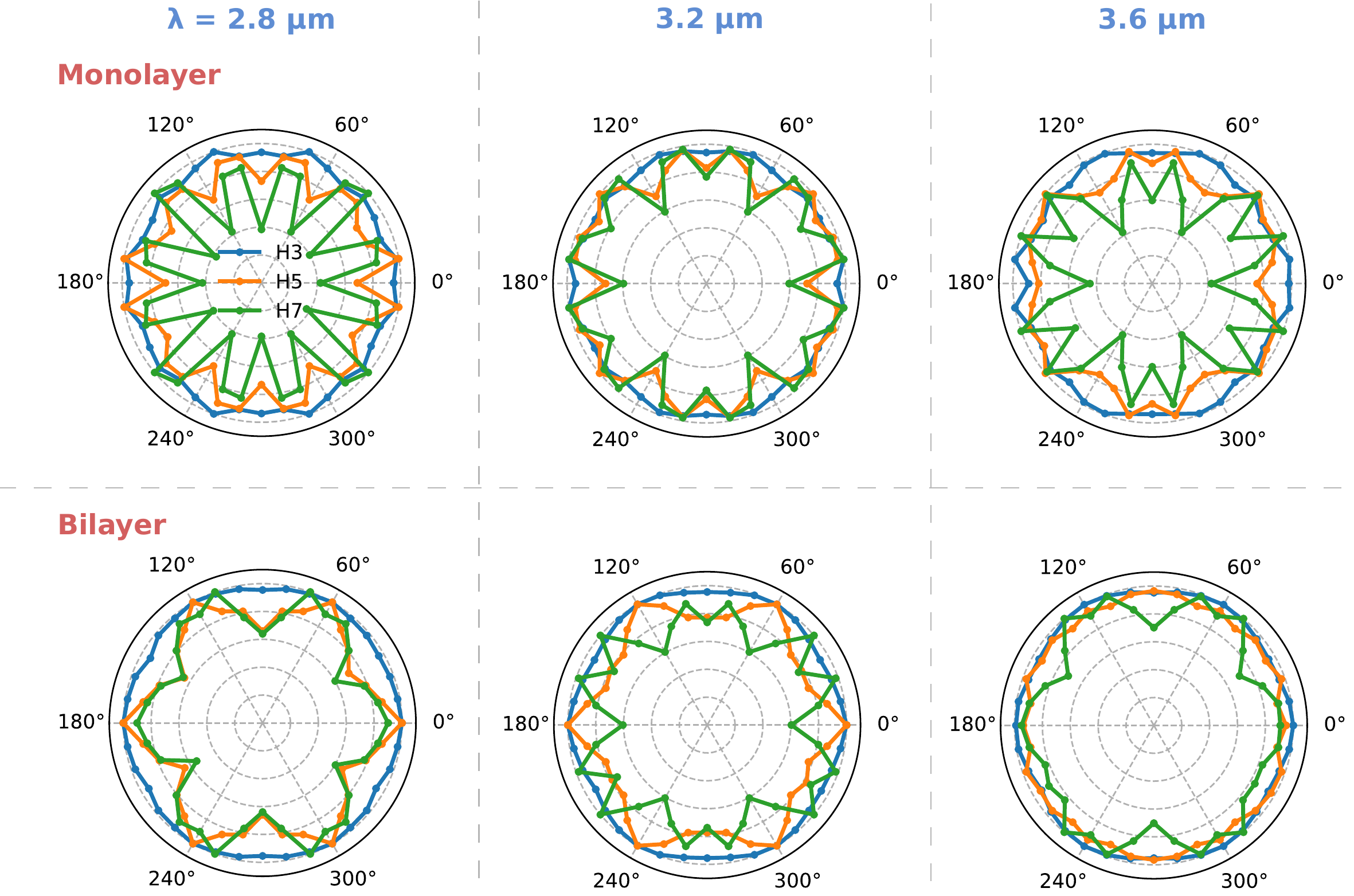}
	\caption{Polarisation dependence of the normalised harmonic yield for monolayer (top-panel) and bilayer (bottom-panel) graphene. Wavelength values of 2.8 $\mu$m, 3.2 $\mu$m, and 3.6 $\mu$m are used.}.\label{fig:w_dep}
\end{figure}

The polarisation-direction dependence on the harmonic yield for  monolayer and bilayer graphenes is
 presented in Figs.~\ref{fig3.4}(a) and (b), respectively. All the harmonics mimic the six-fold symmetry of the graphene lattice.  
As reflected from the figure, H3 exhibits no  significant polarisation sensitivity 
for both monolayer and bilayer graphenes.  
The reason for this isotropic nature can be attributed to the  isotropic nature of the energy 
bands near $\bf{K}$-points. 
However, harmonics higher than H3, show anisotropic behaviour in both  cases.  
Moreover, H5 of monolayer and bilayer graphene shows different polarisation dependence.
The harmonic yield is maximum for angles close to 15$^\circ$ and 45$^\circ$ in monolayer graphene.  

To understand the polarisation dependence of the harmonic yield in monolayer graphene, we employ  a semiclassical explanation as proposed in Refs.~\citep{you2017anisotropic, pattanayak2019direct} by assuming that 
the interband transitions can be translated to a semi-classical real-space model~\citep{kruchinin2018colloquium,parks2020wannier}. 
One-to-one correspondence between interband transition 
and inter-atom hopping in graphene was shown by Stroucken \textit{et al.}~\citep{stroucken2011optical}. 
Here, we assume that an electron can hop between two atoms when the laser is polarised along a direction in which it connects the atoms. The contributions to the harmonic yield from different pairs of atoms drop significantly as the distance between the atoms increases.  
This is in principle governed by the inter-atom momentum matrix elements~\citep{stroucken2011optical}. 
By assuming a finite radius for atoms, farther atoms show sharper intensity peaks as a function of angle of polarisation. 

Figures~\ref{fig3.4}(c) and (d)  show the nearest neighbours of A and B types of atoms in the unit cell, respectively. 
Brown, orange, green and violet coloured atoms are first, second, third and fourth neighbours, respectively. By only considering the nearest-neighbour hopping in graphene, we can see 
that the maximum yield should be for 30$^\circ$ (along $\Gamma$-M direction). However, the maximum yield is near 15$^\circ$ and 45$^\circ$  as seen from  Fig.~\ref{fig3.3}(a). 
This means that the contributions up to the fourth nearest neighbours should be considered to explain the polarisation dependence of H5 and seventh harmonic (H7) 
of monolayer graphene. 

In bilayer graphene, H7 follows the same qualitative behaviour as that of H5 and H7 of monolayer graphene [Fig.~\ref{fig3.3}(b)]. 
In contrast, H5 shows different behaviour and obeys the symmetry of the second nearest neighbour. It is clear from Fig.~\ref{fig3.3}(d) that there are multiple paths for interband transitions for bilayer graphene. 
In bilayer graphene, interband transitions from different pairs of valence and conduction bands can contribute to a particular harmonic, 
and these different transitions interfere. 
This makes the mechanism of harmonic generation from monolayer and AB-stacked bilayer graphene different. 

It is important  to point out that the polarisation dependence is sensitive  to the wavelength of driving laser pulse. We present the polarisation-dependence for laser pulses of wavelengths of 2.8 $\mu$m, 3.2 $\mu$m, and 3.6 $\mu$m in Fig.~\ref{fig:w_dep}. For longer wavelengths, electron dynamics occurs in the isotropic parts of the reciprocal space (close to $\mathbf{K}$-points), and as a result the harmonic spectrum can be entirely isotropic. We have confirmed that the different symmetry observed for monolayer and bilayer graphene is compatible with varying wavelength of the driving laser, and our explanation remains consistent.

\subsubsection{Ellipticity Dependence on the High-Harmonic Spectrum}~\label{sect:ell}
The high-harmonic  spectra for  monolayer and bilayer graphenes corresponding 
	to different polarisation of the driving laser pulse are shown in Fig.~\ref{fig3.5}.
The vector potential corresponding 
to the elliptically polarised pulse with an ellipticity $\epsilon$ is defined as
\begin{equation}
	\textbf{A}(t) = \frac{A_0f(t)}{\sqrt{1+\epsilon^2}}\left[\cos(\omega t)\hat{\textbf{e}}_x + \epsilon \sin(\omega t)\hat{\textbf{e}}_y\right].
\end{equation}
Here, the same laser parameters are used as mentioned in Section~\ref{section:2}. 
Both monolayer and bilayer graphenes show significant ellipticity-dependence in the harmonic yield.  A negligible harmonic yield is obtained for circularly polarised laser pulse. This indicates that using a single colour mid-infrared circular driver is not an appropriate 
	method to generate circularly polarised harmonics from graphene. This has already been  experimentally demonstrated~\citep{taucer2017nonperturbative,yoshikawa2017high}. 
	Recent theoretical studies revealed that efficient generation of circularly polarised harmonics is possible from graphene either by using a near-infrared circular laser pulse~\citep{chen2019circularly} or by using mid-infrared bi-circular counter-rotating laser pulses~\citep{mrudul2021light}. 
	The harmonic spectrum corresponding to bilayer graphene shows the ($6n\pm1$) harmonic orders for circularly polarised laser, as expected from the symmetry considerations~\citep{gupta2003generation}. To have a better understanding about the variation of  the harmonics as a function of ellipticity of the driving laser, we show the integrated harmonic yield below.

\begin{figure}[t!]
	\centering
	\includegraphics[width=0.8\linewidth]{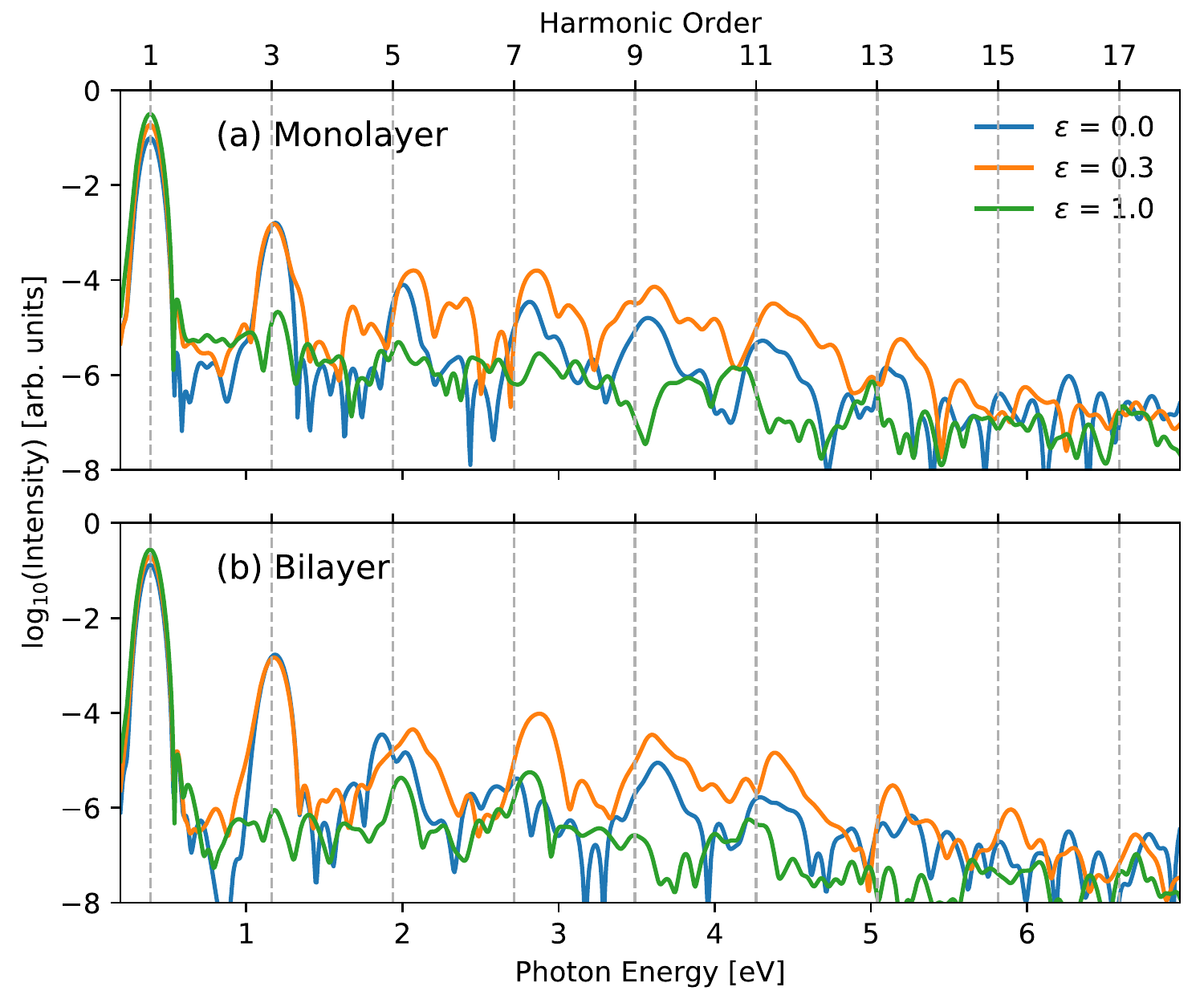}
	\caption{High-harmonic  spectrum for (a) monolayer and (b) bilayer graphenes  
		for different ellipticities of the driving laser pulse. 
		Here, $\epsilon$ = 0 corresponds to a linearly polarised pulse 
		and $\epsilon$ = 1 corresponds to a circularly polarised pulse.} \label{fig3.5}
\end{figure}

\begin{figure}[t!]
	\includegraphics[width=\linewidth]{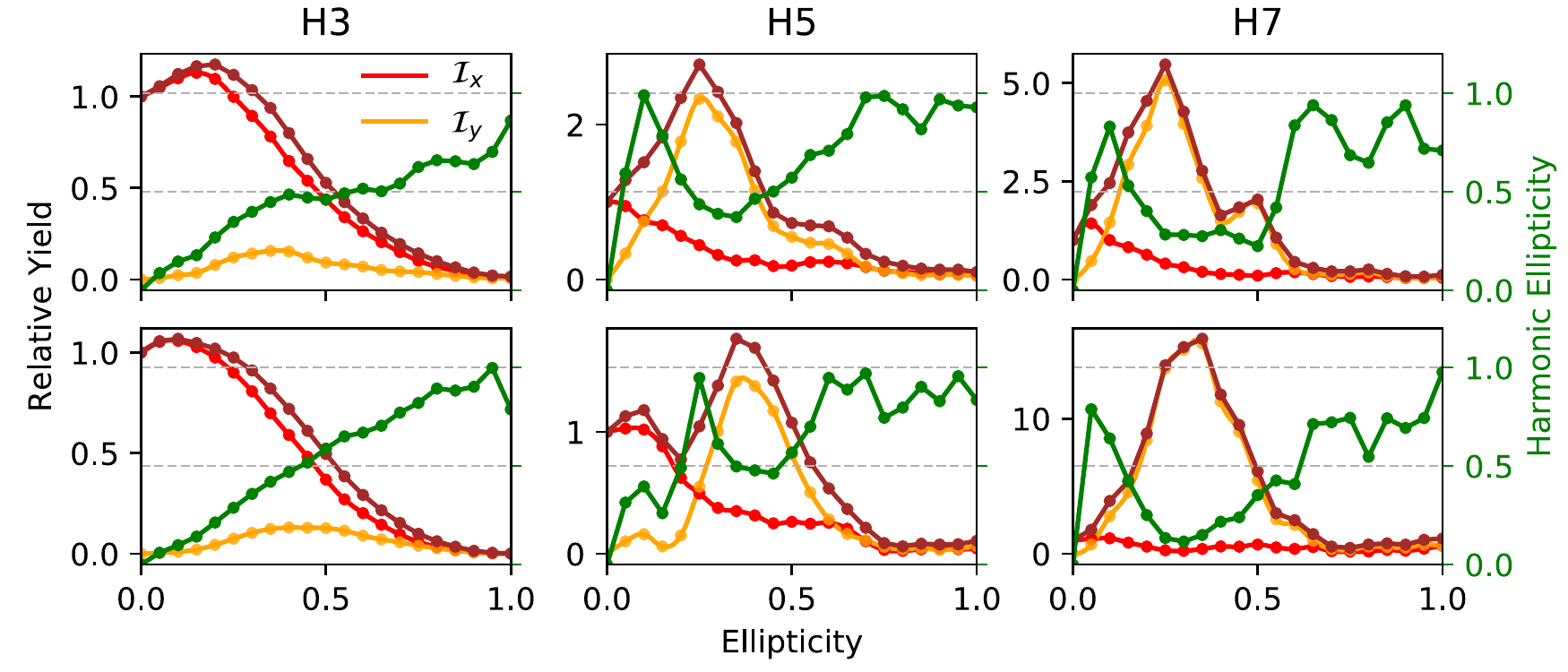}
	\caption{ Ellipticity dependence of the  integrated harmonic yield for 3$^{\textrm{rd}}$ (H3), 5$^{\textrm{th}}$ (H5), and 7$^{\textrm{th}}$ (H7) harmonics of the monolayer (top panel) and bilayer graphenes (bottom panel). 
		The integrated harmonic yield, and its $x$ and $y$-components are shown by brown, red and orange colour lines, respectively. The green line shows the ellipticity dependence on the averaged ellipticity of the emitted harmonics.  
		An elliptically polarised pulse with an intensity of 1$\times$10$^{11}$ W/cm$^2$ is used for HHG. } \label{fig3.6}
\end{figure} 

The harmonic yield as a function of ellipticity for the monolayer (top panel) and bilayer graphenes (bottom panel) are presented in Fig.~\ref{fig3.6}.
The total harmonic yield is normalised with respect to the  harmonic yield for $\epsilon$ = 0. 
The ellipticity dependence of all the three harmonics agrees qualitatively well for monolayer and bilayer graphenes. The atomic-like ellipticity dependence of H3 can be attributed to its isotropic nature [see first column of Fig.~\ref{fig3.6} ]. 
However, H5 and H7 show a characteristic ellipticity dependence. 
The harmonic yield has a maximum for a finite value of the ellipticity and is polarised along the normal direction of the major axis of the ellipse. 
This interesting feature was observed for the monolayer graphene experimentally 
and explained as a consequence of the semi-metallic nature of the monolayer graphene~\citep{yoshikawa2017high}. Since bilayer graphene is also semi-metallic,  it is also expected to exhibit similar ellipticity dependence, 
which we confirm here.

The different qualitative behaviours of the ellipticity dependence of H3 compared to H5 and H7 
are also consistent with the findings that the interband and intraband mechanisms respond differently to the ellipticity of the driving laser~\citep{tancogne2017ellipticity} [see also insets of Figs.~\ref{fig3.2}(b) and (c)]. The characteristic ellipticity dependence of monolayer graphene was shown to be dominated by interband contributions in Ref.~\citep{liu2018driving}. The ellipticity of the maximum yield is different for bilayer graphene as a consequence of interlayer coupling.

The averaged ellipticity of the emitted harmonics as a function of the laser ellipticity shows interesting behaviour as  shown in Fig.~\ref{fig3.6} (see green colour). The averaged ellipticity of H3 of monolayer and bilayer graphene shows monotonically increasing behaviour. On the other hand, the behaviour is highly nonlinear for harmonics higher than H3. It is also interesting to note that harmonics with higher ellipticity can be obtained by a nearly linearly polarised pulse ($\epsilon <$0.3).

\subsection{HHG from Gapped Graphene}\label{section:3.2}

The Hamiltonian for gapped graphene can be modelled by adding different onsite energy at A and B sublattices of graphene [see Fig.~\ref{fig3.1}(a)]. Equation~(\ref{eq:3.1}) modifies as
\begin{equation}
	\hat{\mathcal{H}_g} = -t_0 f(\textbf{k})a_k^\dagger b_k - \frac{\Delta_g}{2}a_k^\dagger a_k + \frac{\Delta_g}{2} b_k^\dagger b_k + \textrm{H.c.}
\end{equation}
Here, $\Delta_g$ is the minimum band-gap energy at $\mathbf{K}$-points in the Brillouin zone. The energy eigenvalues of the gapped graphene are given by 
\begin{equation}
	\mathcal{E}(\textbf{k}) = \pm \sqrt{(\Delta_g/2)^2 + t_0^2|f(\textbf{k})|^2}.
\end{equation}

Adding a small energy gap essentially breaks the inversion symmetry of the lattice. This means that A and B atoms in Fig.~\ref{fig3.1}(a) represent entirely different atoms in this model. An example of such a system is monolayer h-BN.

\begin{figure}[t!]
	\includegraphics[width=\linewidth]{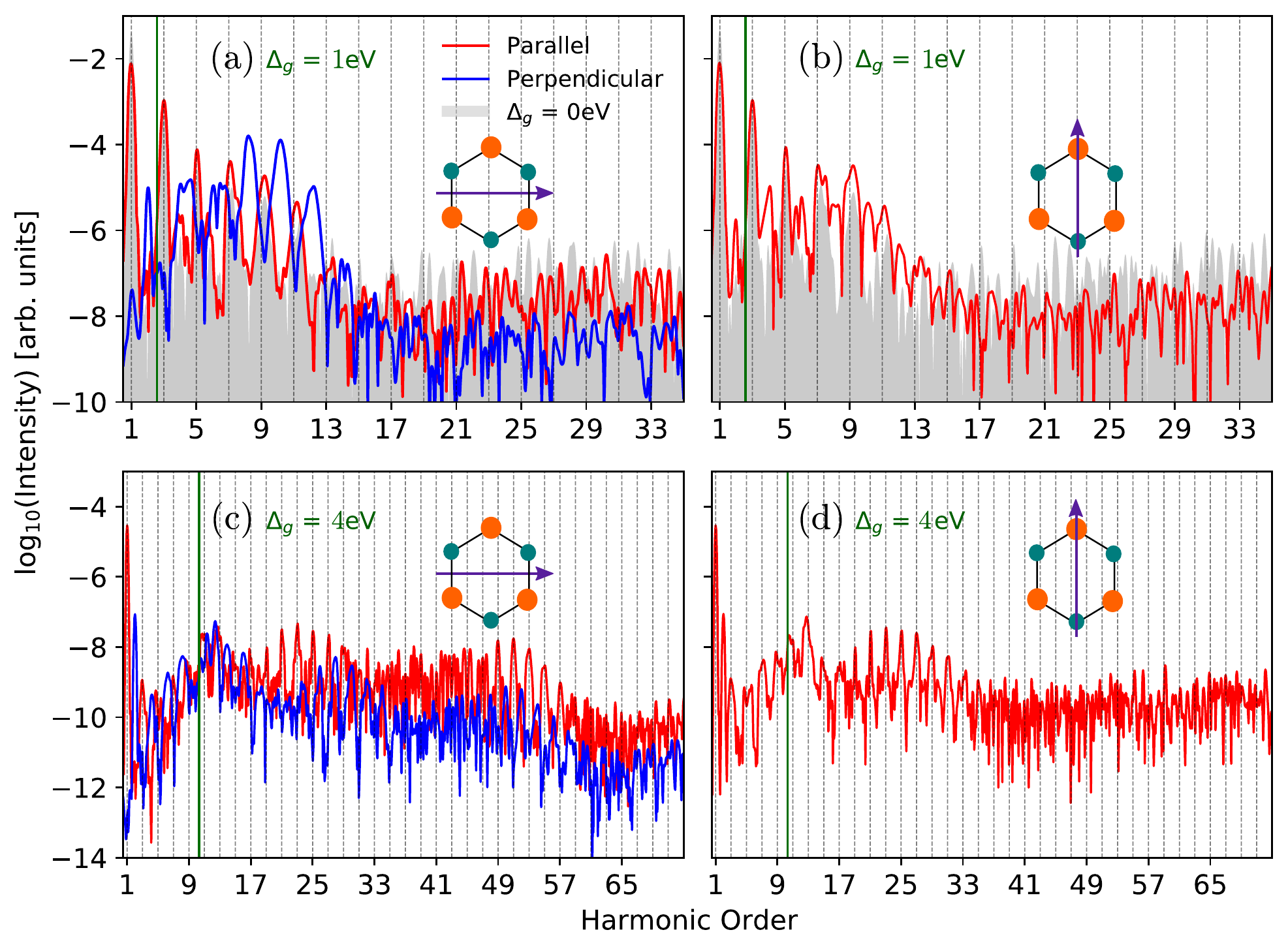}
	\caption{High-harmonic spectrum of gapped graphene with $\Delta_g$ = 1 eV [(a), (b)], and $\Delta_g$ = 4 eV [(c), (d)]. The green vertical line shows the band gap of the gapped graphene. The laser polarisation is along the $x$-axis in (a) and (c), and along the $y$-axis in (b) and (d). Red and blue lines show the harmonic signal parallel and perpendicular to the laser polarisation, respectively. The grey shaded region in (a) and (b) are the harmonic spectra for gapless graphene.} \label{fig3.7}
\end{figure}

In this section, we use the same laser parameters as mentioned in the previous section. A driving laser pulse with an intensity of 1$\times$10$^{11}$ W/cm$^2$ and a wavelength of 3.2 $\mu$m are used. The laser pulse has  eight-cycles in duration with a sin-squared envelope. 

The harmonic spectra corresponding to gapped graphene are presented in Fig.~\ref{fig3.7}. Here, $\Delta_g$ values of 1 eV [Fig.~\ref{fig3.7}(a), (b)] and 4 eV [Fig.~\ref{fig3.7}(c), (d)] are considered. The left (right) panel of Fig.~\ref{fig3.7} shows the spectra calculated when the laser is polarised along the $x$-axis ($y$-axis). Here, the mirror plane of the crystal is along the $y$-axis (see inset of Fig.~\ref{fig3.7}). 

There is a stronger polarisation direction dependence in gapped graphene as evident from figure.
Both even and odd harmonics are present in the harmonic  spectrum. 
Even harmonics are generated perpendicular to the laser field when the laser is polarised along the 
$x$-axis. On the other hand, even and odd  harmonics are generated parallel to the laser field when the laser is along the $y$-axis. Moreover, for $\Delta_g = 1$ eV, the odd harmonics spectrum is comparable to the HHG spectrum for gapless graphene [see Fig.~\ref{fig3.2} (a)]. In contrast, for $\Delta_g = 4$ eV,  the intensity of the harmonics are decreased by at least two orders, and the characteristics of the harmonic spectrum are changed drastically.

\begin{figure}[t!]
	\includegraphics[width=\linewidth]{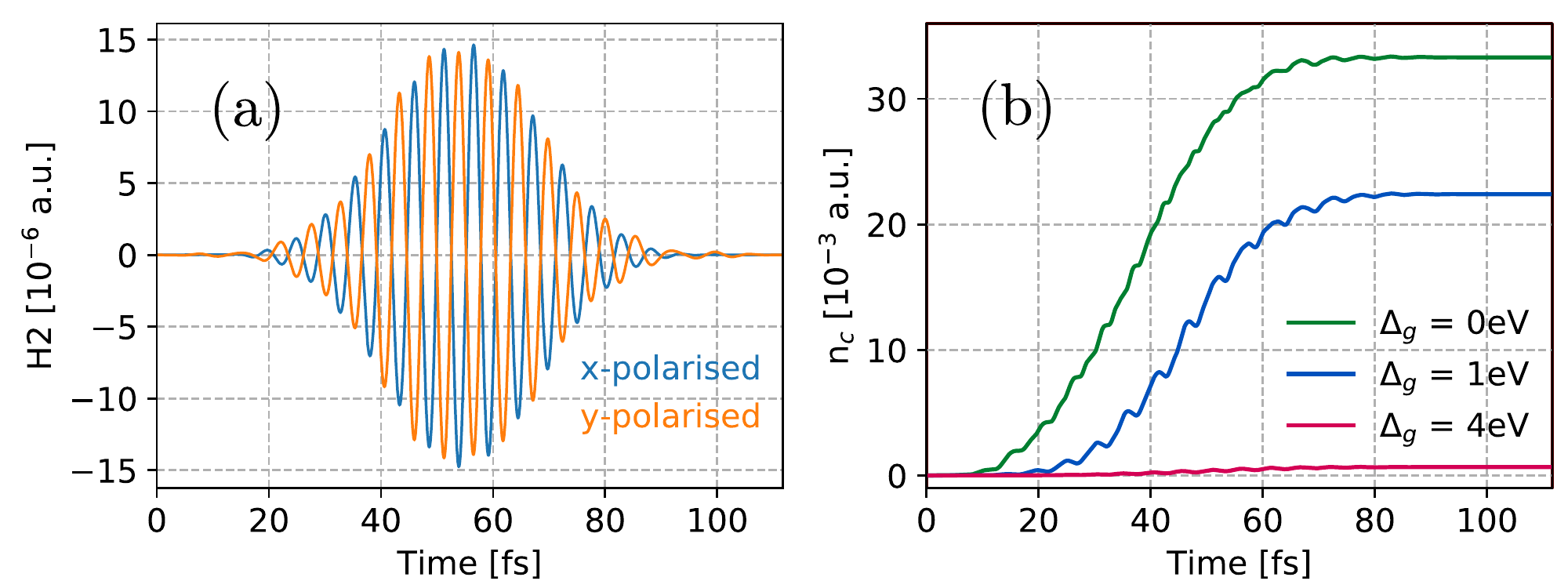}
	\caption{(a) Second-harmonic (H2) generated from gapped graphene ($\Delta_g$ = 4 eV). The laser polarisation is along the $x$-axis for blue line (see Fig.~\ref{fig3.7}(c)), and along the $y$-axis for orange line [see Fig.~\ref{fig3.7}(d)]. (b) Integrated conduction-band population during the laser pulse.} \label{fig3.8}
\end{figure} 

The point group symmetry is reduced from $\mathbf{D}_{6h}$ to $\mathbf{D}_{3h}$ in 
gapped graphene as inversion symmetry is broken. 
As a consequence,  we see even harmonics in the spectrum. 
Here, we present the symmetry of the second-order nonlinear susceptibility of gapped graphene to understand the selection rules for even harmonics. 
For any material, the contributions from the second-order nonlinear susceptibility tensor $\chi^{(2)}_{\mu \nu\lambda}$ to the polarisation is given by

\begin{equation}
	P_\mu^{(2)} = \sum_{\nu,\lambda}\chi^{(2)}_{\mu \nu\lambda}\mathcal{F}_\nu \mathcal{F}_\lambda.
\end{equation}
Here, $\mathcal{F}_\mu$ is the electric field in the $\mu$-direction. 
For a material with inversion symmetry, all elements of the $\chi^{(2)}$ tensor is supposed to be zero. 
In contrast, for a material without inversion symmetry, the non-zero elements of $\chi^{(2)}$ are dictated by the corresponding point group symmetry. 
For $\mathbf{D}_{3h}$ point group,  non-zero elements of $\chi^{(2)}$ hold the following relation:  $\chi^{(2)}_{yyy}$ = -$\chi^{(2)}_{yxx}$ = $\chi^{(2)}_{xxy}$ = -$\chi^{(2)}_{xyx}$~\citep{boyd2020nonlinear}. 
When the laser is polarised along the $x$-axis, contribution to the polarisation vector (dipole moment per unit volume) from $\chi^{(2)}$ is given by 
$P_y^{(2)} = \chi^{(2)}_{yxx}\mathcal{F}_x \mathcal{F}_x$. 
Similarly, for a laser polarised along the $y$-axis, $P_y^{(2)} = \chi^{(2)}_{yyy}\mathcal{F}_y \mathcal{F}_y$. These clearly explain the selection rules of even harmonics in gapped graphene for a linearly polarised electric field. 
To further support our claim, it is expected from $\chi^{(2)}_{yyy} = -\chi^{(2)}_{yxx}$ that  
the resultant second harmonic is $y$-polarised but differ by a sign for laser polarised along $x$- 
and $y$-directions. 
This claim is established by presenting the second harmonic for $x$- and 
$y$-polarised laser in Fig.~\ref{fig3.8}(a).  

We present the integrated conduction-band population  
during the laser to examine the electron dynamics in gapped graphene in Fig.~\ref{fig3.8}(b). 
The conduction band population decreases drastically as the gap increases. Clearly, the interband electron transitions are diminished in gapped graphene. This is also the reason for the less intense harmonic spectrum of gapped graphene [see Fig.~\ref{fig3.7}(c)-(d)].

\begin{figure}[t!]
	\includegraphics[width=\linewidth]{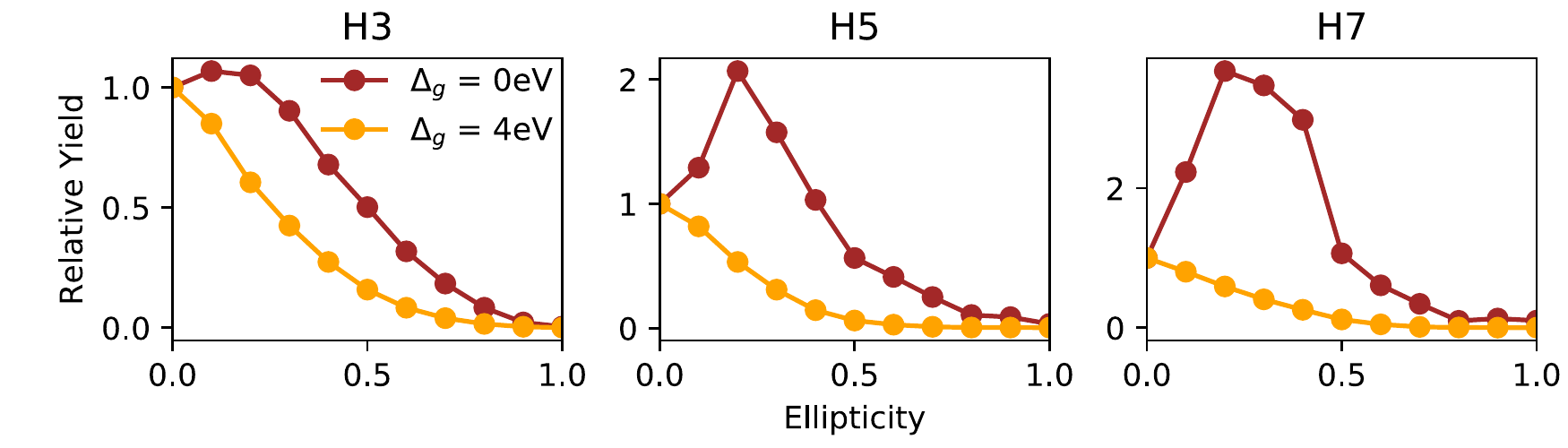}
	\caption{Ellipticity dependence on the  integrated harmonic yield for 3$^{\textrm{rd}}$ (H3), 5$^{\textrm{th}}$ (H5), and 7$^{\textrm{th}}$ (H7) harmonics of gapless (brown) and gapped (orange) graphene. The laser is polarised along $x$-axis.} \label{fig3.9}
\end{figure} 

We can see clean or noisy harmonics in different parts of the harmonic spectrum.  
The reason can be understood as follows: both interband and intraband current contribute to the harmonic spectrum and  their interference results in noisy harmonics. 
When one of the channels is suppressed, we get clean harmonics~\citep{tancogne2017impact}.  For example, in the below bandgap regime, only intraband dynamics are possible, resulting in clean harmonics in this regime.

Ellipticity dependence on the integrated harmonic yield of gapped ($\Delta_g$ = 4 eV) and gapless graphene are presented in Fig.~\ref{fig3.9}. Unlike monolayer or bilayer graphene (see Fig.~\ref{fig3.6}), the low-order harmonics lack anomalous ellipticity dependence in gapped graphene. This further supports the claim that this feature in gapless graphene originates from the semi-metallic nature of strong field interaction~\citep{yoshikawa2017high}. Moreover, we have discussed the significance of interband current on the anomalous ellipticity dependence of gapless graphene in Section~\ref{sect:ell}. On the other hand, the ellipticity dependence on the low-order harmonics of gapped-graphene originates entirely from intraband dynamics, resulting in atom-like ellipticity dependence.

Gapped graphene has non-zero Berry curvature as a consequence of inversion symmetry breaking. The velocity of an electron in $n^{\textrm{th}}$ energy band is given by
\begin{equation}\label{eq3.12}
	\textbf{v}_n(\textbf{k}) = \nabla_{\textbf{k}}\mathcal{E}_n(\textbf{k}) - \mathcal{F}(t)\times\Omega_n(\textbf{k}).
\end{equation}

The first term on the right originates from the band-dispersion and the second term 
stems from the Berry curvature $\Omega_n$. 
This semiclassical electron dynamics in a particular band is directly linked to the intraband current. 
This means that the perpendicular even harmonics in Fig.\ref{fig3.7}(a), (c) arise from the material's Berry curvature. 
In our calculation, the diagonal elements of the dipole matrix elements, $\textbf{d}_{mn}$ gives the Berry connection, 
i.e., $\mathcal{A}_n(\textbf{k}) = \textbf{d}_{nn}(\textbf{k}) = i\left\langle u_{n,\textbf{k}} |\nabla_\textbf{k}| u_{n,\textbf{k}} \right\rangle$. 
Berry curvature and Berry connection are related as $\Omega_n(\textbf{k}) = \nabla_\textbf{k}\times \mathcal{A}_n(\textbf{k})$. Harmonic spectrum with and without Berry connection is presented in Fig.~\ref{fig3.10}. It is evident that exclusion of Berry connection affects HHG spectrum 
considerably and the difference is strong near and below the band-gap energy region. This can be attributed to the fact that Berry curvature affects the intraband current.

\begin{figure}[t!]
	\centering
	\includegraphics[width=0.7\linewidth]{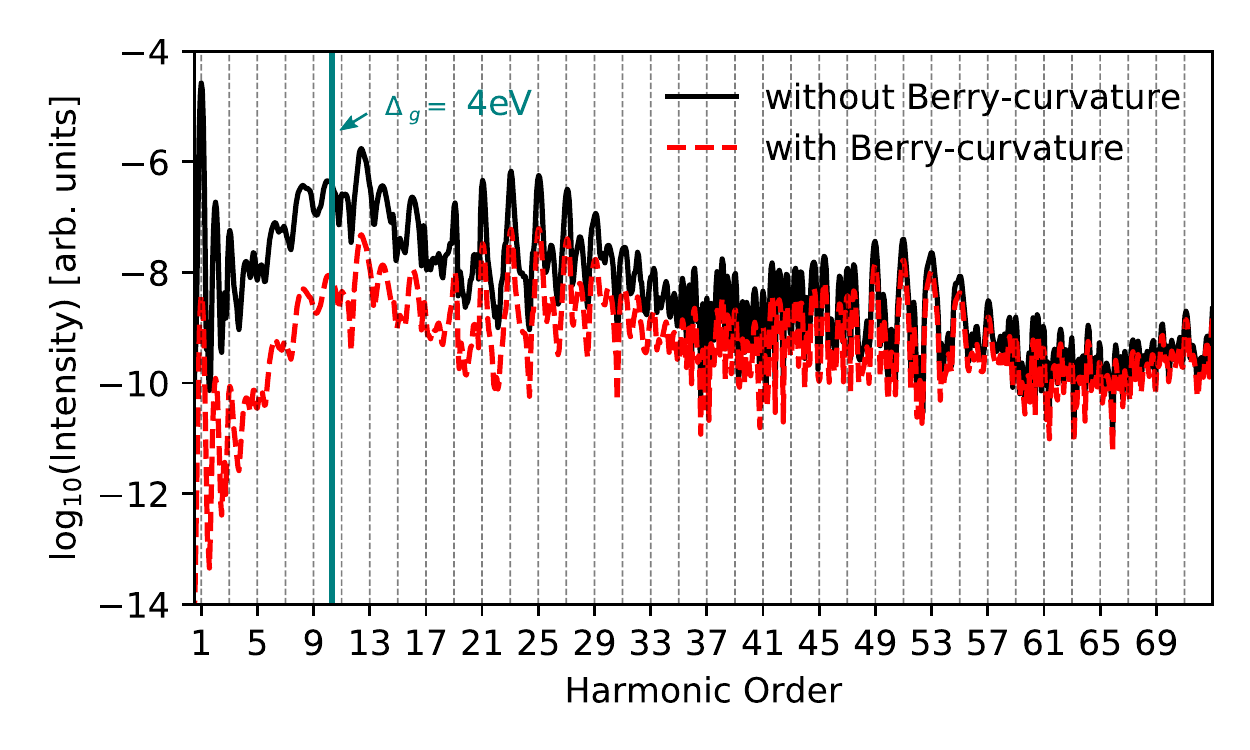}
	\caption{High-harmonic spectrum of gapped graphene with and without including Berry curvature. The laser polarisation is along the $x$-axis.} \label{fig3.10}
\end{figure} 

\section{Summary}\label{section:3}
In summary, HHG from monolayer, bilayer, and gapped graphenes is 
discussed.  Bilayer graphene having AA and AB stacking are considered in this chapter. 
The harmonic spectra of the monolayer and bilayer graphenes are significantly different and exhibit characteristic features of having a vanishing band-gap. 
The HHG spectrum of the bilayer graphene shows signatures of the interlayer coupling, which affects high-order harmonics non-linearly and different harmonics are affected differently. 
The role of interlayer coupling was also found to be stacking dependent, resulting in a similar harmonic spectrum for monolayer and bilayer graphenes with AA stacking.
A strong interplay of the interband and intraband contributions to the total harmonic spectrum is noticed. 
Moreover, distinct polarisation and ellipticity dependence are observed in monolayer and bilayer graphenes. The harmonic spectrum of gapped graphene is significantly different from gapless graphene. Breaking inversion symmetry results in even harmonics, and the selection rules can be explained using the point group symmetry of the material. The polarisation and ellipticity dependence is significantly different in gapless and gapped graphene. We also show the significance of Berry curvature in the harmonic spectrum of gapped graphene.

\cleardoublepage
\chapter{Light-induced Valleytronics in Pristine graphene}\label{Chapter4}

One of the most interesting features of graphene and gapped graphene
materials is the electron's extra degree of freedom, the valley pseudospin,  associated
with populating the local minima $\mathbf{K}$ and $\mathbf{K}^{\prime}$ in 
the lowest conduction band of the Brillouin zone.  
This extra degree of freedom has the potential 
to encode, process and store quantum information, opening the 
field of valleytronics~\citep{vitale2018valleytronics}. 

The monolayer graphene, as opposed to 
gapped graphene materials,
presents a fundamental challenge for valleytronics:  it has zero band-gap and  
zero Berry curvature. These aspects are generally considered to be a major impediment
for valleytronics. In gapped graphene materials, valley selectivity is achieved
by matching the helicity of a circularly polarized pump pulse, 
resonant to the band gap, to the sign of the 
Berry curvature~\citep{schaibley2016valleytronics, mak2012control, jones2013optical, gunlycke2011graphene, xiao2012coupled}. 
Recently demonstrated~\citep{langer2018lightwave} 
sub-cycle manipulation of electron population 
in  $\mathbf{K}$ and $\mathbf{K}^{\prime}$  valleys of 
tungsten diselenide, achieved with the combination of a resonant pump pulse locked to
the oscillations of the THz control pulse, represents a major milestone. Precise sub-cycle 
control over the driving light fields opens new opportunities for valleytronics,
such as those offered by the new concept of a topological 
resonance, discovered and analysed in Refs.~\citep{motlagh2019topological,motlagh2018femtosecond,kelardeh2016attosecond}.
Single-cycle pulses with the controlled phase of carrier oscillations 
under the envelope offer a route to valleytronics in gapped graphene-type materials 
even when such pulses are linearly, not circularly, polarized~\citep{jimenez2021sub}.
It is also possible to avoid the reliance on resonances in gapped graphene-type materials, 
breaking the symmetry between the $\mathbf{K}$ and $\mathbf{K}^{\prime}$ 
valleys via a light-induced topological phase transition, closing the 
gap in the desired valley~\citep{jimenez2019lightwave}.

Thus, with its zero band-gap, zero Berry curvature, and identical
dispersion near the bottom of the valleys, pristine
graphene appears unsuitable for valleytronics -- a disappointing conclusion in view of its 
exceptional transport properties. In this chapter, we discuss how 
this generally accepted conclusion is not correct, and that the 
preferential population of a desired valley
can be achieved by tailoring the polarization state of the driving light pulse to
the symmetry of the lattice. 
Present proposal offers an all-optical route to valleytronics in pristine graphene, 
complementing approaches based on  
creating a gap by using a heterostructure of graphene with hexagonal boron nitride~\citep{gorbachev2014detecting, 
	yankowitz2012emergence, hunt2013massive, rycerz2007valley}, or by adding strain and/or defect 
engineering~\citep{grujic2014spin, settnes2016graphene, faria2020valley, xiao2007valley}.  

We also show valley selectivity of harmonic generation in graphene and demonstrate
it with the same field as we use for valley-selective electronic excitation. This aid us to understand how the underlying electron dynamics are different. Last but not least, we describe an all-optical method for measuring valley-polarisation in graphene with a weak probe-pulse.

The key idea of our approach is illustrated in Fig.~\ref{fig4.1}, which shows graphene in real (a) and reciprocal (b)
space, together with the structure of the incident electric field (a) and the corresponding 
vector potential (b). The field is made by superimposing two counter-rotating 
circularly polarized colours at the frequencies $\omega$ and $2\omega$. The Lissajous
figure for the total electric field is a trefoil, and its orientation is controlled by the 
relative two-colour phase $\phi$, i.e., the sub-cycle two-colour delay measured in terms of $\omega$.  
In the absence of the field, the two carbon atoms A and B, in real space,
are related by the inversion symmetry. When the field is turned on, this inversion symmetry
is broken: the electric field in Fig.~\ref{fig4.1}(a) always points from the atom A to the one of the 
three neighbouring atoms B during the full laser cycle, but not the other way around.
If the centre of the Lissajous figure is placed on the atom B, the field points
in the middle between its neighbours.
One can control this symmetry breaking by rotating the trefoil, interchanging the
roles of the atoms A and B. Thus, the bi-circular field offers a simple, all-optical, ultrafast 
tool to break the inversion symmetry of the graphene lattice in a controlled
way. Such controlled symmetry breaking  naturally controls the relative 
excitation probabilities induced by the same laser field in the 
adjacent $\mathbf{K}$ and $\mathbf{K}^{\prime}$ valleys of the Brillouin zone. 

\begin{figure}[t!]
	\centering
	\includegraphics[width= 12cm]{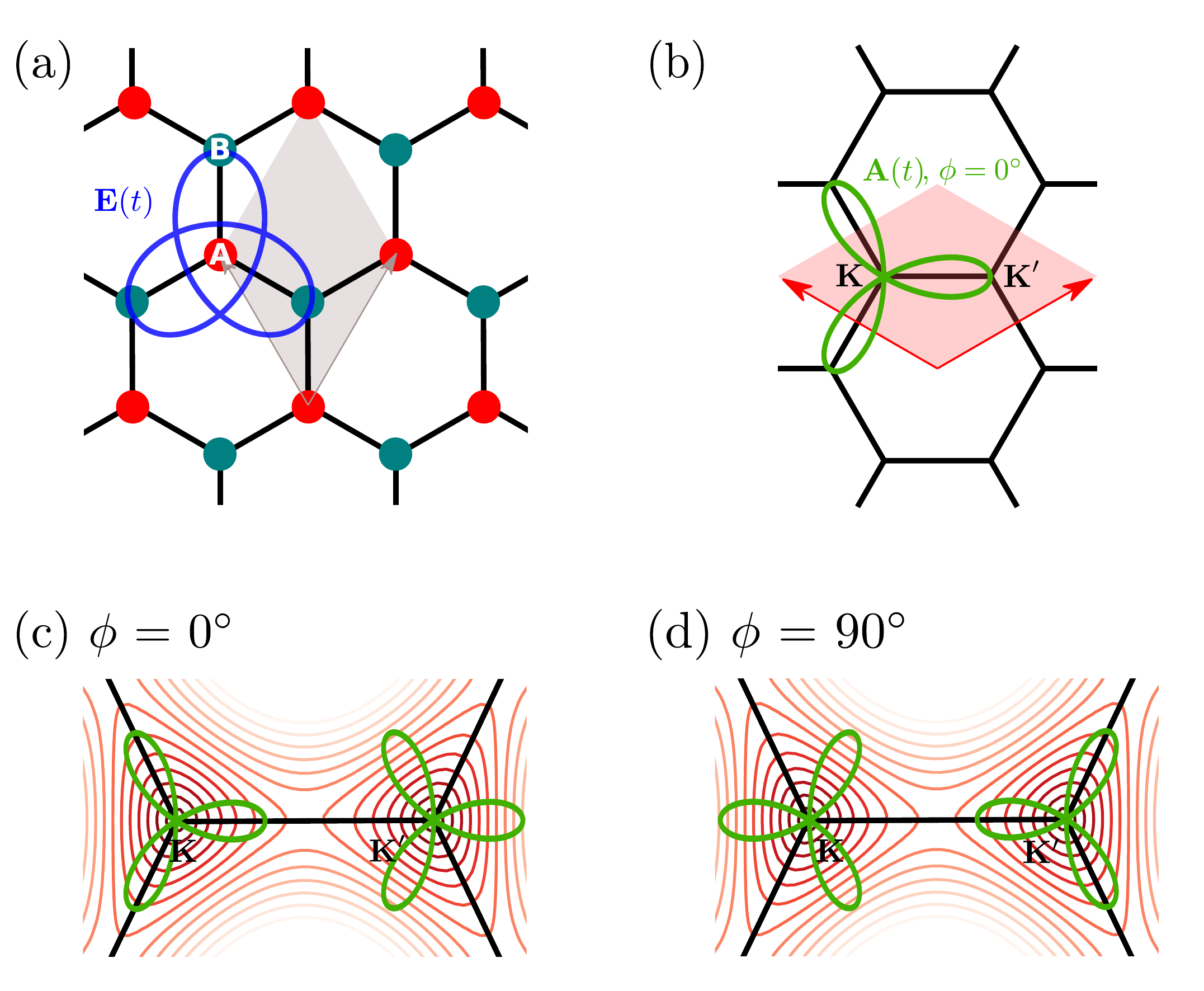}
	\caption{   
		(a) Graphene lattice in the coordinate-space, with the Lissajous 
		figure of the bicircular electric field breaking the symmetry between otherwise 
		identical carbon atoms  A and B; two-colour phase $\phi = 0^{\circ}$.
		(b) Optically induced symmetry breaking viewed 
		in the momentum space, the vector potential of the 
		bicircular field is shown for $\phi = 0^{\circ}$. (c, d) Close-up images of the 
		valleys show their asymmetry, leading to different laser-driven 
		dynamics and different excitation rates
		once the electron leaves the bottom of the valley. Here, the red contours show the conduction band energy in the reciprocal space.} 
	\label{fig4.1}
\end{figure} 

Figure~\ref{fig4.1}(b) provides
the complementary reciprocal-space perspective. In the laser field, the electron crystal momentum
follows  ${\bf k}(t)={\bf k}_i+\mathcal{A}(t)$\footnote{This is popularly known as acceleration theorem.}, where ${\bf k}_i$ is the initial crystal momentum and $\mathcal{A}(t)$ is the vector potential, 
shown in Fig.~\ref{fig4.1}(b) for the 
electric field shown in Fig.~\ref{fig4.1}(a). The asymmetry between the two valleys with respect
to the vector potential is immediately visible. 

Figures \ref{fig4.1}(c) and \ref{fig4.1}(d) provide additional support to this qualitative picture. 
One can observe that the two field-free 
valleys, $\mathbf{K}$ and $\mathbf{K}^{\prime}$, are only identical near their very bottoms. As soon
as one moves away from the bottom, the valleys start to develop 
the trefoil structure, with $\mathbf{K}$ and $\mathbf{K}^{\prime}$ being the mirror images of each other. 
How the symmetry of the vector potential fits into the symmetry of the 
valley away from their bottoms will control the dynamics and the excitation
probability. 

In sufficiently strong fields, excitation happens not only
at the Dirac point where the gap is equal to zero, but also
in its vicinity. For the vector potential in Fig.~\ref{fig4.1}(c), the average
gap seen by the electron when following the vector potential from
the Dirac point in the ${\bf K}$ valley, i.e., moving
along the trajectory ${\bf K}+\mathcal{A}(t)$,  
is less than in the $\mathbf{K}^{\prime}$ valley, i.e., when 
following the trajectory $\mathbf{K}^{\prime}+\mathcal{A}(t)$. 
In sufficiently strong and low-frequency fields, such that 
the band gap along the trajectories ${\bf K}+\mathcal{A}(t)$ and  
$\mathbf{K}^{\prime}+\mathcal{A}(t)$ quickly exceeds the photon energy, the 
excitation probability should therefore be higher in Fig.~\ref{fig4.1}(c). For the same reason, rotating the vector potential as shown in Fig.~\ref{fig4.1}(d) should favour population of the $\mathbf{K}^{\prime}$ valley. Here, ``quickly" means ``within a fraction of one-third of the laser cycle", which is relevant time-scale for the bicircular field. In this context lower laser frequencies leading to higher vector-potential are best suited to meet this requirement

\section{Numerical Methods}
In the simulations, we used the nearest-neighbour tight-binding approximation to obtain the ground state of graphene with a 
hopping-energy of 2.7 eV~\citep{reich2002tight}. The lattice parameter of graphene is chosen to be 2.46 \AA. 
The resultant band-structure has zero band-gap with linear dispersion near the two points in the Brillouin zone 
known as $\mathbf{K}$ and $\mathbf{K}^{\prime}$ points as shown in Fig.~\ref{fig3.1}. The density matrix approach was used to follow the electron dynamics in graphene. 
Time-evolution of density matrix element, $\rho^{\textbf{k}}_{mn}$,  was performed using SBE
within the Houston basis $| n, \textbf{k} + \mathcal{A}(t) \rangle $~\citep{golde2008high, floss2018ab}, as discussed in Chapter 2 [see Eq.~(\ref{eqn:SBE})]. 
We sampled the Brillouin zone with a 180$\times$180 k-point grid. SBE is solved using fourth-order Runge-Kutta method with a time-step of 0.8 a.u. 

\section{Results and Discussion}
The valley-selective electron-dynamics under the tailored field is confirmed by our numerical simulations. In the simulations, graphene is exposed to the bicircular field with the 
vector potential 
\begin{equation}\label{eq4}
	\mathcal{A}(t) = \frac{A_0 f(t)}{\sqrt{2}} \left(\left[\cos(\omega t + \phi) + \frac{\mathcal{R}}{2} \cos(2\omega t)\right]\hat{\textbf{e}}_x 
	+  \left[\sin(\omega t + \phi) - \frac{\mathcal{R}}{2} \sin(2\omega t)\right]\hat{\textbf{e}}_y\right) .
\end{equation}
Here, $A_0=F_{\omega}/\omega$ is the amplitude of the vector potential
for the fundamental field, $F_{\omega}$ is its strength, $f(t)$ is the temporal envelope of the driving field, 
$\phi$ is the sub-cycle phase difference between the two fields, and $\mathcal{R}$ is the ratio
of the electric field strengths for the two fields, leading to $\mathcal{R}/2$ ratio for the 
amplitudes of the vector potentials.  
The amplitude of the fundamental field was varied up to $F_{\omega}$ = 15 MV/cm, leading to 
the maximum fundamental intensity 3$\times$10$^{11}$ W/cm$^2$, with the fundamental wavelength  
$\lambda = 6~\mu$m. This laser intensity is weaker than the one used in experiments for monolayer graphene~\citep{yoshikawa2017high,heide2018coherent,higuchi2017light} and is below the damage threshold. 
We have also
varied $\mathcal{R}$, using  $\mathcal{R} = 1$ and $\mathcal{R} = 2$.   
A pulse duration of 145 femtosecond with sin-square envelope is employed in this work. Our findings 
are valid for a broad range of wavelengths and field intensities. 
To obtain  the total population of the
different valleys, we have integrated the momentum-resolved population  
over the sections shown in Fig.~\ref{fig4.2}(a). 

\begin{figure}[t!]
	\includegraphics[width=12cm]{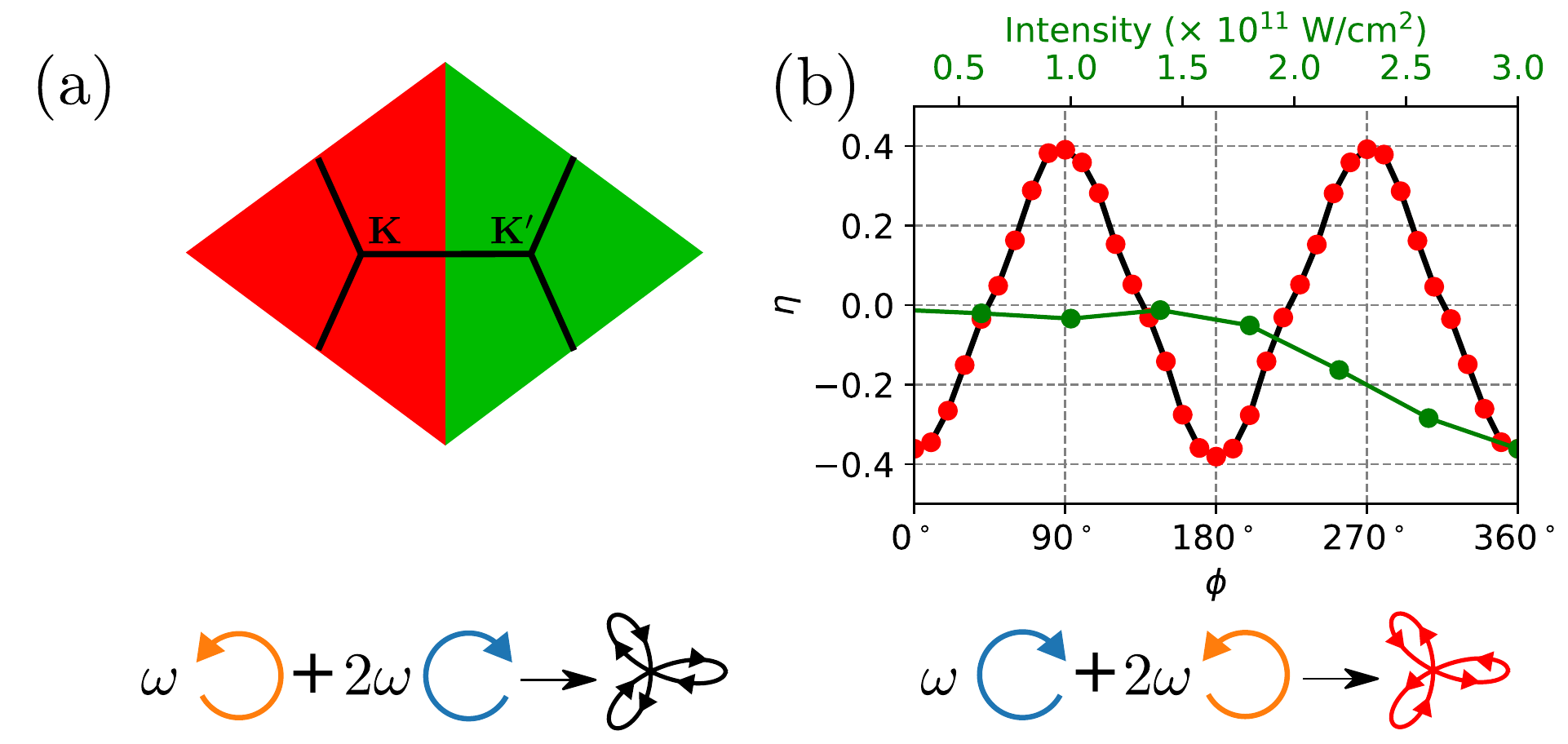}
	\centering
	\caption{ (a) Separation of the Brillouin zone into the 
		$\mathbf{K}$ and $\mathbf{K}^{\prime}$ valleys. (b) Asymmetry in the 
		valley-resolved populations in the conduction band  as a function of $\phi$ (black line with an intensity 3$\times$10$^{11}$ W/cm$^2$), and laser intensity (yellow line with $\phi = 0^{\circ}$) for 
		a laser with wavelength of 6 $\mu$m, $\mathcal{R}$ = 2, and a dephasing time  $T_{2}$ 
		of 10 $fs$. 
		The red dots show the asymmetry as a function of $\phi$ after switching the helicities of both the fields as shown in the bottom panel.} 
	\label{fig4.2}
\end{figure}

\subsection{Light-Induced Valley-Polarisation}

\begin{figure}[h!]
	\includegraphics[width=13cm]{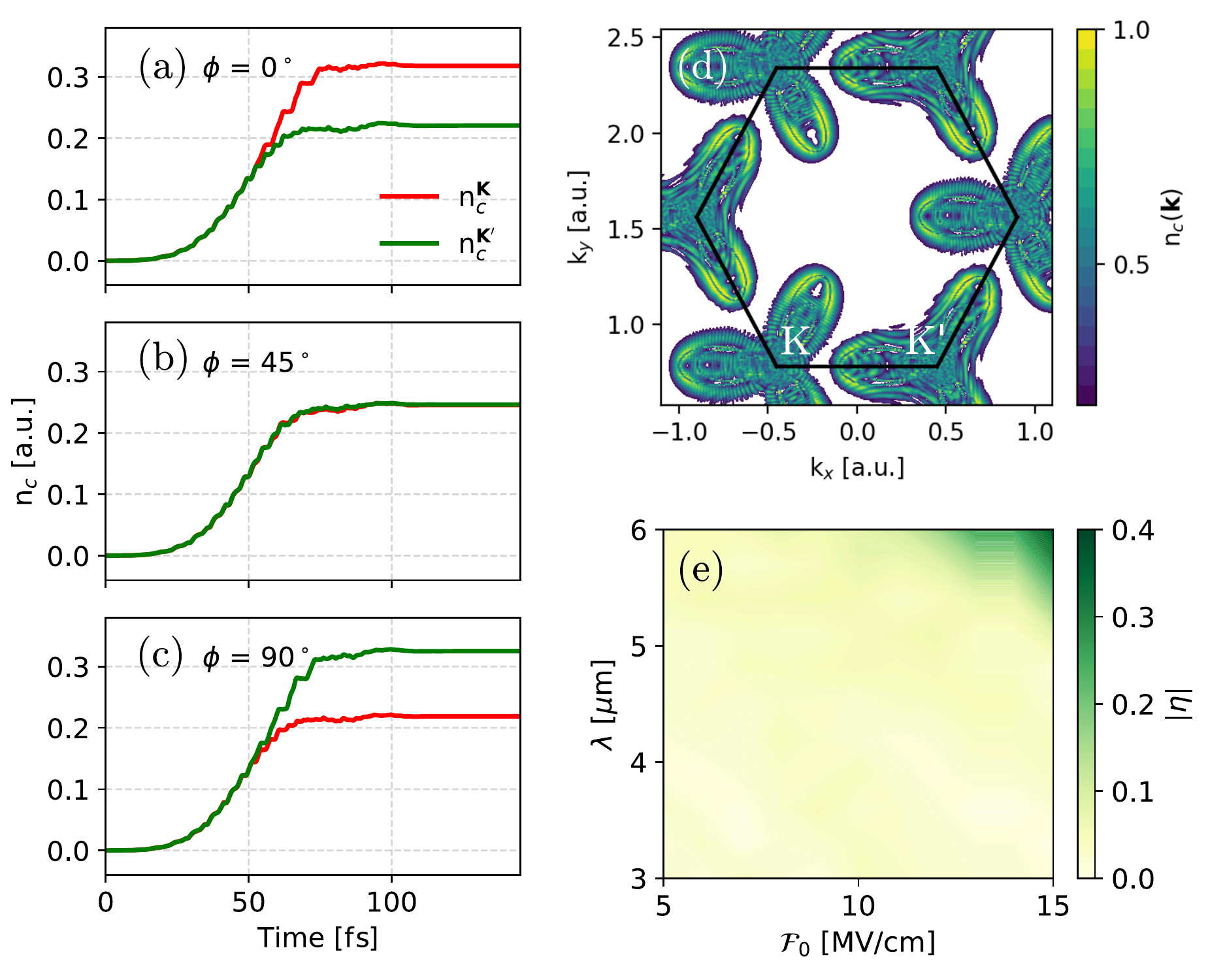}
	\centering
	\caption{(a)-(c) Excitation dynamics during the laser pulse, respectively, for $\phi$ = 0$^\circ$, 45$^\circ$, and 90$^\circ$. (d) Conduction-band population in the Brillouin zone at the peak of the laser pulse. (e) Valley-asymmetry as a function of wavelength ($\lambda$) and electric-field amplitude ($\mathcal{F}_0$).} 
	\label{fig4.3}
\end{figure}

To quantify the amount of the valley polarisation, the valley asymmetry  parameter is defined 
as 
\begin{equation}\label{eq1}
	\eta = \frac{n_{c}^{\mathbf{K}^{\prime}} - n_{c}^{\mathbf{K}}}{(n_{c}^{\mathbf{K}^{\prime}} + n_{c}^{\mathbf{K}})/2}, 
\end{equation}
where $n_{c}^{\mathbf{K}}$ and $n_{c}^{\mathbf{K}^{\prime}}$ are electron populations at the end of the 
laser pulse in the  conduction band 
around   $\mathbf{K}$ and $\mathbf{K}^{\prime}$ valleys, respectively. 

Figure~\ref{fig4.2}(b) shows the asymmetry in the populations of the ${\bf K}$ and $\mathbf{K}^{\prime}$ valleys as a function
of the two-colour phase $\phi$ (black line), for different values of the fundamental field
amplitude and $\mathcal{R} = 2$. Substantial contrast between the two valleys is achieved with values as high as $\pm 36 \%$ for $\phi  = 0^{\circ}, 90^{\circ}$, and no asymmetry for 
$\phi  = 45^{\circ}$. Here, each 180$^\circ$ change in $\phi$ results in a  120$^\circ$ rotation of the trefoil, resulting in an equivalent configuration. This is the reason for the two-fold symmetry of the valley-asymmetry in Fig.~\ref{fig4.2}(b). 
The same results are obtained when the  helicities of both driving fields are exchanged 
simultaneously [see red dots in Fig.~\ref{fig4.2}(b)].

The higher-populated valley is the one where the  vector potential ``fits'' better into the 
shape of the valley, minimizing the band gap along the electron trajectory
in the momentum space. 
The asymmetry in the valley population is negligible up to an intensity of $2 \times 10^{11}$ W/cm$^2$ 
and gradually increases with intensity [see green line in Fig.~\ref{fig4.2}(b)].

To further explore the underlying mechanism, we show the excitation dynamics for three different cases in Figs.~\ref{fig4.3}(a)-(c). Analysing Figs.~\ref{fig4.3}(a)-(c) along with Fig.~\ref{fig4.2}(b) suggests that the valley asymmetry is observed only when the laser pulse is able to drive the electrons to the anisotropic part of the conduction band. We notice that the valley-excitation dynamics is reversed for $\phi$ = 0$^\circ$ and $\phi$ = 90$^\circ$. Since the electron-dynamics in both the valleys are equivalent for $\phi$ = 45$^\circ$, there is no valley-polarisation. 
The $\bf{k}$-resolved excitation dynamics for $\phi$ = 0$^\circ$ is presented in Fig.~\ref{fig4.3}(d). The populations near the valleys show signatures of energy contours [Fig.~\ref{fig4.1}(c)] of the respective valley. Figure~\ref{fig4.3}(e) shows wavelength and electric-field amplitude dependence on the valley-asymmetry. A reasonable valley asymmetry is achieved for longer wavelengths and stronger electric-field amplitudes. This is directly linked to the fact that the extend an electron travel in the reciprocal space is proportional to the amplitude of the vector-potential, as a consequence of the acceleration theorem.

\subsection{Valley-Polarised HHG}

Figure~\ref{fig4.4}(a) shows polarisation-resolved high-harmonic spectrum. The
$(3n+1)$ harmonics follow the polarisation of the $\omega$ pulse (left-handed circular polarisation), whereas 
$(3n+2)$ harmonics follow the polarisation of the $2 \omega$ pulse (right-handed circular polarisation), while  
$3n$ harmonics are missing, just like in atomic media~\citep{fleischer2014spin,neufeld2019floquet, dixit2018control, ansari2021controlling}.
In this context, we note that while harmonic generation with
single-colour circularly polarized drivers is forbidden in atoms, 
such selection rules do not generally arise in solids~\citep{ghimire2011observation}. 
However, the ellipticity-dependence 
studies on graphene show very weak harmonic yield for drivers with higher ellipticity~\citep{yoshikawa2017high, taucer2017nonperturbative, liu2018driving}, as we have  
also observed and discussed in the previous chapter. 
As in atoms, application of bicircular fields allows for efficient 
generation of circularly polarized harmonics in graphene.

In general, the interband and intraband harmonic emission mechanisms in graphene are 
coupled~\citep{taucer2017nonperturbative,liu2018driving,al2014high}, except 
at low electric fields~\citep{al2014high}, leading to a complex interplay between the 
interband and intraband emission mechanisms. We present the interband and intraband resolved HHG spectrum in Fig.~\ref{fig4.4}(b). In the present case, the intraband mechanism show dominancy. This further supports our analysis based on the acceleration theorem, which is based on intraband electron dynamics. 

\begin{figure}[t!]
	\centering
	\includegraphics[width= \textwidth]{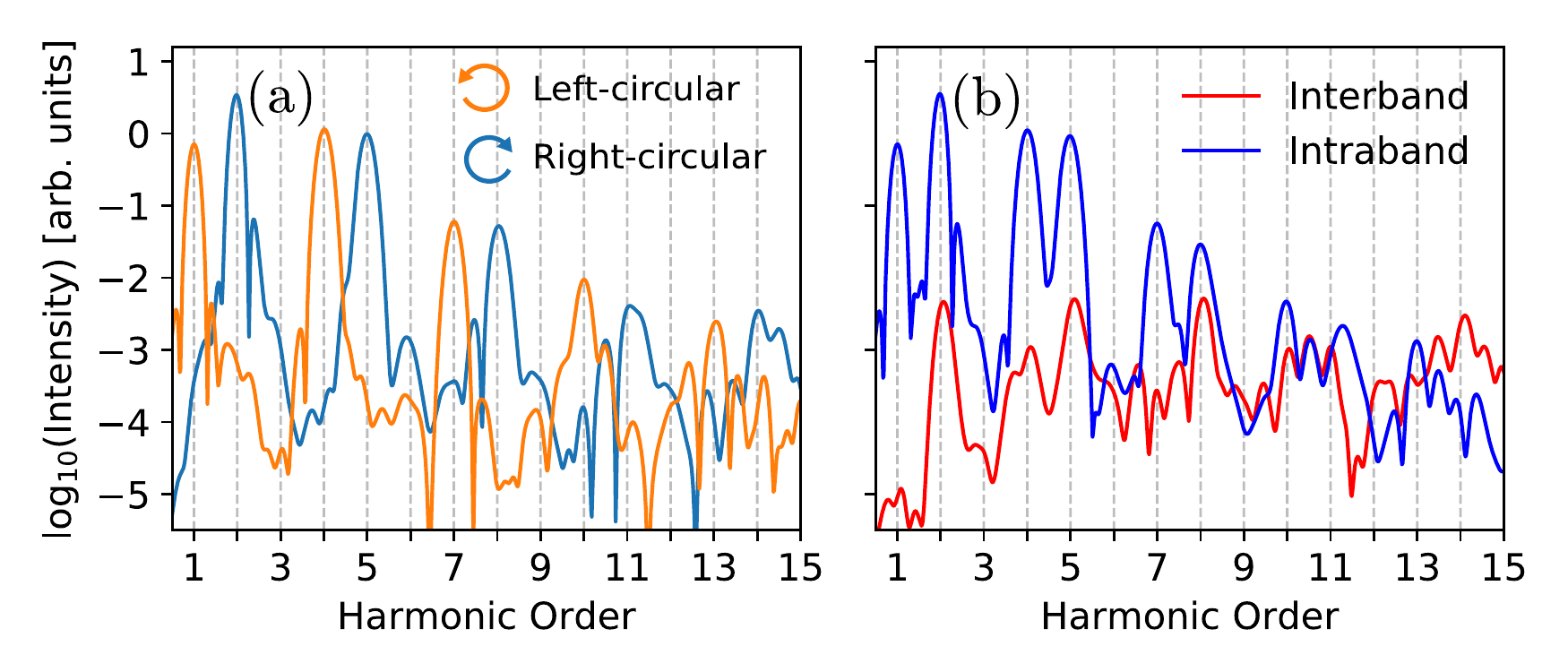}
	\caption{(a) Polarisation-resolved high-harmonic emission, 
		for left-handed circularly polarised 
		fundamental  (orange) and right-handed circularly polarised 
		second harmonic (blue). Alternating harmonics follow alternating 
		helicities: $3n+1$ follow the fundamental,  
		$3n+2$ follow the second harmonic,  the $3n$ harmonics 
		missing due to symmetry. (b) Harmonic spectrum resolved into interband and intraband contributions ($\phi = 0^{\circ}$).} 
	\label{fig4.4}
\end{figure} 

The current generated by the electron injection into the conduction band valleys is 
accompanied by harmonic radiation 
and makes substantial contribution to the lower-order harmonics, 
such as H4 and H5 for our two-colour driver. These harmonics are stronger 
whenever the dispersion $\epsilon({\bf k})$ is
more nonlinear. In this respect, for the electrons following the trajectories ${\bf K}+\mathcal{A}(t)$ and  
$\mathbf{K}^{\prime}+\mathcal{A}(t)$, the current-driven HHG from 
the  $\mathbf{K}^{\prime}$ valley 
is preferred for the vector potential in Fig.~\ref{fig4.1}(c).  Conversely, for the 
vector potential shown in Fig.~\ref{fig4.1}(d) low-order, current-driven 
harmonic generation should be preferred from the ${\bf K}$ valley. Indeed, 
following the vector potential, the
electron is driven against the steeper walls in the $\mathbf{K}^{\prime}$ valley in Fig.~\ref{fig4.1}(c) and 
against the steeper walls in the ${\bf K}$ valley in Fig.~\ref{fig4.1}(d).

These qualitative expectations are also confirmed by our 
numerical simulations, shown in Fig.~\ref{fig4.5}.
The same laser parameters used in Fig.~\ref{fig4.2} are used in this simulation.

Fig.~\ref{fig4.5} shows the dependence of harmonic generation on the orientation of
the vector potential relative to the structure of the Brillouin zone.
As expected, the total harmonic yield is modulated as the trefoil is
rotated, with the lower-order current-driven harmonics H4 and H5 following the expected
pattern, maximizing when the electrons are driven into the steeper walls
in either the $\mathbf{K}^{\prime}$ or ${\bf K}$ valley.

\begin{figure}[t!]
	\centering
	\includegraphics[width= 12 cm]{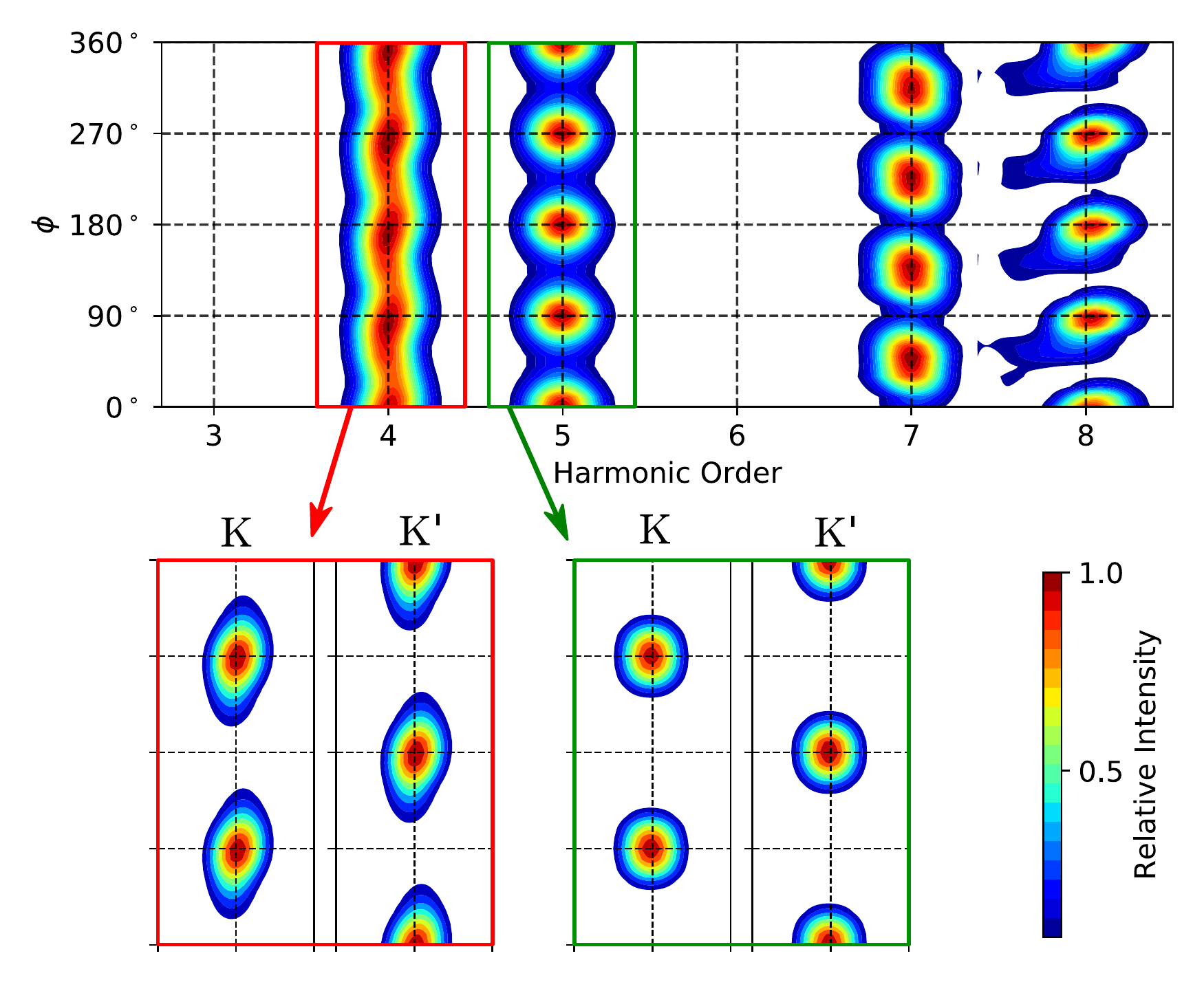}
	\caption{Harmonic emission as a function of the 
		two-colour phase $\phi$. Valley polarization of H4 and H5 as a function
		of the two-colour phase is shown in the bottom panel.
		The total field is identical for $\phi = 0^{\circ}$ and $\phi = 180^{\circ}$ owing to the threefold symmetry.} 
	\label{fig4.5}
\end{figure} 

The contribution of different valleys to H4 and H5 as a function of
the field orientation is presented in the bottom panel of Fig.~\ref{fig4.5}. Consistent with the qualitative analysis above, 
maximum harmonic contribution of the $\mathbf{K}^{\prime}$
valley corresponds to the vector potential orientations such as shown in Fig.~\ref{fig4.1}(c),
while the maximum contribution of the ${\bf K}$
valley corresponds to the vector potential orientations such as shown in Fig.~\ref{fig4.1}(d). Therefore, we are able to control the valley-polarisation of 
the harmonics by controlling the two-colour phase $\phi$. We have also checked
that these results do not depend on the specific directions of rotation of the 
two driving fields. That is, we find the same results when simultaneously changing 
the helicities of both driving fields. 

\subsection{Probing Valley-Polarisation}

To read out the induced valley polarization, we employ a probe
pulse of frequency 3$\omega$ linearly polarized along the $x$-direction.
The amplitude of the probe field is 1.5 MV/cm (F$_\omega$/10). Since
 $\mathbf{K}$ and $\mathbf{K}^{\prime}$ valleys in graphene are related by space inversion, the
even-order harmonics generated by individual (asymmetric) valleys are equal in magnitude but opposite in phase [Fig.~\ref{fig4.6}(a), red and
green lines]. In the absence of valley polarization, their interference leads to the complete cancellation of even harmonics [Fig.~\ref{fig4.6}(a)]
(full Brillouin zone signal). In the presence of valley polarization,
the symmetry is broken, the cancellation of even harmonics is quenched, and even harmonics signal scales proportional to valley
polarization [see also Refs.~\citep{jimenez2021sub,golub2014valley}]. The phase of the even harmonics follows the dominant valley. Importantly, 3$n \omega$ harmonics are absent in the spectra generated by the bicircular $\omega$ - 2$\omega$ field
[Fig.~\ref{fig4.4}(a)]. Thus, even harmonics generated by the 3$\omega$ probe
provide a background-free measurement of valley polarization.
Figure~\ref{fig4.6}(b) shows the generation of the second harmonic of the 3$\omega$
probe pulse (labelled H6) for the two-colour phases of the bicircular
pump $\phi$ = 0$^\circ$ (red curve) and 90$^\circ$ (green curve), which switches
the valley polarization between $\mathbf{K}$ and $\mathbf{K}^{\prime}$ valleys. While the H6
intensity measures valley polarization, its phase clearly identifies
the dominant valley. This phase can be measured by interfering
the signal with the reference second harmonic of 3$\omega$ generated, e.g., from a beta barium borate (BBO) crystal. Controlling the delay of the reference second harmonic generated in the BBO crystal, we can map the phase of H6 generated by graphene on the amplitude modulation of their interference.

\begin{figure}[t!]
	\centering
	\includegraphics[width= \textwidth]{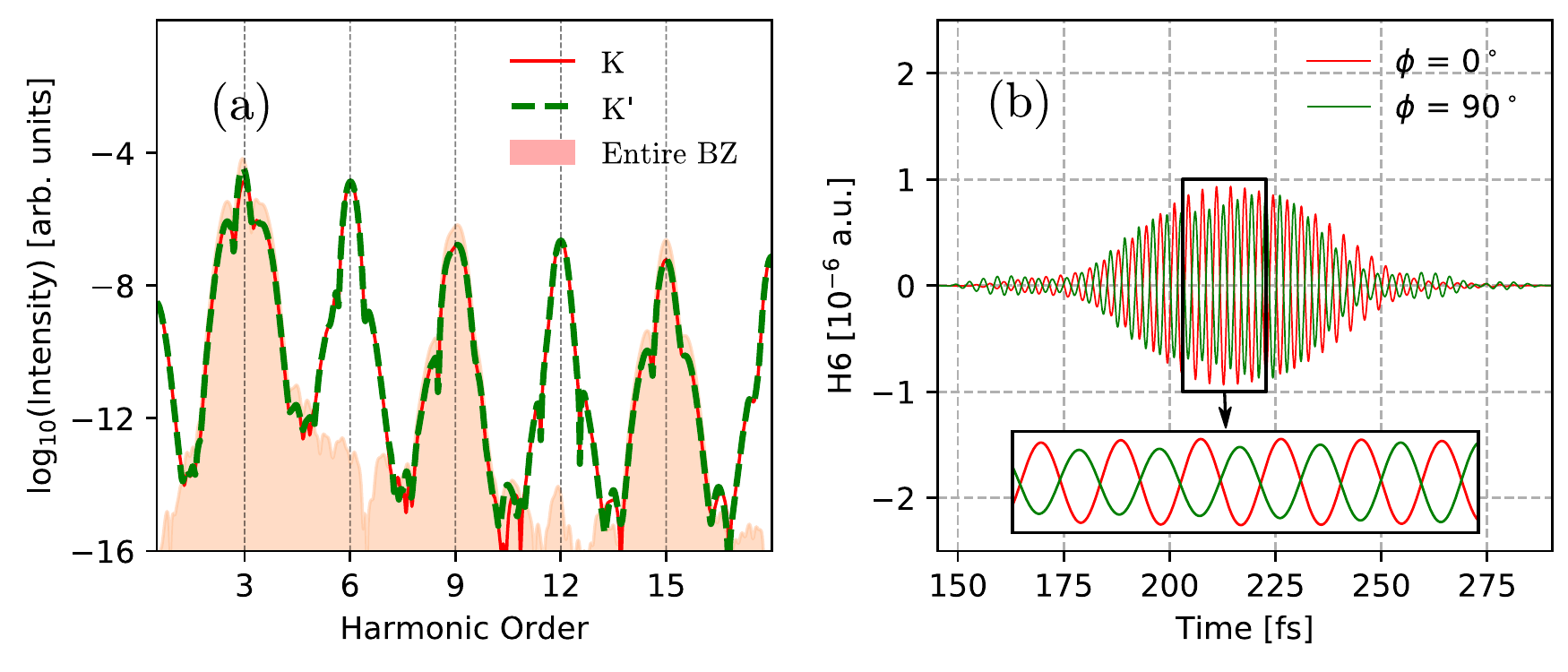}
	\caption{(a) High-harmonic spectrum corresponding to a linearly polarised field with amplitude of 1.5 MV/cm and frequency of 3$\omega$. The red and green lines show the valley-resolved HHG-spectrum. (b) Sixth harmonic (H6) generated by the 3$\omega$ pulse after the bicircular field broke the inversion symmetry in graphene.}
	\label{fig4.6}
\end{figure} 

While the light configuration we use is similar to that used in Ref.~\citep{jimenez2019lightwave} for finite band-gap materials, the physical mechanism of valley-selective excitation in a zero band-gap centerosymmetric material is quite different. In gapped materials, valley-polarisation is achieved by selectively reducing the effective band gap in one of the valleys~\citep{jimenez2019lightwave}. Here, valley-polarisation is achieved only when the light-driven electrons explore the anisotropic region in the Brillouin zone.

\section{Summary}
In summary, valley polarisation in both electronic excitation and harmonic generation
can be achieved in pristine graphene by 
tailoring the Lissajous figure of the driving pulse to the symmetry of the graphene
lattice. This allows one to both break the inversion symmetry between the adjacent carbon atoms and also 
exploit the anisotropic regions in the valleys, taking advantage of
the fact that the energy landscape of the valleys are mirror images of each other.  
Present work opens an avenue for a new regime of valleytronics in pristine graphene and similar materials with zero 
band-gap and zero Berry curvature.  

\cleardoublepage
\chapter{HHG from Spin-polarised Defects}


Due to the growth processes, defects are inevitable in real solids~\citep{hayes2012defects}. Defects in materials can  appear in the form of vacancies, impurities, interstitials (all of these can be neutral as well as charged), dislocations, etc.
Defect-induced microscopic modifications in material 
have important impacts on its macroscopic properties~\citep{wilson1931theory}.  
Electronic, optical, vibrational, structural and diffusion  properties of solids with defects have been thoroughly reviewed over the past century~\citep{
	barker1975optical, pantelides1978electronic, queisser1998defects, van2004first, bockstedte2010many, 
	alkauskas2011advanced, freysoldt2014first}. 
Defect engineering is 
used  to achieve desirable characteristics for materials such as doping, which has revolutionised the field of electronics~\citep{bardeen1949physical}. 
Defects can also be highly controlled, and it is possible to create isolated defects such as nitrogen-vacancy defects  in diamond~\citep{doherty2013nitrogen, maze2008nanoscale} or single photon emitters in two-dimensional materials~\citep{tran2016quantum}.

The  influence of defects in solids  is not well explored in  strong-field physics. 
In this chapter, we aim to address interesting questions such as 
Is it possible to observe defect-specific fingerprints in strong-field driven electron dynamics?
Or 
does the electron-electron interaction play a different role for defects and bulk materials?   
Moreover, some defects are also spin polarised in nature and therefore leads to a 
question of the possibility to control the electron dynamics for different spin channels independently in a non-magnetic host material. In this chapter,  we will discuss how  HHG is a unique probe for defects in solids.

There are theoretical predictions to 
understand HHG from doped semiconductors,  
using one-dimensional  model Hamiltonians~\citep{yu2019enhanced, huang2017high, pattanayak2020influence}. 
In Ref.~\citep{yu2019enhanced}, 
by using TDDFT, it is predicted  that there are several orders of magnitude enhancement  in the efficiency of HHG in a donor-doped semiconductor.
In contrast to this in Ref.~\citep{huang2017high}, with a single-active electron model,  
it is found that the efficiency of the second plateau from the doped semiconductor is enhanced. 
Similar calculation within a tight-binding Anderson model indicates that disorder 
might lead to well-resolved peaks in  HHG~\citep{orlando2018high}. 
The conceptual idea for tomographic imaging  of shallow impurities in solids, within a one-dimensional hydrogenic model,  has been developed by Corkum and co-workers~\citep{almalki2018high}. 
Even though these pioneer works have shown that defects can influence HHG, many points remain elusive. So far, only model systems in one-dimension have been considered, and no investigation of realistic defects (through geometry optimisation and relaxation of atomic forces) has been done. 
Beyond the structural aspect, several other essential aspects need to be investigated in order to have a better understanding of HHG in defected-solids such as the importance of electron-electron interaction (that goes beyond single-active electron and independent-particle  approximations),  the role of the electron's spin, the effect of the symmetry breaking due to the defects, etc. 
This chapter aims to shed light on some of these crucial questions. For that, 
we need to  go beyond  one-dimensional model Hamiltonian approach. 

Monolayer hexagonal boron nitride (h-BN) 
is an interesting material to study  electronic and  optical properties. 
h-BN is a promising candidate for light-emitting devices in  
far UV region due to the strong exciton emission~\citep{bourrellier2014nanometric, bourrellier2016bright}. 
Due to  technological importance, 
several experimental and theoretical studies have been carried out for h-BN with defects~\citep{tran2016quantum, tran2016robust, huang2012defect, zobelli2007vacancy, alem2009atomically, jimenez1996near, suenaga2012core, liu2014direct, thomas2015temperature, gilbert2017fabrication, alem2009atomically, alem2011probing,  wong2015characterization, pierret2014excitonic,
	bourrellier2016bright, attaccalite2011coupling, azevedo2009electronic, orellana2001stability, mosuang2002influence}. Different kinds of defects in h-BN can be classified as vacancy (mono-vacancies to cluster of vacancies), antisite, and impurities. In particular, defects like monovacancies of boron and nitrogen atoms in h-BN are one of the most commonly realised defects.  Recently,  Zettili and co-workers have shown the possibility of engineering a cluster of vacancies and characterising them using ultra-high-resolution  transmission electron microscopy~\citep{gilbert2017fabrication, alem2009atomically, wong2015characterization}. Signatures of defects in h-BN are identified by  analysing    cathodoluminescence and photoluminescence spectra ~\citep{pierret2014excitonic, bourrellier2014nanometric}. It is shown that the emission band around 4 eV is originating from the transitions including deep defect levels~\citep{attaccalite2011coupling}. Ultra-bright single-photon emission from a single layer of h-BN with nitrogen vacancy is experimentally realised, which was used for large-scale nano-photonics and quantum information processing~\citep{tran2016quantum, tran2016robust}. 
In Refs.~\citep{huang2012defect, attaccalite2011coupling, azevedo2009electronic,
	liu2007ab, orellana2001stability}, 
different kinds of defects in h-BN are modelled, and their role on the electronic and optical properties  have been thoroughly investigated.  

In this chapter, 
using the well-established supercell  approach to model the defects in h-BN, 
we analyse the influence of defects in HHG. Due  to the partial ionic nature of its bonds, h-BN is a wide band-gap semiconductor with an experimental band-gap of 6 eV, which also makes it interesting for generating high-order harmonics without damaging the material. Moreover, 2D material like h-BN enables us to visualise the induced electronic density and localised defect states easily. The presence of boron or nitrogen vacancy in h-BN acts as a spin-polarised defect. These factors make h-BN an ideal candidate for exploring spin-resolved HHG in defected, non-magnetic solids. Below we will discuss HHG in h-BN with and without spin-polarised monovacancies. 

\section{Numerical Methods}
Here, we employ \textit{ab initio} TDDFT to simulate the 
strong-field driven electron dynamics in defected-solid. This allows  
to come up with theoretical predictions relevant for real experiments on defected-solid without empirical inputs.

\subsection{Geometry Relaxation}
Geometry optimisation was performed using the DFT code Quantum ESPRESSO \citep{giannozzi2009p,giannozzi2017p}. We allow for both atomic coordinate and lattice constant relaxation. 
Forces were optimised to be below 10$^{-3}$ eV/\AA. We used an energy cutoff of 150 Ry., and a k-point grid of 10$\times$10$\times$1  $\mathbf{k}$-points. We used a vacuum region of more than 20 \AA~ to isolate the monolayer from its periodic copies.
We obtained relaxed lattice constants of 12.60 \AA~, 12.57 \AA~ for 
h-BN with a single boron vacancy ($V_B$), and
12.48 \AA~ for h-BN with a single nitrogen  vacancy ($V_N$) 
while the lattice constant for pristine h-BN was found to be 12.56 \AA.
The structure of $V_B$ relaxes by  lowering the local threefold symmetry~\citep{huang2012defect}. The lowering of the symmetry with the boron vacancy is attributed to the Jahn-Teller distortion, which is found to be independent of the defect concentration~\citep{huang2012defect,liu2007ab,attaccalite2011coupling}.

\subsection{TDDFT Simulations}
By propagating the Kohn-Sham equations within TDDFT, the evolution 
of the time-dependent current is computed using the 
Octopus package~\citep{andrade2015real, castro2004propagators}. The harmonic spectrum of pristine h-BN calculated within local density approximation (LDA) and generalised gradient approximation (GGA) are presented in Fig.~\ref{fig5.1}(a). The HHG spectrum is consistent with the choice of pseudo-potential. TDDFT  within GGA with exchange and correlations of Perdew-Burke-Ernzerhof (PBE)~\citep{perdew1996generalized} is used for all the simulations presented here. The adiabatic approximation is used for all the time-dependent simulations. 
We used norm-conserving pseudo-potentials. The real-space cell was sampled with a grid spacing of 0.18 \AA, and we used a 6$\times$6 $\mathbf{k}$-points grid to sample the 2D Brillouin zone. The semi-periodic boundary conditions are employed.  A simulation box of 74.08 \AA~ along the nonperiodic dimension, including 21.17 \AA~of absorbing regions on each side of the monolayer is used. The absorbing boundaries are treated using the complex absorbing potential method, and the cap height h is taken as h = -1 atomic units to avoid the reflection error in the spectral region of interest~\citep{de2015modeling}. Harmonic spectrum is calculated using equations mentioned in section~\ref{section:2.3.2}. Spin-polarised calculations are used to address the spin-polarised  vacancies. 

\begin{figure}[t!]
	\centering
	\includegraphics[width= \textwidth]{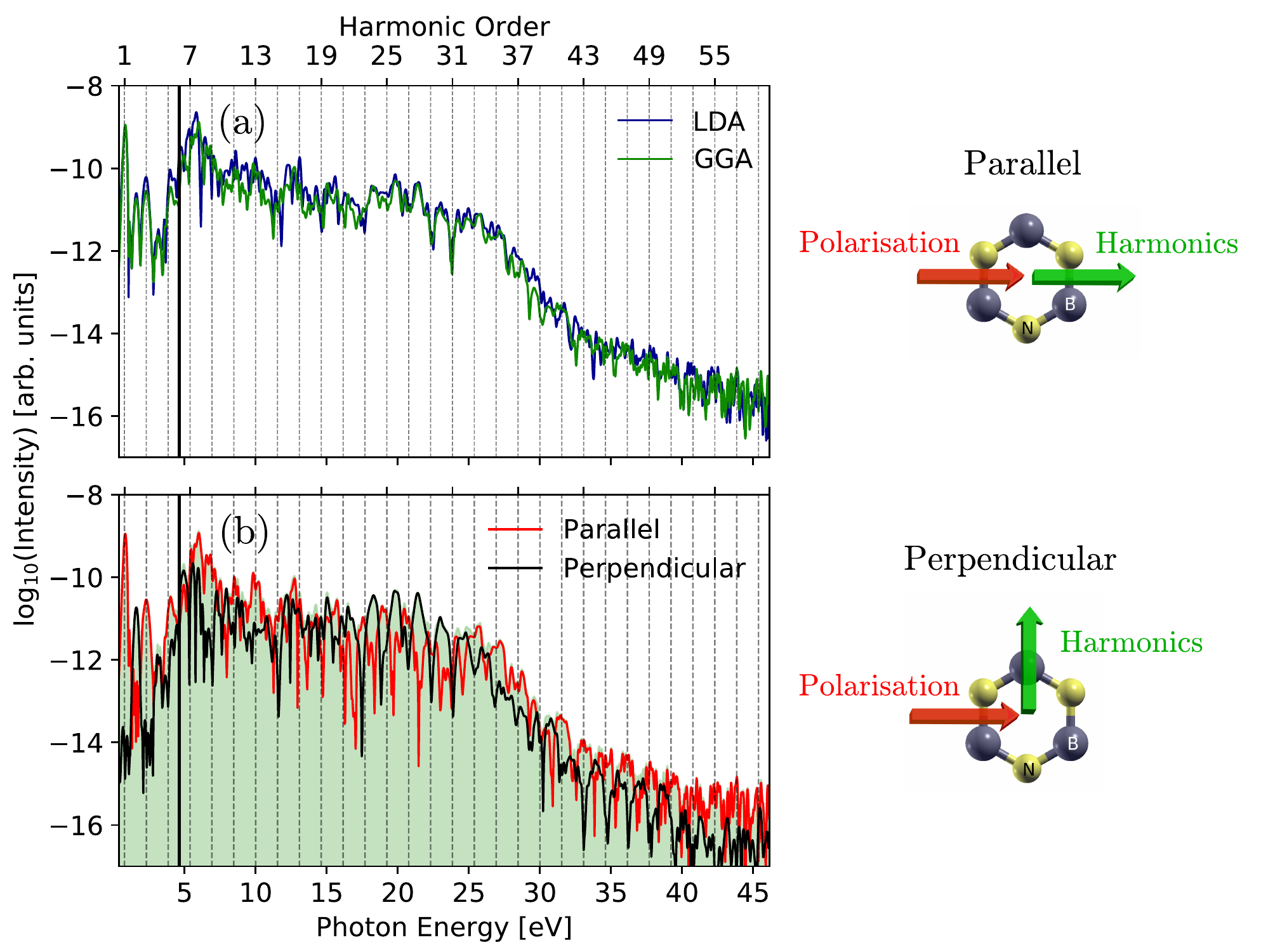}
	\caption{(a) High-harmonic  spectrum of h-BN calculated within local density approximation (LDA), and generalised gradient approximation (GGA). (b) Harmonic spectrum in the direction parallel and perpendicular to the laser polarisation. Here, GGA is used to simulate the spectra. The two configurations are presented in the right panel. The black vertical line is the band-gap of h-BN.} 
	\label{fig5.1}
\end{figure}

The total number of excited electron is obtained by projecting the time-evolved wavefunctions ($|\psi_n(t)\rangle$) on the basis of the ground-state wavefunctions ($|\psi_{n'}^{\mathrm{GS}}\rangle$) 
\begin{equation}
	N_{\mathrm{ex}}(t) = N_e - \frac{1}{N_\mathbf{k}}\sum_{n,n'}^{\mathrm{occ.}}\sum_{\mathbf{k}}^{\mathrm{BZ}} |\langle \psi_{n,\mathbf{k}}(t) | \psi_{n',\mathbf{k}}^{\mathrm{GS}} \rangle|^2,
\end{equation}
where $N_e$ is the total number of electrons in the system, and $N_{\mathbf{k}}$ is the total number of  $\mathbf{k}$-points used to sample the Brillouin zone. The sum over the band indices $n$ and $n'$ run over all occupied states.

\subsection{Effective Band-Structure}

When the energy band-structure is calculated for a supercell  of the primitive unit-cell, 
energy-bands fold into the Brillouin zone of the supercell. An effective band-structure within the  primitive unit-cell can be extracted from the eigenvalues and the eigenvectors of supercell~\citep{ku2010unfolding, popescu2012extracting, medeiros2014effects}. Let $\left| \textbf{k}n\right\rangle$ and $\left| \textbf{K}m\right\rangle$ be the eigenstates of the primitive unit-cell and supercell,  respectively. A spectral weight function can be defined as
\begin{equation}
	P_{\textbf{K}m}(\textbf{k})=\sum_n\left|\left\langle \textbf{K}m | \textbf{k}n\right\rangle\right|^2.
\end{equation}

The spectral function or the effective band-structure corresponding to supercell can be expressed as a function of energy ($\mathcal{E}$) as
\begin{equation}
	A(\textbf{k},\mathcal{E}) = \sum_m P_{\textbf{K}m}(\textbf{k})\delta(\mathcal{E}_m-\mathcal{E}).
\end{equation}

We have implemented a utility for unfolding band-structure in the Octopus package~\citep{andrade2015real, castro2004propagators,mrudul_unfold}. In the present work effective single-particle band-structure or spectral function for the vacancy structures are visualised by unfolding the band-structure of 5$\times$5  supercell with 50 atoms.   

\section{Results}
h-BN has a  two-atoms primitive cell. We model the vacancy in a 5$\times$5 (7$\times$7) supercell with 50 atoms (98 atoms) following the method in  Ref~\citep{huang2012defect}. The size of the 5$\times$5 supercell is large enough to separate the nearest defects with a distance larger 
than 12 \AA. This avoids the interaction between the nearby 
defects~\citep{huang2012defect,liu2007ab}, and the defect wavefunctions are found to be 
well-localized within the supercell~\citep{azevedo2009electronic}. 
In the present work, 
we are not considering more than one point defect.  
Both of our vacancy configurations (single boron as well as single nitrogen vacancy) has a total magnetic moment of +1$\mu_B$, consistent with the results reported in the literature ~\citep{huang2012defect,liu2007ab,azevedo2009electronic}. 
For the low defect concentration considered here (2\% and $\sim1\%$), the strength of the total magnetic moment is independent of the defect concentrations~\citep{liu2007ab}. We have found that the 7$\times$7 and 5$\times$5 supercells converged to similar ground-states~\citep{huang2012defect}. 

In all the present calculations, we consider a laser pulse of 15 femtosecond duration at  full-width half-maximum with sine-squared envelope and a peak intensity of $13.25$ TW/cm$^2$ in the matter [for an experimental in-plane optical index $n$ of $\sim 2.65$~\citep{PhysRev.146.543}]. The carrier wavelength of the pulse is 1600 nm, which corresponds to a photon energy of 0.77 eV. The polarisation of the laser is linear, and its direction is normal to the mirror plane of h-BN. For pristine h-BN, 
the band-gap is found out to be 4.73 eV within DFT-PBE level. The energy band-gap falls near the sixth harmonics of the incident photon energy, and laser parameters are well-below the damage threshold of the pristine. The symmetry of the pristine h-BN permits to observe harmonics in the parallel and perpendicular directions to the laser polarisation. The direction resolved analysis of the HHG spectrum is shown in Fig.~\ref{fig5.1}(b). There are odd harmonics parallel to and even harmonics perpendicular to the laser polarisation. The perpendicular even harmonics are due to the anomalous current originating from the Berry curvature of the material.

\subsection{HHG from Hexagonal Boron Nitride with a Boron Vacancy}

\begin{figure}[t!]
	\includegraphics[width= 13 cm]{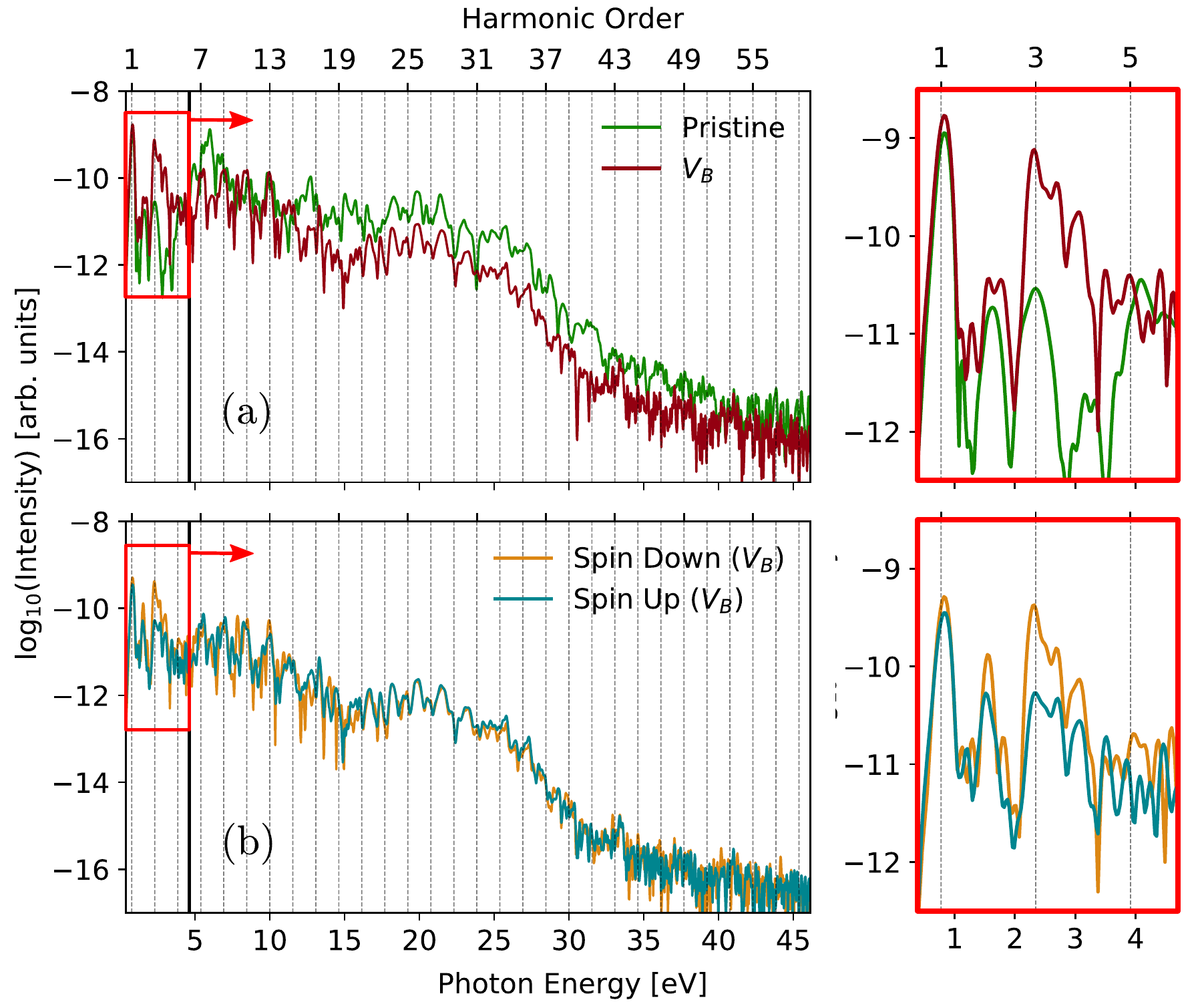}
	\caption{(a) High-harmonic  spectrum of boron-vacant h-BN ($V_B$) and pristine h-BN.  (b) Spin-resolved harmonic spectrum of $V_B$.  The below band-gap portion of (a) and (b) are zoomed in the right panel. The black vertical line presents the energy band-gap of h-BN.} 
	\label{fig5.2}
\end{figure}

We start our analysis by comparing the HHG from pristine h-BN  and h-BN with a boron vacancy ($V_B$). Removal of a boron atom makes the system spin-polarised~\citep{huang2012defect,liu2007ab}.  
The high-harmonic spectrum of  $V_B$ and its 
comparison with pristine h-BN is presented in Fig.~\ref{fig5.2}(a). 
The spectrum of $V_B$ is different from the pristine h-BN as evident from the figure. 
There are two distinct differences: First, the below band-gap harmonics correspond 
to  $V_B$ are significantly enhanced. 
Second,  harmonics have a much lower yield for $V_B$ 
in comparison to the pristine close to the energy cutoff.

\begin{figure}[t!]
	\centering
	\includegraphics[width= 0.65\textwidth]{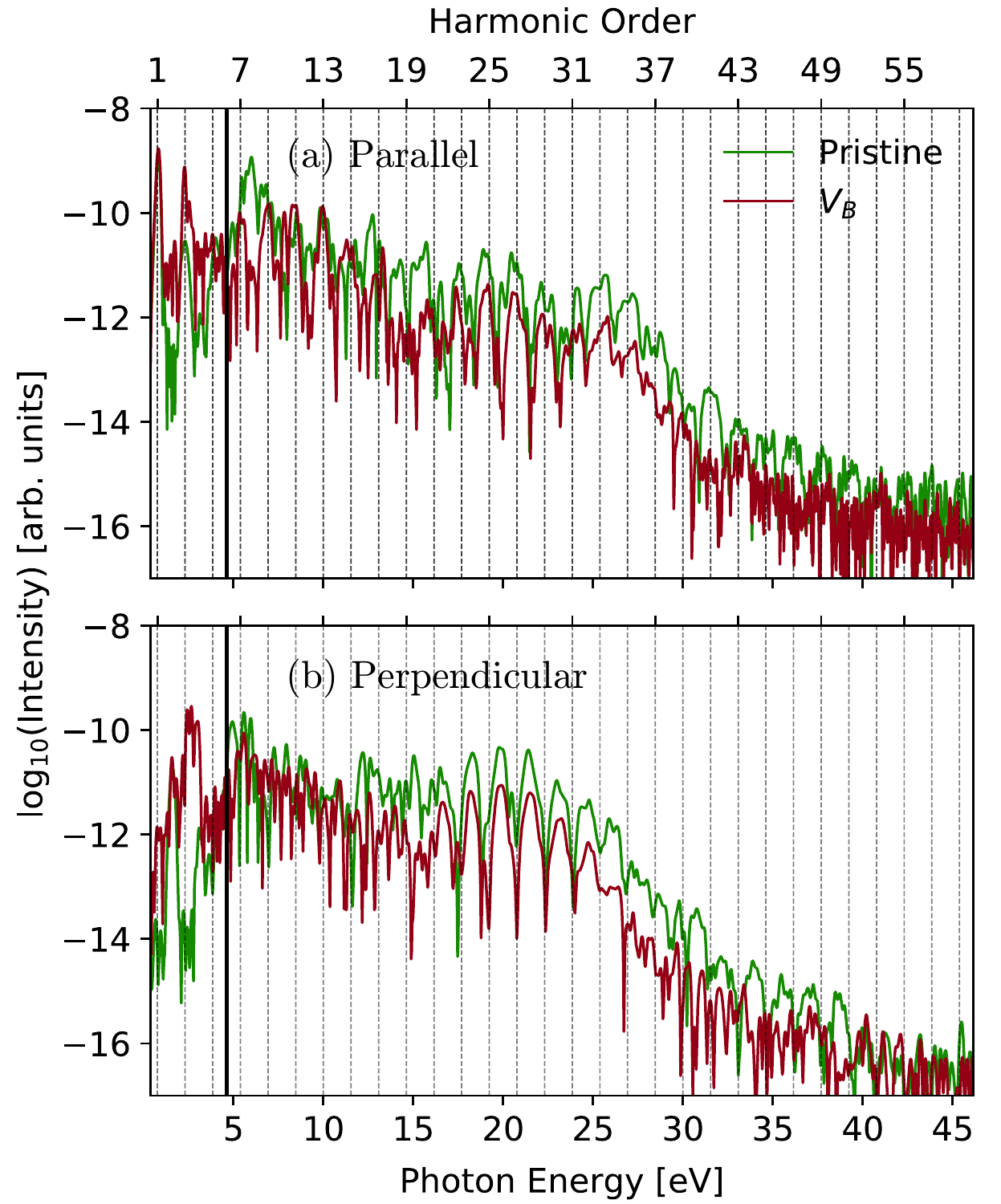}
	\caption{High-harmonic spectrum of boron-vacant h-BN ($V_B$) and pristine h-BN in the direction parallel (a) and perpendicular (b) to the laser polarisation. The black vertical line is the band-gap of h-BN.} 
	\label{fig5.3}
\end{figure}

An interesting aspect for the spin-polarised defects in a non-magnetic host is that the effect of the defect states, compared to bulk states, can be identified clearly by 
examining the spin-resolved spectrum. 
The harmonics correspond to spin-up, and spin-down channels in $V_B$ are 
shown in Fig.~\ref{fig5.2}(b). As reflected in the figure, the below band-gap harmonics are different for both the channels. 
The strength of the third harmonic corresponding to the spin-down channel is much stronger (by an order of magnitude)
in comparison to  the associated spin-up channel. This  indicates that the increase in the yield observed in Fig.~\ref{fig5.2}(a) dominantly originates from the spin-down channel.
At variance, the decrease in harmonic yield in higher energies matches well for the two spin-channels.  This is a strong indication that the defects states do not play a direct role in this part of the HHG spectrum of $V_B$, but that bulk bands predominantly affect this spectral region, as we will discuss below. The features described in the harmonic spectrum are consistent in both parallel and perpendicular configurations as presented in Fig.~\ref{fig5.3}.

\subsection{Gap States and Electron Dynamics}

To understand the difference in the harmonics yield associated with spin-up and spin-down channels, let us analyse the ground-state energy band-structure.
The unfolded band-structures of $V_B$ for spin-up and spin-down channels are shown in Figs.~\ref{fig5.4}(a), and ~\ref{fig5.4}(d), respectively. 
The band-structure within the band-gap region is zoomed for both the channels and shown in Figs.~\ref{fig5.4}(b), and \ref{fig5.4}(c).  
As visible from the figure, 
there is one spin-up  [labeled  as 1 in Fig.~\ref{fig5.4}(b)] and two spin-down  
[labeled  as 2 and 3 in Fig.~\ref{fig5.4}(c)]
defect levels within the band-gap of pristine h-BN.   One defect state corresponding to spin-up channel is pushed within the valence bands [see Fig.\ref{fig5.4}(a)]. 
All three defect states within the band-gap are found to be unoccupied. These factors make the $V_B$ a triple acceptor.  
The corresponding wavefunctions for these defect states are presented 
in Fig.~\ref{fig5.4}(e)-(g). The wavefunctions are found to be localized around the vacancy
as expected for dispersion-less states.
The $p_{x}$ and $p_{y}$ states of the nitrogen atoms in the vicinity of a boron vacancy are contributing to these defect-states, giving a  $\sigma$-character to the vacancy wavefunctions~\citep{liu2007ab}.

\begin{figure}[t!]
	\includegraphics[width=\textwidth]{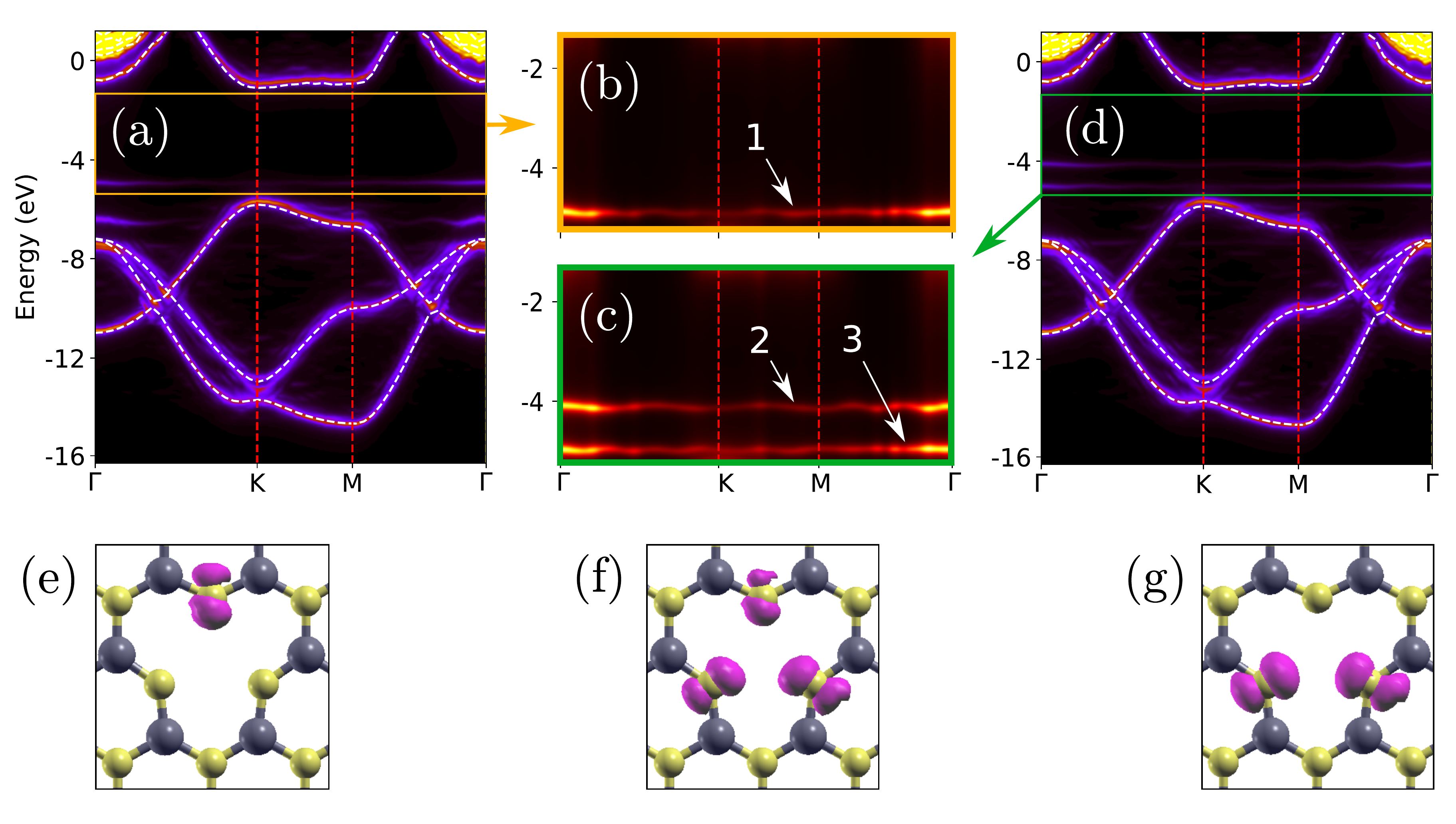}
	\caption{Spectral function for (a) spin-up and (d) spin-down components of $V_B$,  The pristine band-structure is plotted with white dotted lines for reference. The in-gap portion of the spectral function for spin-up and spin-down components are zoomed respectively in (b) and (c), where vacancy states are recognised  by 1 (for spin-up); and by 2 and 3 
		(for spin-down).
		(e)-(g) The absolute wavefunctions of the in-gap vacancy states as recognised in (b) and (c), respectively.} 
	\label{fig5.4}
\end{figure} 

Unlike the pristine h-BN, spin-up and spin-down electrons in $V_B$ see a
different band-structure near the band-gap, as the spin-resolved in-gap states are different. Therefore they evolve differently in the presence of laser pulse.
It means that interband transitions and ionization involving the defects states will contribute differently to the spectrum. 
Hence, the spectral enhancement of the third harmonic can be understood as follows:
There is an  additional defect state near the valence band as visible from the spin-down band-structure, which allows spin-down electrons to be ionised or to recombine
through multiple channels and contribute more to the third harmonic [see Fig.~\ref{fig5.2} ].

\begin{figure}[t!]
	\centering
	\includegraphics[width=0.8\textwidth]{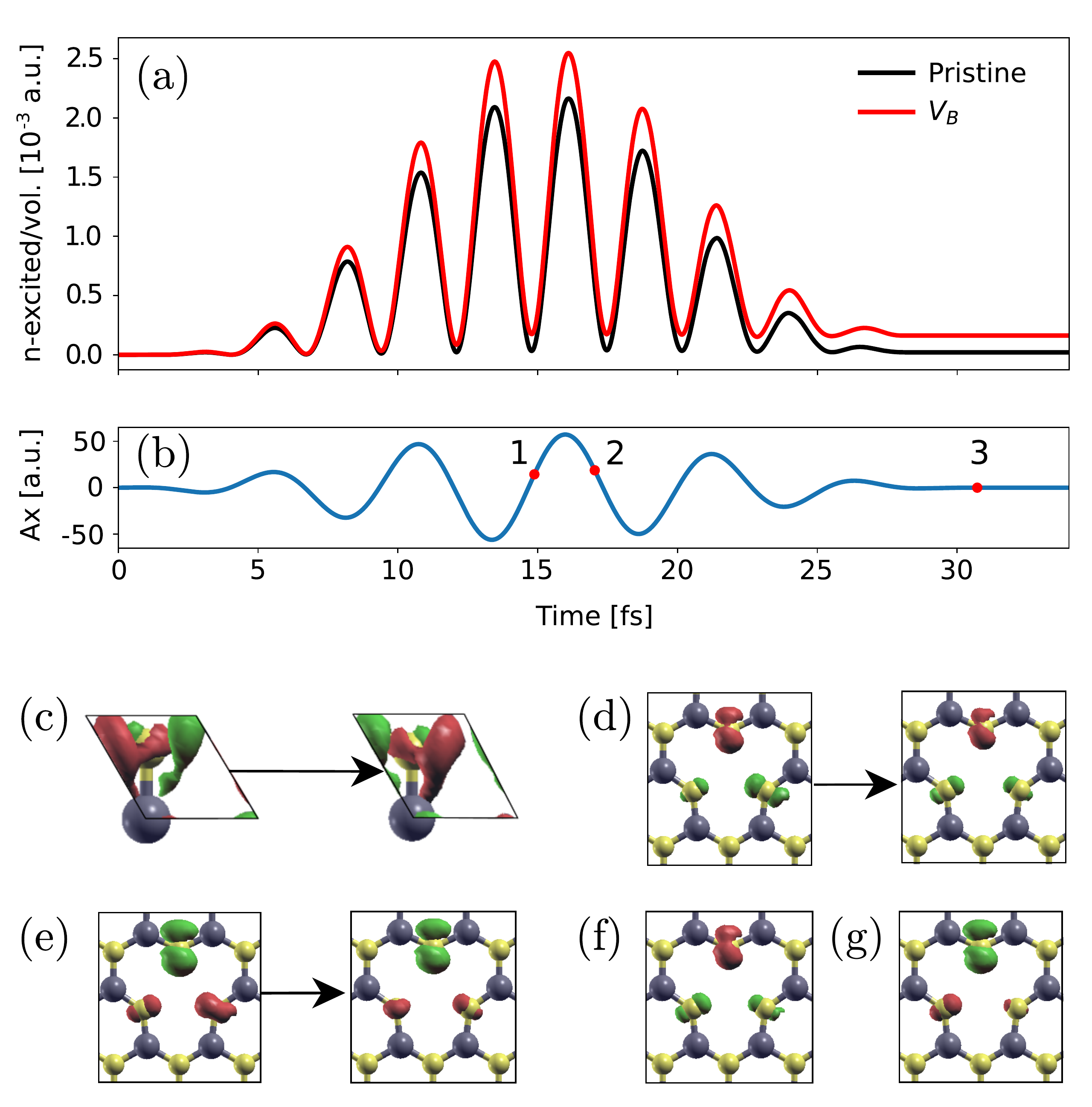}
	\caption{(a) Number of excited electrons per unit volume in pristine h-BN (black colour) and $V_B$ (red colour), and (b) vector potential of the driving laser pulse.
		Snap shots of the time-evolving induced electron density ($n_{ind}$) near the peak of the vector potential [marked as 1 and 2 in (b)] for (c)  pristine h-BN, (d)  spin-up, and (e)  spin-down channels 
		in $V_B$. $n_{ind}$ for (f)  spin-up and (g)  spin-down channels 
		in $V_B$ after the end of the vector potential [marked as 3 in (b)]. Green and red colours in the induced electron density stand for positive and negative values, respectively. } 
	\label{fig5.5}
\end{figure}

It is evident from the band structure that additional defect-states 
effectively reduce the minimal band-gap needed to reach the conduction bands. 
Due to the relaxation of atoms neighbouring to the vacancy, 
the pristine bands get also slightly modified. However, this modification 
is found here to be negligible compared to the photon energy of the laser and is not further discussed.
In order to understand how the presence of the defect states influence the interband tunneling, we evaluated  the number of excited electrons during the laser pulse [see Fig.~\ref{fig5.5}(a)].
In the presence of defect states, there are mostly two possible ways in which ionization can be modified compared to  the bulk material. A first possibility is to have direct ionization of the defect states if they are occupied, or filling if they are originally unoccupied. Another possibility is a double sequential ionization, in which the defect states play an intermediate role in easing the ionization to the conduction bands.
In $V_B$, there is a finite probability of finding the electrons  in conduction 
bands even after the laser pulse is over [see the red curve  in Fig.~\ref{fig5.5}(a)]. 
In contrast to this,  the pulse is not able to promote a significant portion of the valence electrons to the conduction bands permanently in the case of pristine h-BN as the band-gap 
is significantly large [see the black curve in Fig.~\ref{fig5.5}(a)]. 
More precisely, for the $5\times5$ supercell, we found that 1.6 electrons are ionised, compared to 0.25 for the case of pristine for the same cell. 

To have a better understanding of possible ionization mechanisms, we also consider the induced electron density ($n_{ind}$) at two different times near the peak of the vector potential  
[marked as 1 and 2 in Fig.~\ref{fig5.5}(b)] and at the end of the laser field [Fig.~\ref{fig5.5}(f)-(g)].
As reflected in Figs.~\ref{fig5.5}(d), and \ref{fig5.5}(e);
the spin-polarised induced densities of $V_B$  have 
a pronounced localised component near the defects and 
resembles the spatial structure of the initially unoccupied defect wavefunctions [see Fig.~\ref{fig5.4}(e)-(g)]. 
The induced densities at the end of the vector potential show that electrons remain in the 
defect states even after the laser pulse is over [see Figs.~\ref{fig5.5}(f), and \ref{fig5.5}(g)]. 
Considering that the three defect states are originally unoccupied, we cannot conclude that more electrons are ionised to the conduction bands. It is most likely that in-gap defect states get filled during the laser excitation as ~1.6 electrons are excited and $V_B$ is a triple acceptor.

Overall, these results show that the electron dynamics in acceptor-doped solids implies a net transfer of population to the originally unoccupied gap states, but that for a wide band-gap host no more electrons are promoted to conduction bands. This explains why only the low-order third harmonics is directly affected by the presence of spin-polarised defect states (as evidenced by the spin dependence of the spectrum). 
The low density of these defect states helps only fewer photons to get absorbed. 
We note that the irreversible population change, assisted by the defect states, ultimately implies that more energy is absorbed by the defected solid, which leads to a lower damage threshold in comparison to the pristine. However, the intensity considered here is low enough not to see such effect.

\subsection{Effect of Electron-Electron Interaction}

We found so far that the increase of the low-order harmonic yield is compatible with the presence of the defects states in the band-gap of h-BN. This is an independent-particle vision, in which we used the ground-state band-structure of $V_B$ to explain the observed effect on the HHG spectrum.  We now turn our attention to the higher-energy harmonics, 
for which the harmonic yield is decreased. This seems not to be compatible with a simple vision in terms of single-particle band structure, especially with the fact that more electrons are excited by the laser pulse, as shown in Fig.~\ref{fig5.6}(a). 

To understand this, let us investigate  the effect of the electron-electron interaction on the electron dynamics in $V_B$  and pristine h-BN. 
Within dipole approximation, the HHG spectrum can be expressed for the many-body system as 
~\citep{tancogne2017impact, stefanucci2013nonequilibrium} [see Eq.~(\ref{eq:current_der})]
\begin{equation}
	\textrm{HHG}(\omega)  
	\propto  \left| \mathcal{FT} \left[ \int d^3\textbf{r} ~\left\lbrace n_{ind}(\textbf{r},t) + n_0(\textbf{r}) \right\rbrace \nabla v_0 (\textbf{r}) \right] + N_e \textbf{E}(\omega)  \right|^2.
	\label{current}
\end{equation}
Here, $n_{ind}(\textbf{r},t)$ is the induced electron density, 
$v_0$ is the electron-nuclei interaction potential, $N_e$ is the total number of electrons, and $\textbf{E}$ is the applied electric field. $n_{ind}(\textbf{r},t)$   is the difference of  the total 
time-dependent electron density $n(\textbf{r},t)$ 
and the ground-state electron density $n_0(\textbf{r})$, 
i.e., $n_{ind}(\textbf{r},t) = n(\textbf{r},t) - n_0(\textbf{r})$.  Also, $n(\textbf{r},t)$ is  decomposed in 
spin-polarised fashion as  
$n(\textbf{r},t)$ = $n_{\uparrow}(\textbf{r},t)$ + $n_{\downarrow}(\textbf{r},t)$.
If one analyses this expression, it is straightforward to understand how the introduction of  
vacancy can change the harmonic spectrum, through a change in the local potential structure near the defect. This results in the change in gradient of the electron-nuclei interaction potential $v_0$, which is independent of having electron interacting among themselves or not. 
Apart from this explicit source of change, it is clear that the dynamics of the induced density, evolving from a different ground-state also lead to the modifications in 
the harmonic spectra. The fact that the ground states are different, and hence $n_0(\textbf{r})$ is different, does not affect the harmonic spectrum because of the absence of time-dependence in both $n_0$ and $v_0$. The possible difference between the HHG spectra can therefore be understood in terms of independent particle effects (originating from the structural change of the nuclear potential $v_0$) and interaction effects through the induced density $n_{ind}$.

\begin{figure}[t!]
	\centering
	\includegraphics[width=0.65\textwidth]{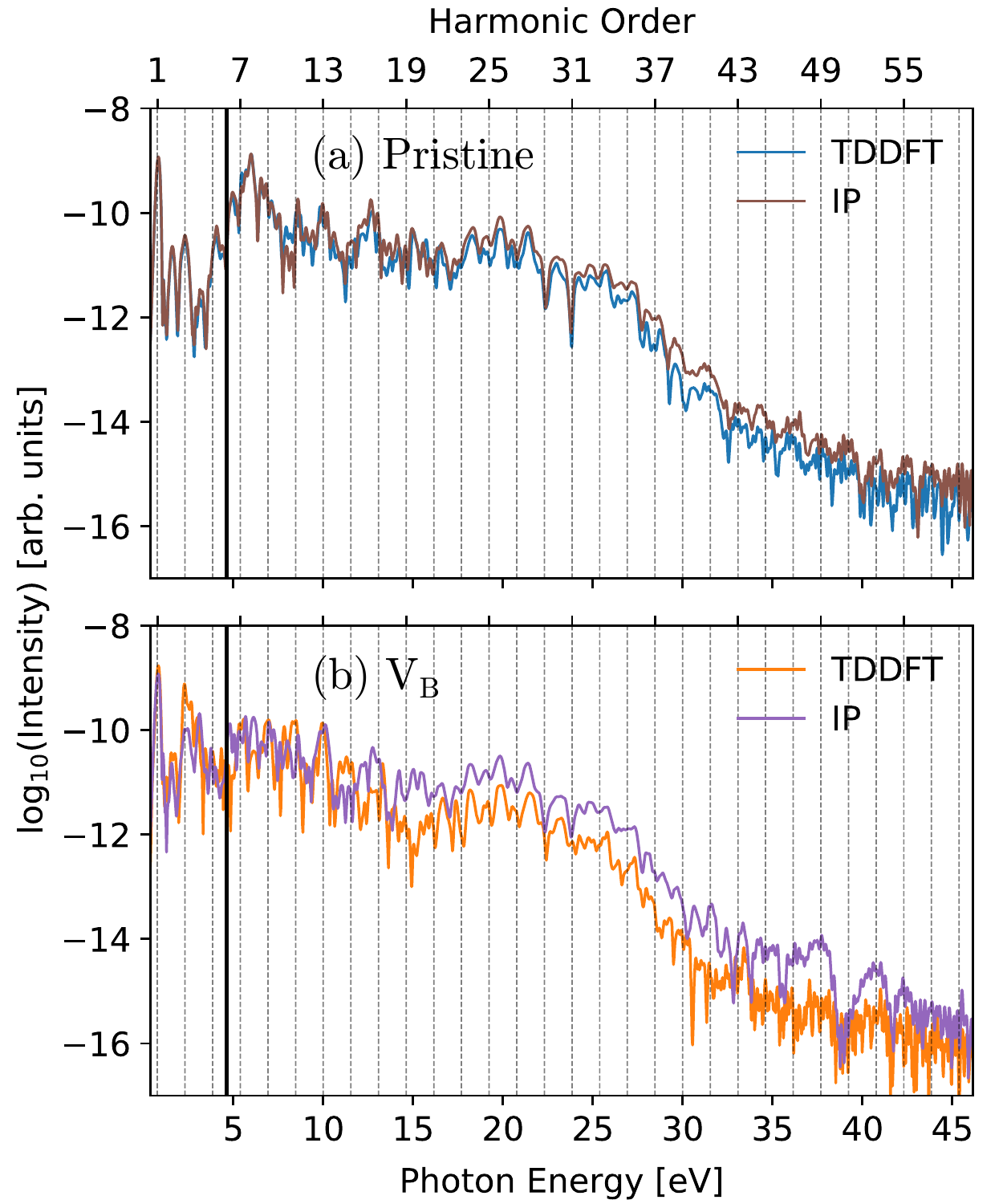}
	\caption{ (a) High-harmonic spectrum for pristine h-BN calculated using TDDFT (blue) and the independent particle (IP) approximation (brown). (b) The harmonic spectrum for $V_B$ calculated using TDDFT (orange) and IP approximation (violet). The black vertical line represents the energy band-gap of pristine h-BN.} 
	\label{fig5.6}
\end{figure} 

In order to disentangle these two sources of differences between pristine and defected h-BN, we simulate  the harmonic spectra 
within an independent particle (IP) approximation by freezing the Hartree and exchange-correlation potentials to the ground-state value. The harmonic spectra for pristine h-BN and $V_B$ within TDDFT and IP approximation are compared in Fig.~\ref{fig5.6}.  
In the case of pristine h-BN, the HHG spectra  obtained by IP approximation and 
TDDFT are similar. Hence, there is no significant many-body effect in HHG from pristine h-BN with an in-plane laser polarisation, at least as described by PBE functional used here. 
A similar finding has been reported for Si~\citep{tancogne2017impact} and MgO~\citep{tancogne2017ellipticity}, within LDA.
Only the high-order harmonics display small differences and there the electron-electron interaction reduces here the harmonic yield as found in Ref.~\citep{yu2019enhanced}.
In contrast, the HHG spectra obtained by TDDFT and IP approximation 
 are  significantly different for $V_B$ as reflected in Fig.~\ref{fig5.6}(b).  
This indicates that the electron-electron interaction is essential for HHG in defected-solids.
 
For the case of $V_B$,  $n_{ind}$ is displayed in Fig.~\ref{fig5.5}. This helps us to understand how the spatial structure of the defect states influences the harmonic spectrum.
Figure~\ref{fig5.5}(c) indicates that the spatial density oscillations are 
responsible for HHG in pristine h-BN. It is clear that the induced density is different in the two systems.
The substantial  difference in the harmonic spectra obtained by two approaches, 
TDDFT and IP approximation,  for $V_B$ is due to  the so-called local field effects, which is explored in detail in the following paragraphs. As we will show, it is this difference which is responsible for the decrease of the harmonic yield for $V_B$.

In the presence of an external electric-field, the localised induced-charge  acts as 
an oscillating dipole near the vacancy. The dipole induces a local electric field, which 
screens the effect of the external electric field. This is usually referred to as local field effects. The same mechanism is responsible for the appearance of a depolarization field at the surface of a material driven by an out-of-plane electric field~\citep{le2018high, tancogne2018atomic}. It is important to stress that this induced dipole is expected to play a significant role here, due to the $\sigma$-character of the vacancy wavefunctions.
The induced electric-field is directly related to the electron-electron interaction term 
as clearly shown in Ref.~\citep{le2018high}. 
As shown in Fig.~\ref{fig5.6}(b), the harmonic yield at higher energies is increased
if we neglect local field effects, i.e., we treat electrons at the IP approximation.  
Moreover, the electron-electron interaction also affects the third harmonic, as shown in  Fig.~\ref{fig5.6}(b), but lead to an increase of the harmonic yield. We, therefore, attribute this effect not to local field effects but to correlation effects.
It is important to stress that in Maxwell equations, the source term of the induced electric field is the induced (summed over spin) density. Therefore, the induced electric field acts equally on the HHG from both spin channels, which is why the decrease of the yield occurs equally on both spin channels, see Fig.~\ref{fig5.2}(b). 

We conclude that the modifications in the HHG spectrum, due to a point defect, are originating from a complex interplay of two important contributions: One due to the electronic transitions including the in-gap defect states, and second arising from the electron-electron interaction.  
This indicates that HHG in defected-solids  can not be fully addressed within 
IP approximation, as this can, for some cases at least, lead to a wrong qualitative prediction. 

\begin{figure}[t!]
	\centering
	\includegraphics[width= 0.65\textwidth]{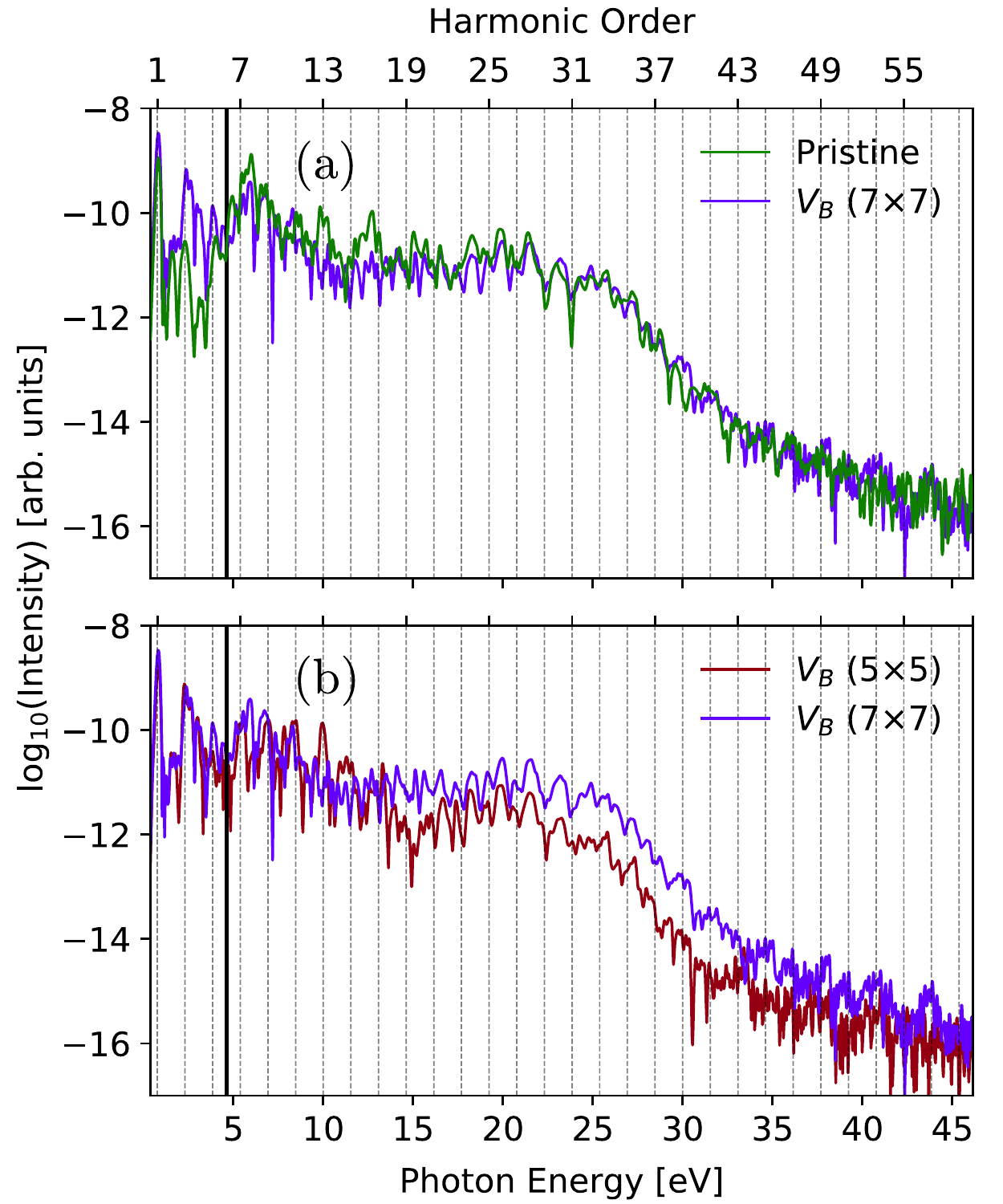}
	\caption{ High-harmonic spectrum of $V_B$ in a 7$\times$7 supercell is compared with the harmonic spectrum of (a) pristine h-BN and (b) $V_B$ in a 5$\times$5 supercell. The black vertical line presents the energy band-gap of pristine h-BN.} 
	\label{fig5.7}
\end{figure}

Finally, we discuss the dependence of the defect concentration on the HHG spectrum by computing the HHG for  a 7$\times$7 supercell with a boron vacancy, which corresponds to a $\sim$1\% doping concentration. The harmonic spectrum is presented in Fig.~\ref{fig5.7}. The third harmonic enhancement persists with comparable intensity even with a lower defect
concentration [Fig.~\ref{fig5.7}(b)], whereas the higher-energy region of the harmonic spectrum is matching well with the
pristine spectrum [Fig.~\ref{fig5.7}(a)]. This is consistent with the observation made that the
higher-energy spectrum  is dominated by the bulk bands. The effects due to electron-electron interaction diminishes for weaker defect concentration, while the contributions from the gap-states are consistent. This indicates that some of the effects we found here depend on the defect concentration and might not be observed below a certain concentration threshold.

\subsection{HHG from Hexagonal Boron Nitride with a Nitrogen Vacancy}

\begin{figure}[t!]
	\includegraphics[width=13 cm]{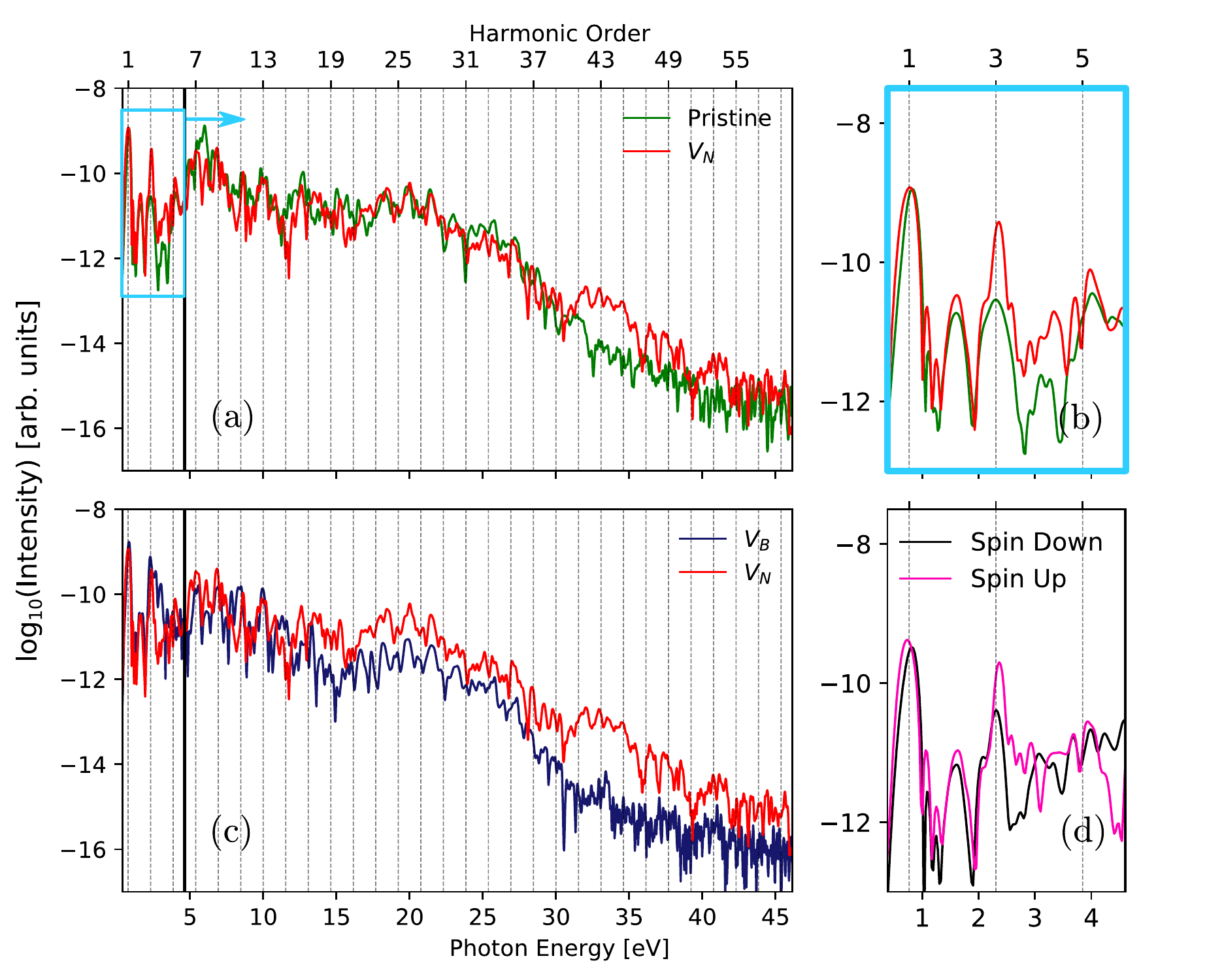}
	\caption{ High-harmonic spectrum of (a) nitrogen vacant h-BN ($V_N$) and pristine
		h-BN, (b) $V_N$ and $V_B$, and (c) below band-gap harmonic spectrum of $V_N$ and  
		pristine h-BN. (d) Spin-resolved harmonic spectrum for $V_N$ in the below band-gap 
		region. The black vertical line represents the energy band-gap of pristine h-BN. } 
	\label{fig5.8}
\end{figure}

\begin{figure}[t!]
	\includegraphics[width=15 cm]{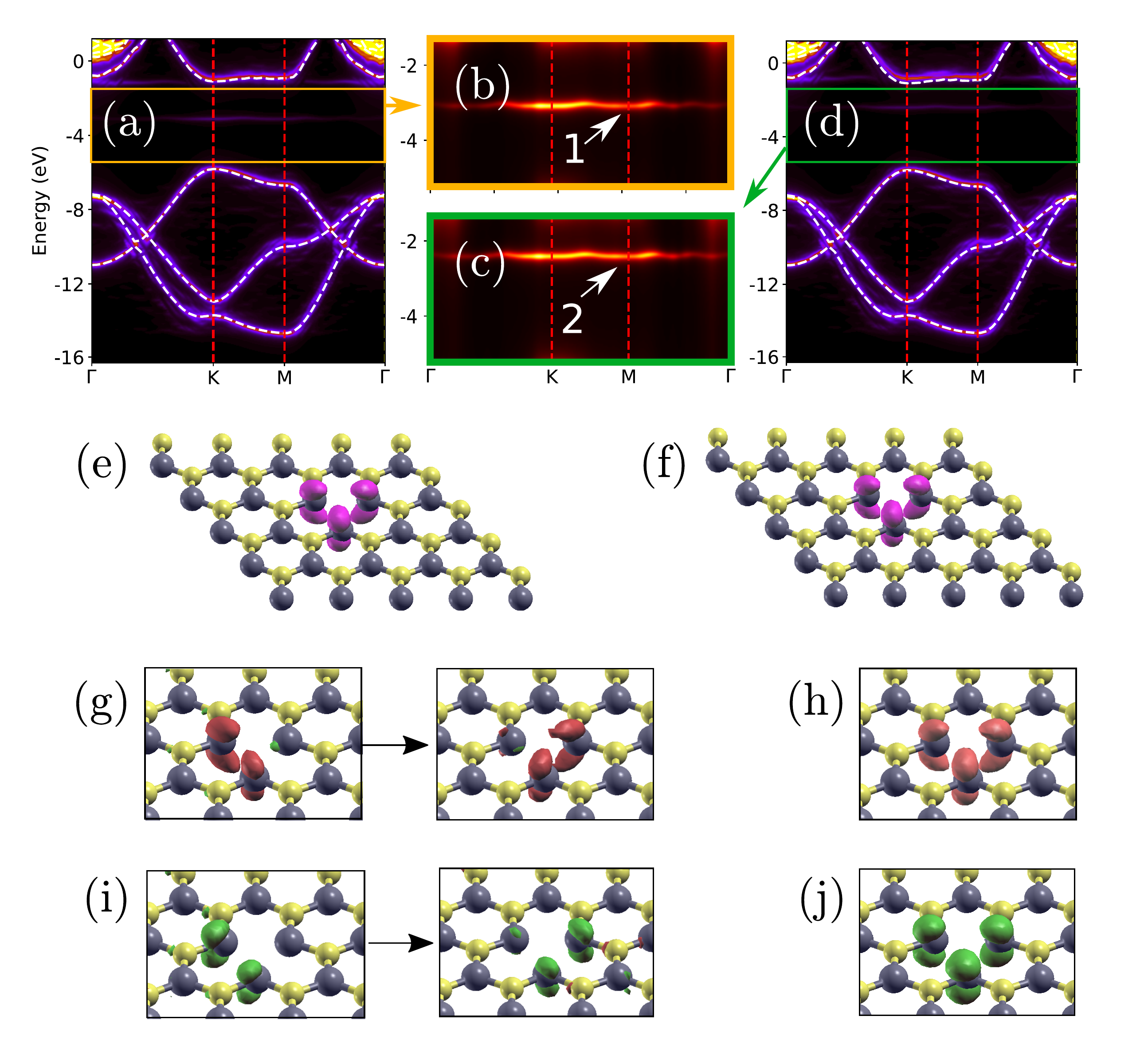}
	\caption{ 
		Spectral function for (a) spin-up and (d) spin-down components of $V_N$, The pristine band-structure is plotted with white dashed lines for reference. The in-gap portion of the spectral function for spin-up and spin-down components are zoomed respectively in (b) and (c)  
		where in-gap vacancy states are recognised  by 1 (for spin-up); and by 2 
		(for spin-down). 
		(e)-(f) The absolute wavefunctions of the in-gap vacancy states 
		1 and 2 in (a) and (b), respectively.
		Snap shots of the time-evolving $n_{ind}$ near the peak of the vector potential for (g) spin-up and (i) spin-down channels; and 
		after the end of the vector potential  for (h) spin-up and (j) spin-down channels 
		in $V_N$. The  vector potential is shown in Fig.~\ref{fig5.5}(b). Green and red colours in the induced electron density stand for positive and negative values, respectively.} 
	\label{fig5.9}
\end{figure} 

\begin{figure}[t!]
	\centering
	\includegraphics[width= 0.65\textwidth]{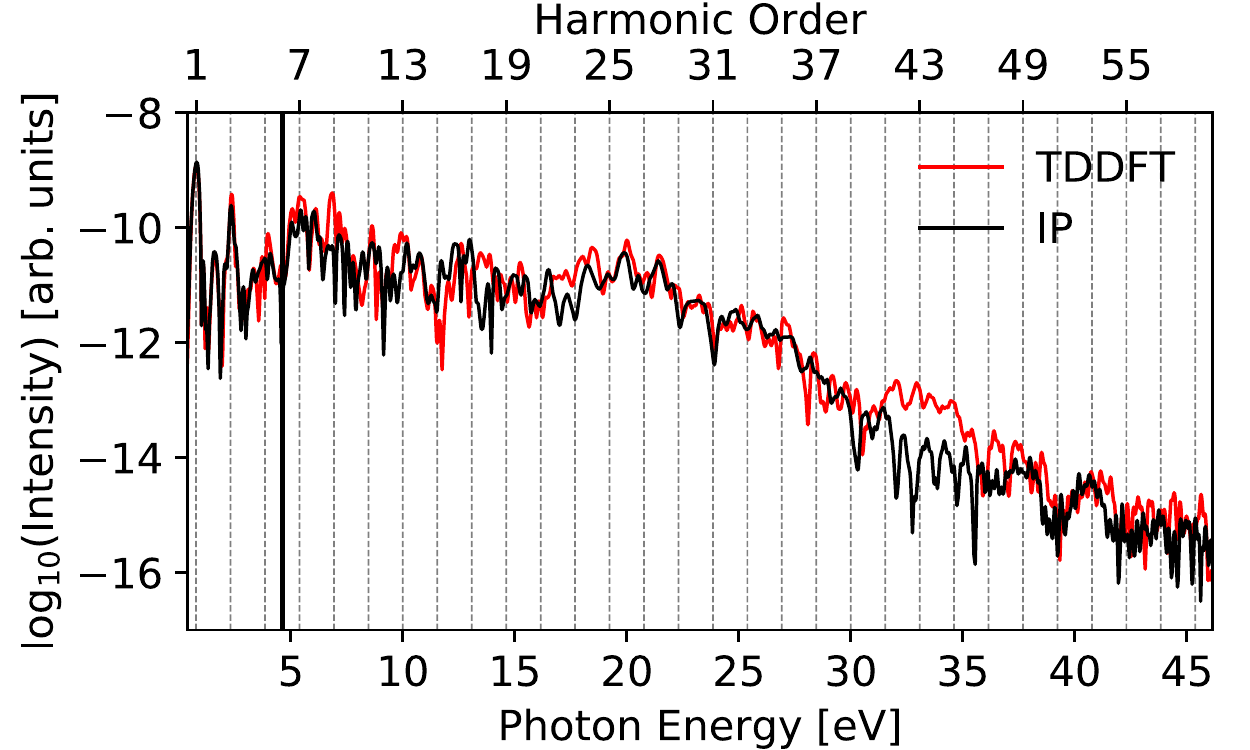}
	\caption{High-harmonic spectrum within TDDFT and IP approximation 
	are compared for nitrogen-vacant h-BN ($V_N$). 
	The black vertical line presents the energy band-gap of pristine h-BN.} 
	\label{fig5.10}
\end{figure}

So far, we have discussed  HHG in $V_B$. 
We now explore HHG in h-BN with a nitrogen vacancy ($V_N$).
Similar to the $V_B$ case, $V_N$ is also spin polarised.
Fig.~\ref{fig5.8}(a) presents HHG spectrum 
of $V_N$ and 
its comparison with the spectrum of pristine h-BN. 
As apparent from the figure, the spectrum  of $V_N$ 
resembles more to the pristine spectrum except in the below band-gap regime, and for an increase of the yield between 30 to 35 eV.
All the laser parameters are identical to the previous case of $V_B$. 
Naively, the spectra of $V_N$ and $V_B$ could be expected to look similar as one atom 
from the pristine h-BN has been removed in both the situations. 
However, this is not the case, as evident from Fig.~\ref{fig5.8}(b) and we note that the high-energy part of the spectrum of $V_N$ is much closer to the one of pristine than $V_B$, except above 30 eV.  The spectra of 
$V_N$ and $V_B$ are also fairly different. 
On close inspection of the  below band-gap spectrum, one finds that the 
third harmonic in  $V_N$ is significantly enhanced  with respect to  the pristine case
[see Fig.~\ref{fig5.8}(c)]. HHG in $V_B$ portrayed the identical observation. 
To know whether the reason behind this enhancement is identical to $V_B$ or not, let us 
analyse the spin-resolved harmonics in $V_N$. 
The below band-gap spin-resolved spectrum 
reveals that the third harmonic has more contribution from the spin-up electron than the spin-down electron [see Fig.~\ref{fig5.8}(d)]. This is completely opposite to the findings in $V_B$ where major contribution originated from the spin-down electron. 
To understand this behaviour, let us explore the ground-state band-structure of $V_N$. 

The unfolded energy band-structure of $V_N$ is presented in 
Fig.~\ref{fig5.9}. 
Each of the three boron atoms has an unpaired electron after the removal of a nitrogen atom 
from pristine h-BN.   
One spin-up and one spin-down vacancy states within the band-gap are emerging and are 
closer to the conduction band [see Fig.~\ref{fig5.9}(a)-(d)]. 
One more defect state  is further pushed towards the conduction band in each case. The spin-up defect state is found to be occupied. This makes $V_N$ a single donor \citep{huang2012defect}. The $p_{z}$ states of the boron atoms, which are in the vicinity of a nitrogen vacancy,  contribute to the defect states.  
This gives a $\pi$-character to the defect wavefunctions  [Fig.~\ref{fig5.9}(e),(f)]~\citep{huang2012defect,azevedo2009electronic}.  
Similar to $V_B$, only gap states are analysed. 
The spin-up defect level is 
occupied and close to the conduction band, which explains 
the  major contribution of the spin-up electron to the third harmonic in $V_N$.  
Electrons in this state can easily get ionised to 
the conduction bands and add more spectral weight  to the third harmonic. 
Note that, unlike $V_B$, the local symmetry in $V_N$ is preserved, which also explains why the HHG spectrum from $V_N$ is close to the HHG from pristine h-BN. 

The unfolded band-structures of  $V_N$ and $V_B$ mainly
explain the significant enhancement of the third harmonic and its different spin-polarised nature.
To explain the overall difference in the harmonic spectrum,  
the snap shots of $n_{ind}$ in $V_N$ at different times along the vector potential are presented in Fig.~\ref{fig5.9}(g)-(j). 
In  real-space picture, defect contribution is coming from the localised induced density, which
can be seen from the integral in Eq.~(\ref{current}). 
The depletion of the spin-up defect state as well as induced electron density in the spin-down defect state can be observed at the end of the pulse [see Fig.~\ref{fig5.9}(h),(j)].

In comparison to $V_B$, the effect of screening due to the local field effects is weaker in the case of $V_N$. The weaker impact of local field effects on $V_N$, compared to $V_B$ can be attributed to the following two reasons: 1) the induced charge density has a pronounced localised component around the nitrogen vacancy. As evident from Figs.~\ref{fig5.9}(e), and \ref{fig5.9}(f), the wavefunctions of spin-up and spin-down electrons have similar spatial structure, whereas the corresponding induced charge densities have opposite sign [see Fig.~\ref{fig5.9}(g)-(j)]. This partial cancellation of the spin-resolved induced charge density makes the total induced charge to be much lower and results in a weaker local field effects. 2) For the in-plane laser polarisation, the induced dipole due to the $\pi$-like defect states get much less polarised than for the  $\sigma$-like defect states. This results in weaker screening in $V_N$ than $V_B$.

The weaker local field effects in $V_N$ is fully consistent with the HHG spectra of $V_N$ within TDDFT and IP approximation as presented in Fig.~\ref{fig5.10}. Here, the third harmonic enhancement is well captured within IP approximation, but the increase in the yield between 30\,eV to 35\,eV is found to 
originate from the electron-electron interaction. 
Finally, the defect states plays a role in the HHG spectrum of $V_N$ even at higher-orders, 
though these effects are feeble.

The total spectrum includes the contributions  from the energy-bands of  
pristine h-BN as well as from transitions including gap states in  $V_B$ and $V_N$ case.  
In both cases, we found the significant and different role of the electron-electron interaction.
The harmonic spectrum of the defected solid preserves some piece of information of the pristine structure in the higher energy regime 
along with the characteristic signatures of the defect in the near band-gap regime, or close to the cutoff energy for $V_N$.

\section{Summary}
In summary, we have investigated the role of vacancy-defect in solid-state HHG. 
For this purpose, h-BN with a boron or a nitrogen atom vacancy is considered. 
In a simple minded picture, 
one may think  that h-BN with a boron atom vacancy 
or with a nitrogen atom vacancy would exhibit similar HHG spectra since a 
single atom from h-BN has been removed. However, this is not the case as boron and 
nitrogen vacancies lead to qualitatively different electronic structures, and this is visible from their corresponding 
gap states. It has been found that by removing  an atom from h-BN, either boron or nitrogen, the system becomes spin-polarised with non-zero magnetic moment near the vacancy. As a consequence, the defect-induced gap states are found to be different for each spin channel and for each vacancy, which we found to be strongly reflected in the low-order harmonics.  These contributions are strongly spin-dependent, depending on the ordering and occupancy of the defect states. Altogether, we can understand the role of the defect states by analysing the spin-polarised spectra, and the findings are in accordance with the spin-polarised band-structure. This establishes one aspect of the role of defect states in strong-field dynamics in solids.

In addition, the vacancy wavefunctions of $V_B$ and $V_N$ show $\sigma$ and $\pi$-characters, respectively, which lead to different qualitative changes in the harmonic spectra of vacancies, due to the local-field effects and electron correlations. 
These different behaviours are stemming 
through the creation or not of an induced dipole, which may counteract the driving electric field, and directly depends on the spatial shape of the defect-state wavefunctions.
Moreover, the electron-electron interaction also manifests itself in the decrease 
of the harmonic yield close to the energy cutoff in the $V_B$ case, whereas this effect is completely absent in the $V_N$ or pristine cases.
This implies that the nature of the vacancies in $V_B$ and $V_N$ are entirely different, and this is reflected in their HHG spectra, even at a defect concentration as low as 2\%. 
The HHG spectrum of $V_N$ is similar to the pristine h-BN, 
whereas the spectrum of $V_B$ differs significantly. These effects essentially imply that some defects are more favourable to modify the HHG spectra of the bulk materials. Thus, opening the door for tuning HHG by defect engineering in solids.

From the present work, we can also speculate what would be the effect of other known defects in h-BN.
If one considers the doping impurity instead of vacancy, e.g., carbon impurity, the band-structure of h-BN remains spin-polarised in nature near the 
band-gap~\citep{attaccalite2011coupling,huang2012defect}.  If a boron atom is replaced by a carbon atom, one occupied spin-up and one spin-down defect levels appear. Both the defect levels are near the conduction band with the wavefunctions contributed from the $p_{z}$ orbitals of carbon and nearby nitrogen atoms. Therefore, the carbon doping defect is expected to show qualitatively similar behaviour in the HHG spectrum as we have seen in the case of $V_N$. On the other hand, two defect states with occupied spin-up and unoccupied spin-down states appear near the conduction bands in the case when nitrogen is replaced with a carbon atom.  The wavefunction in this case is contributed mostly from the $p_{z}$ orbitals of the carbon as well as nearby boron atoms. In this case, the  enhancement in the below band-gap harmonics  is expected due to the defects states near the valence bands similar to $V_B$. However, the effect of screening is expected to be lower compared to $V_B$ as the nature of wavefunctions here is similar to that of $ V_N$.

In the case of bi-vacancy in h-BN, there are occupied defect state near the valence band and two unoccupied defect-states near the conduction band  as presented in Ref.~\citep{attaccalite2011coupling}. In this situation, if the separation between defect-states and nearby bands is small compared to the photon energy, then no significant changes in the below band-gap harmonics are expected.

Let us finally comment on the possibility of performing imaging of polarised defects in solids using HHG. As spin channels are not equivalent in the studied defects, one might think about using circularly polarised pulses to probe each spin channels independently.
However, in the limit of dilute magnetic impurities, as studied here, a crystal will host as 
many defects with positive magnetic moments and negative ones, and the signal for up or down spin channel will appear after macroscopic average as identical.

\cleardoublepage
\chapter{Conclusions and Future Directions}

In this thesis, we have presented how HHG can be used as a method of choice to understand  
static and dynamical electronic properties in 2D materials. Also, we have proposed how ultrafast laser can be used for novel applications in fundamental science and technology. 
We have considered 2D materials with hexagonal symmetry in this thesis, and  
laser polarisation is deemed in-plane.		

The non-perturbative electron dynamics in the presence of an intense laser field is modelled by solving TDSE with reliable approximations. 
Semiconductor Bloch Equation  is used to solve TDSE within a single-active electron approximation. We have also solved many-electron TDSE incorporating electron-electron interaction within the framework of TDDFT. 

In \textbf{Chapter 5}, we have demonstrated  electron-electron interaction is less critical for pristine h-BN when the laser polarisation is in-plane. Moreover, a recent work offered the same conclusion for graphene, a semimetal~\citep{li2021ab}. 
This adds reliability  to the single-active electron approximation, which is also widely employed 
for HHG from atoms. 
It is important to note that TDDFT calculations are  computationally expensive. 
To get an idea about computational cost, TDDFT calculation for h-BN requires approximately 7000 core hours, whereas the same for a two-band model of gapped-graphene takes less than a core hour.  
Nevertheless, it is essential to list out the cases in which electron-electron interaction is critical. 
One such situation we reported in this thesis is the electron-dynamics in a defective material.

The electron dynamics in semimetals such as monolayer and bilayer graphene shows qualitatively similar behaviour compared to gapped-graphene, as discussed in \textbf{Chapter 3}. 
A signature of interlayer coupling in  HHG from bilayer graphene is established. 
HHG spectrum of monolayer graphene is comparable to that of AA-stacked bilayer graphene, 
whereas it is significantly different for  AB-stacked bilayer graphene. 
This signifies that the high-harmonic spectrum depends on the electronic band structure and the interband coupling between different energy-bands. 
We confirm that one of the signatures of electron-dynamics in a semimetal is the anomalous driver-ellipticity dependence.

The high-harmonic spectrum of gapped graphene can be easily distinguished from gapless graphene, with the presence of even harmonics. 
This is a consequence of the broken inversion symmetry in gapped graphene. Here, the contribution of the Berry-connection term is essential, especially close to the band-gap region. When the laser is polarised along the $\Gamma$-$\mathbf{K}$ direction, even harmonics are generated perpendicular to the laser polarisation, due to non-zero Berry curvature. 
On the other hand, when the laser is polarised along the $\Gamma$-$\mathbf{M}$ direction, even harmonics are parallel to the laser polarisation. 
The selection rules for even harmonics can be understood using simple symmetry considerations. Moreover, as the band-gap increases, the number of excited electrons in the conduction band decrease due to the inverse relation of interband transition and band-gap. 

Understanding the electron dynamics enabled us to propose new methods for technological applications in \textbf{Chapter 4}. Hexagonal 2D materials have conduction band minima in two inequivalent points in the reciprocal space, termed $\mathbf{K}$ and $\mathbf{K}^{\prime}$ valleys.  In gapped-graphene materials, valley-selective excitations can be achieved by circularly polarised light, resonant with the band-gap. 
This property is the basis of valleytronics. Due to vanishing band-gap and Berry curvature, 
resonant excitation by chiral light is not applicable to pristine graphene. 
We proposed a light-induced approach of valleytronics in pristine graphene, using a tailored field.  
In ultrafast optics, tailored field is used to obtain the desired control over electron dynamics in solids~\citep{heinrich2021chiral}. 

We have used a combination of counter-rotating circularly polarised fields with frequencies $\omega$ and 2$\omega$ to achieve valley-polarisation in pristine graphene.  
The resultant total field has a trefoil structure, which breaks the inversion symmetry of graphene. The energy bands near adjacent $\mathbf{K}$-points are related by mirror inversion. 
Therefore, the driving field leads to different electron dynamics in different $\mathbf{K}$-points, resulting in valley-selective excitation and HHG. The valley-polarisation can be controlled by 
changing the sub-cycle phase delay between the two pulses as the trefoil is rotated. 
We observed considerable valley-polarisation for strong and long-wavelength laser fields as electron traverse the anisotropic regime of the Brillouin zone. Moreover, a third harmonic probe pulse is used to measure the valley-polarisation. Here, even harmonics of the probe pulse is a measure of broken inversion symmetry, whereas which valley is polarised is embedded in the phase of these 
even harmonics.  

Due to semimetallic nature of graphene, there is a strong interplay of interband and intraband contributions to the total harmonic spectrum. 
Interestingly, relative contributions of interband and intraband terms in graphene strongly depend on the polarisation of driving laser.  As we have seen in \textbf{Chapter 3}, the characteristic driver-ellipticity dependence in HHG originates from the interband contribution. On the other hand, the dominating contribution is intraband for a bicircular driving field, as shown in \textbf{Chapter 4}.  
Therefore, it is interesting to explore the driver-polarisation dependence on the underlying mechanism for different tailor fields. 
Furthermore, electron dynamics and dynamical symmetries in 2D materials can be investigated using  tailored fields in future. 

So far, we have considered different aspects of electron dynamics in pristine materials. 
In \textbf{Chapter 5}, we analysed the effect of spin-polarised defects on HHG. 
This study is of fundamental importance in experiments. In particular, we have investigated the impact of vacancy defects in monolayer hBN. 
The HHG spectrum exhibits signatures of the defect as well as preserves most of the bulk properties of pristine h-BN. 
As shown in \textbf{Chapter 2}, using a periodic model potential, the harmonic spectrum and 
band structure are related. For real materials, correlating HHG directly with band structure is challenging. However, any change in the effective band structure should be imprinted in the harmonic spectrum. 
One such instance is mono-vacancy defect (boron or nitrogen vacancy) in h-BN. 
As a result of vacancy defect in h-BN, there are spin-polarised defect states within the band-gap. 
These defect states result in the enhancement of the below band-gap harmonics, 
which are also spin-polarised in nature. 

We have shown that electron-electron interaction in essential to describe electron dynamics in 
h-BN with vacancy defect. 
Interband transitions to (or from) defect states result in the induced charge near vacancy.  
Local-field effects stemming from this induced charge is the reason for the importance of 
the electron-electron interaction. 
This effect also depends on the shape of the defect wavefunctions and defect concentration. 

Our work opens interesting perspectives for further studies on strong-field electron dynamics in 2D and extended systems, especially involving isolated defects. Further works should address the possibility to monitor the electron-impurity scattering using HHG, more complex defects such as bi-vacancies, and a practical scheme for imaging buried defects in solids.

\cleardoublepage
\end{spacing}
\begin{spacing}{1.3}
\addcontentsline{toc}{chapter}{Bibliography}
\bibliographystyle{abbrvnat.bst}
\bibliography{solid_HHG}
\cleardoublepage
\end{spacing}
%
\cleardoublepage \cleardoublepage
\addcontentsline{toc}{chapter}{List of Publications}
\markboth{List of Publications}{List of Publications}
\newpage
\thispagestyle{empty}
\begin{center}
\vspace*{-0.4cm} {\LARGE {\textbf{List of Publications}}}
\end{center}
{\setlength{\baselineskip}{8pt} \setlength{\parskip}{2pt}
\begin{spacing}{1.5}
\vspace*{.7cm} \noindent{\bf \large A. Part of this thesis} \vspace*{.7cm}
\begin{enumerate}
	\item  \textbf{M. S. Mrudul}, and Gopal Dixit, ``High-harmonic generation from monolayer and bilayer
		graphene'', {\bf Physical Review B}
	103, 094308 (2021).
	\item \textbf{M. S. Mrudul}, Alvaro Jimenez-Galan, Misha Ivanov, and Gopal Dixit,    ``Light-induced
	valleytronics in pristine graphene'' \textbf{Optica} 8, 422 (2021).
	\item \textbf{M. S. Mrudul}, Nicolas Tancogne-Dejean, Angel Rubio, and Gopal Dixit, ``High-
	harmonic generation from spin-polarised defects in solids'', \textbf{npj Computational Materials}
	6, 1 (2020).
	\item \textbf{M. S. Mrudul}, Adhip Pattanayak, Misha Ivanov, and Gopal Dixit, ``Direct numerical
	observation of real-space recollision in high-order harmonic generation from solids''
	\textbf{Physical Review A} 100, 043420 (2019).
	\item \textbf{M. S. Mrudul}, and Gopal Dixit, ``A detailed understanding of high-order harmonic
	generation from gapped-graphene materials'' (under preparation).
\end{enumerate}

\vspace*{.7cm} \noindent{\bf \large B. Not part of this thesis} \vspace*{.7cm}
\begin{enumerate}
	\item Adhip Pattanayak, \textbf{M. S. Mrudul}, and Gopal Dixit. ``Influence of vacancy defects in solid high-
	order harmonic generation'' \textbf{Physical Review A} 101, 013404 (2020).
	\item Irfana N. Ansari, \textbf{M. S. Mrudul}, Marcelo F. Ciappina, Maciej Lewenstein, and Gopal Dixit.
	`` Simultaneous control of harmonic yield and energy cutoff of high-order harmonic generation
	using seeded plasmonically enhanced fields'' \textbf{Physical Review A} 98, 063406 (2018).
\end{enumerate}

\newpage
\thispagestyle{empty}
\begin{center}
	\vspace*{-0.4cm} {\LARGE {\textbf{Conferences Attended}}}
\end{center}

\vspace*{.7cm} \noindent{\bf \large A. International}
\vspace*{.7cm}

\begin{enumerate}
	\item \textbf{PALM International School}, Paris, France (28 May - 01 June, 2018).
	\item \textbf{ATTO2019}, Szeged, Hungary (01 - 05 July, 2019).
	\item \textbf{QUTIF International Conference}, online (22 - 25 February 2021).
	\item \textbf{CLEO/Europe-EQEC 2021}, online (21 - 25 June 2021).
\end{enumerate}

\vspace*{.7cm} \noindent{\bf \large B. National}
\vspace*{.7cm}

\begin{enumerate}
	\item \textbf{UFS 2017}, University of Hyderabad (02 - 04 November, 2017).
	\item \textbf{13$^\textrm{th}$ AISAMP}, IITB \& TIFR, Mumbai (03 - 08 December 2018).
	\item \textbf{Symphy 2019}, IITB (16 - 17 March 2019).
	\item \textbf{22$^\textrm{nd}$ NCAMP}, IIT Kanpur (25 - 28 March 2019).
	\item \textbf{UFS 2019}, IITB (07 - 09 November, 2019).
\end{enumerate}

\end{spacing}

\cleardoublepage
\end{document}